\documentclass{article}
\usepackage[margin=1in]{geometry}
\usepackage{setspace}
\usepackage{lineno}
\usepackage{authblk}
\usepackage{caption}
\usepackage{xcolor}
\definecolor{NMblue}{HTML}{025E8D}
\usepackage[
  colorlinks   = true,
  linkcolor    = NMblue,
  citecolor    = NMblue,
  urlcolor     = NMblue
]{hyperref}
\usepackage{amsfonts,amssymb,amsmath,mathrsfs}
\usepackage{graphicx}
\usepackage{soul}
\usepackage{bm}
\usepackage{enumerate,algorithmicx,algorithm}
\usepackage{changepage}
\usepackage{comment}

\usepackage{tabularx}
\usepackage{hhline}
\usepackage{diagbox}

\usepackage[numbers,super,sort&compress]{natbib}
\newcommand{\autocite}[1]{\cite{#1}}

\usepackage{booktabs}

\usepackage{siunitx}
\newcommand{\be}{\begin{equation}}
\newcommand{\ee}{\end{equation}}
\newcommand{\ba}{\begin{aligned}} 
\newcommand{\ea}{\end{aligned}}
\newcommand{\bea}{\begin{eqnarray}} 
\newcommand{\eea}{\end{eqnarray}}

\usepackage{pdfpages}
\begin{document}
%TC:ignore 
\title{Prolate spheroidal wave functions enable fast and exponent-aware long-range machine learning interatomic potentials}

\author[1,3]{Jiuyang Liang}
\author[1]{Libin Lu}
\author[2,3]{Yajie Ji}
\author[1,*]{Shidong Jiang}

\affil[1]{Center for Computational Mathematics, Flatiron Institute, Simons Foundation, New York, New York 10010, USA}
\affil[2]{Department of Statistics and Data Science, Yale University, New Haven, CT 06511, USA}
\affil[3]{School of Mathematical Sciences, Shanghai Jiao Tong University, Shanghai, 200240, China}

\date{\today}

\maketitle

\begingroup
\renewcommand{\thefootnote}{\fnsymbol{footnote}}
\footnotetext[1]{Correspondence: sjiang@flatironinstitute.org}
\endgroup

\begin{abstract}
Long-range interactions such as electrostatics and dispersion remain a central bottleneck for machine learning interatomic potentials (MLIPs), especially in ionic, polar and interfacial systems. Ewald-based reciprocal-space mechanisms provide a physically grounded route for capturing these nonlocal effects, but often require dense Fourier grids and can become memory-limited at scale. This problem is particularly pronounced in molecular dynamics, where high efficiency requirements make accurate long-range modelling particularly costly. Here we introduce PSWF-LR, an exponent-aware long-range framework based on prolate spheroidal wave functions (PSWFs) that can be easily incorporated into existing model architectures. Its core components are PSWF-based mollification and atom-grid spreading, which enable compact and efficient representation of arbitrary inverse-power channels $1/r^p$ while treating the decay exponent as a physical prior. Across diverse long-range benchmarks, PSWF-LR reduces Fourier-mode requirements, improves energy and force accuracy, accelerates production-level simulations by about threefold, and extends long-range MLIP simulations beyond the memory limits of conventional MLIPs. 

\end{abstract}

Machine learning interatomic potentials (MLIPs) have transformed atomistic simulation by providing first-principles-level energy and force predictions at substantially lower cost than ab initio methods, enabling molecular dynamics (MD) across chemistry, physics and materials science~\cite{kalita_machine_2025,friederich2021machine,unke2021machine,deng2026potential}. Most existing MLIPs, however, are built on a local-neighborhood approximation~\cite{kohn1996density,zhang2018deep}: each atomic energy is inferred from its environment within a finite cutoff. This assumption is effective for short-range (SR) bonding and local structural relaxation, but it becomes limiting in ionic, polar and interfacial systems~\cite{ko2023recent}, where electrostatics, induction and dispersion decay slowly and can shape the potential energy surface (PES) over long distances. Capturing these nonlocal effects therefore requires not only accurate local descriptors, but also a physically grounded and computationally efficient mechanism for representing long-range (LR) interactions.

Existing strategies address this issue in several ways. Early approaches add empirical force-field corrections with fixed point charges~\cite{bartok2010gaussian,niblett2021learning}, but fixed charges cannot generally describe environment-dependent charge transfer, and such corrections are not readily available for many systems. Message-passing neural networks (MPNNs)~\cite{schutt2017schnet,batatia2022mace,batzner20223,batatia2025design} extend the receptive field through repeated communication across interaction shells, but slowly decaying interactions can require large cutoffs or many rounds of message passing. Another important class of methods predicts charges, dipoles or related latent variables, often under global charge constraints or charge equilibration~\cite{unke2019physnet,ko2021fourth,shaidu2024incorporating,gong2025predictive,rappe1991charge}. Examples include 4G-HDNNP~\cite{ko2021fourth}, BAMBOO~\cite{gong2025predictive} and CHGNet~\cite{deng2023chgnet}, which learn charge-informed representations, and DPLR~\cite{zhang2022deep} and SCFNN~\cite{gao2022self}, which introduce virtual sites or self-consistent electronic variables. After such LR variables are obtained, their LR interactions are usually evaluated by Ewald summation~\cite{ewald1921ap}, fast Fourier transform (FFT)-accelerated particle-mesh variants~\cite{darden1993jcp,loche2025fast,nufft6,barnett2019sisc}, or self-consistent procedures~\cite{ewald1921ap,darden1993jcp}. Other approaches avoid explicit charge prediction: Ewald message passing~\cite{kosmala2023ewald} uses learnable reciprocal-space filters, LODE~\cite{grisafi2019incorporating,huguenin2023physics} and density-based methods~\cite{faller2024density} encode LR information through inverse-power descriptors, latent Ewald summation (LES)~\cite{cheng2025latent,king2025machine} learns effective LR variables from energies and forces, and more flexible LR tails can be represented through explicit polarization models~\cite{gao2025foundation} or learned with sum-of-Gaussians neural networks~\cite{ji2025machine,ji2026accurate}.

Despite these advances, accurately and efficiently modeling LR interactions in MLIPs remains far from settled. In many LR MLIPs, once charges, virtual sites or latent LR variables are predicted, the nonlocal electrostatic contribution is often included through an Ewald-type mechanism, in which Gaussian screening softens the singular $1/r$ interaction and leaves a smooth reciprocal-space term to be computed on Fourier grids. Ewald-type treatments are physically interpretable because they retain the Coulomb LR decay, and particle-mesh variants reduce the asymptotic cost to near $O(N\log N)$. In practical MLIP workflows, however, local-environment cutoffs are often short, typically about $4$--$6~\mathring{\mathrm{A}}$, while the MD systems can be much larger than the training configurations. This mismatch can force dense reciprocal grids, costly particle–mesh operations and high memory demand, thereby slowing training and inference and limiting scalability~\cite{li2025scaling}. In addition, many LR MLIPs focus mainly on charge--charge interactions, whereas realistic atomistic systems can also include charge--dipole ($1/r^2$), dipole--dipole ($1/r^3$), induction-like and dispersion ($1/r^6$) contributions. These limitations point to the need for a compact, differentiable LR mechanism that can represent multiple physically motivated asymptotic decays within a common framework.

Here we introduce PSWF-LR, an exponent-aware framework for LR MLIPs based on prolate spheroidal wave functions (PSWFs). In PSWF-LR, an upstream SR model encodes local atomic environments and supplies suitable LR variables, such as latent variables, explicit partial charges, or ML-parameterized force field terms. Then, these variables are fed into inverse-power channels $1/r^p$, whose LR contributions are then evaluated by PSWF-based mollification and atom-grid spreading. PSWFs are eigenfunctions of the truncated Fourier transform and provide optimally band-concentrated representations on finite intervals~\cite{slepian1961prolate}. We use this property to replace the Gaussian screening used in Ewald-type treatments with compact PSWF-based mechanisms. This construction exploits the spectral concentration of the smooth LR component, allowing it to be represented with substantially fewer Fourier modes while retaining the prescribed inverse-power decay. 
In this formulation, the decay exponent $p$ becomes a tunable physical prior rather than a fixed consequence of the Gaussian Ewald ansatz.

We validate PSWF-LR across systematic accuracy benchmarks and production MD simulations. In synthetic LR systems, molecular dimers and polar dipeptides, PSWF-LR reduces Fourier-mode requirements and shows that the optimal decay exponent depends on the target observable. In charge-transfer benchmarks, it captures nonlocal charge redistribution and improves energy and force accuracy. We further compare the PSWF-based LR evaluation layer with differentiable Torch-PME and JAX-PME backends~\cite{loche2025fast}, and evaluate it in production MD workflows based on CACE with an Ewald LR treatment~\cite{king2025machine} and ByteFF-Pol~\cite{zheng_bridging_2026} with particle-mesh Ewald (PME)~\cite{darden1993jcp} in OpenMM~\cite{eastman2023openmm}. These tests show that PSWF-LR substantially accelerates LR evaluation and MLIP-based production MD while preserving the tested interfacial structure, charge response and electrolyte transport observables.

\section*{Results}
\subsection*{Overview of the PSWF Framework}

\begin{figure}[!ht]
\centering
\includegraphics[width=0.95\linewidth]{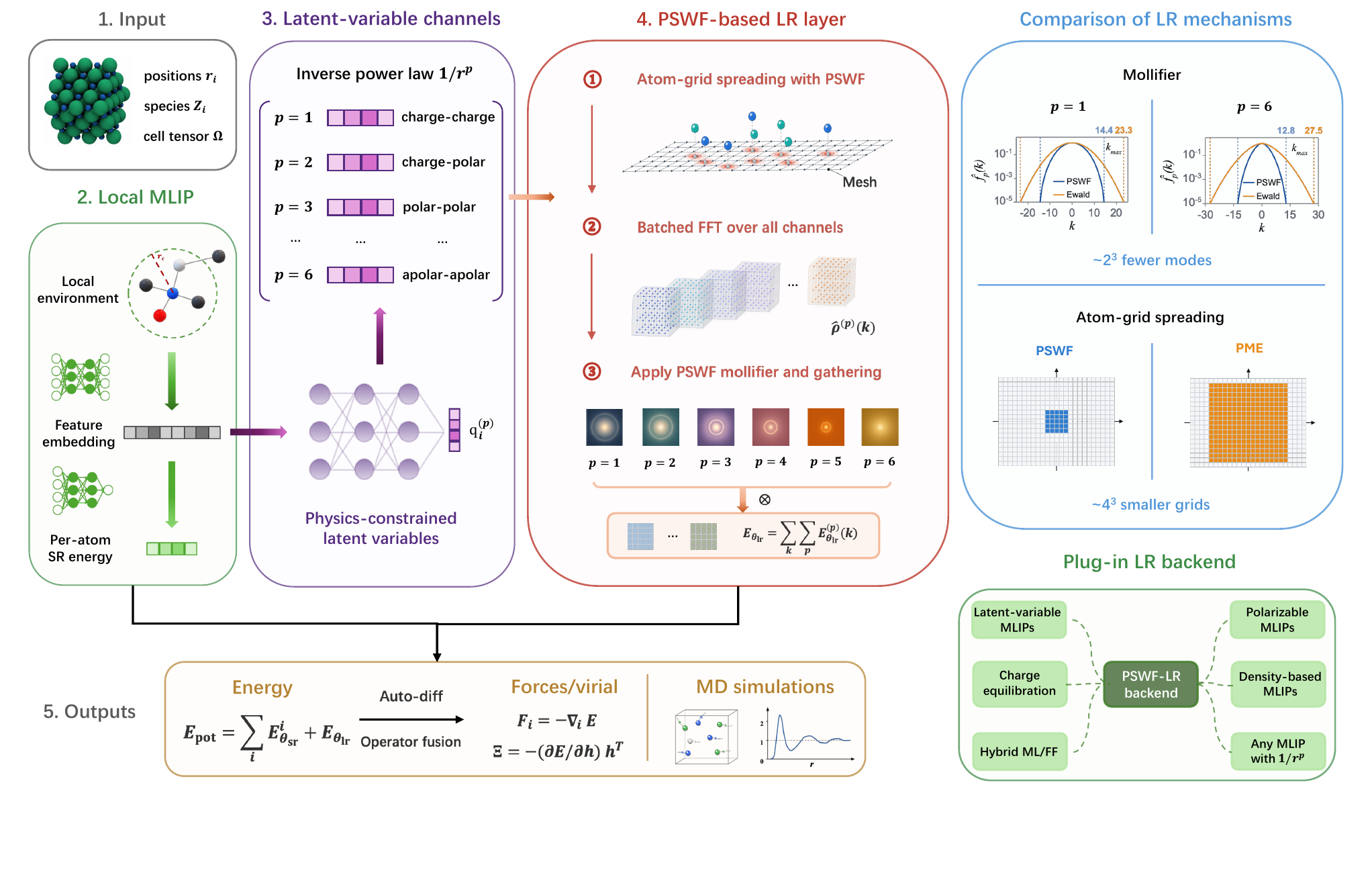}
\vspace{-3mm}
\caption{\sf Schematic overview of the PSWF-LR framework for exponent-aware long-range machine-learning interatomic potentials. Starting from atomic positions $\bm r_i$, species $Z_i$, and the cell tensor $\Omega$, a local short-range MLIP encodes each atomic environment and predicts short-range atomic energies together with features used to infer physics-constrained latent variables. These latent variables define inverse-power long-range channels $q_i^{(p)}$ associated with kernels $1/r^p$, covering charge--charge, charge--polar, polar--polar, and higher-order apolar interactions. The PSWF-LR backend spreads all latent variables onto a mesh with PSWF atom-grid windows, evaluates the reciprocal-space contributions through a single batched FFT over all channels, applies the corresponding PSWF mollifiers, and gathers the results to obtain the long-range energy. The total energy combines short-range and long-range terms, while forces and virials are obtained by automatic differentiation for molecular dynamics simulations. The upper-right panel compares PSWF-LR with Ewald/PME-style backends: PSWF mollifiers show faster reciprocal-space decay and PSWF spreading kernels are more localized, enabling fewer Fourier modes and substantially smaller grids when atom-grid spreading and FFT acceleration are used together. The lower-right panel highlights the modularity of PSWF-LR, which can be coupled to diverse LR-augmented MLIP workflows without changing their upstream representations.}
\label{fig:MLIPSketchMap}
\end{figure}

The PSWF framework is illustrated in Fig.~\ref{fig:MLIPSketchMap}. We consider periodic systems with Cartesian coordinates $\{\bm{r}_i\}$ and chemical species $\{Z_i\}$, and seek a model for the total potential energy $E_{\text{pot}}$ and forces $\bm{F}_i=-\partial E_{\text{pot}}/\partial \bm{r}_i$. As in many LR MLIPs, we decompose the total energy into a local SR contribution and a nonlocal LR correction,
\begin{equation}
E_{\text{pot}}=\sum_{i=1}^{N}E_{\bm{\theta}_{\text{sr}}}^{i}(\bm{D}_i)+E_{\bm{\theta}_{\text{lr}}}.
\end{equation}
The SR term is a sum of atomic energies predicted from local descriptors $\bm{D}_i$ within a cutoff $r_c$; these descriptors may be explicit, as in ACE~\cite{drautz2019atomic} and CACE~\cite{cheng2024cartesian}, or learned end to end, as in deep potential~\cite{zhang2018deep}, Allegro~\cite{musaelian2023learning}, and NequIP~\cite{batzner20223}.

For the LR contribution, we write the energy as a sum over inverse‑power channels,
\begin{equation}\label{eq::realLR}
E_{\bm{\theta}_{\text{lr}}}
=\sum_{p\in\mathcal{S}} E_{\bm{\theta}_{\text{lr}}}^{(p)}
=\frac{1}{2}\sum_{p\in\mathcal{S}}
\sum_{\bm{n}\in\mathbb{Z}^3}\sum_{i,j=1}^{N} q_i^{(p)} q_j^{(p)}\, 
\frac{\mathcal{G}_p(|\bm{r}_{ij}+\bm{n}\cdot\bm{L}|)}{|\bm{r}_{ij}+\bm{n}\cdot\bm{L}|^p}.
\end{equation}
Here $\mathcal{S}$ is a prescribed set of exponents, $\bm{r}_{ij}=\bm{r}_i-\bm{r}_j$, and $\bm{n}\in\mathbb{Z}^3$ enumerates periodic images. Each channel $p$ has its own latent variables $q_i^{(p)}$, predicted by a dedicated module (e.g., a multilayer perceptron that maps $\bm{D}_i$ to $\bm{q}_i$), and its own mollifier $\mathcal{G}_p(r)$, which regularizes the $1/r^p$ singularity. In a conventional Ewald treatment, $\mathcal{G}_p(r)$ is an incomplete‑gamma screening function (Eq.~\eqref{eq::screening} in Methods). In PSWF-LR, we instead derive $\mathcal{G}_p(r)$ from the order-zero PSWF $\psi_0^c(r)$
and its derivatives. For example, for charge-charge and dispersion channels ($\mathcal{S}=\{1,6\}$), we use
\begin{equation}\label{eq::G1}
\mathcal{G}_{1}(r)=\frac{1}{C_0}\int_{0}^{r/r_c}\psi_{0}^{c}(x)\,dx, 
\qquad C_0:=\int_{0}^{1}\psi_{0}^{c}(x)\,dx,
\end{equation}
and
\begin{equation}\label{eq::G6}
\mathcal{G}_{6}(r)=1-\psi_{0}^{c}(r/r_c)
+\frac{5r}{8r_c}\psi_{0}^{c}{}^{\prime}(r/r_c)
-\frac{r^2}{8r_c^2}\psi_{0}^{c}{}^{\prime\prime}(r/r_c)
\end{equation}
where $c>0$ is the mollifier parameter. The full construction, including the cases $p\neq 1,6$, is given in Methods.

To avoid the explicit lattice sum in Eq.~\eqref{eq::realLR}, we evaluate the LR term $E_{\bm{\theta}_{\text{lr}}}$ in Fourier space:
\begin{equation}\label{eq::Fourier}
E_{\bm{\theta}_{\text{lr}}}
=\frac{1}{2V}\sum_{p\in\mathcal{S}}\sum_{\bm{k}\neq \bm{0}}
\widehat{\Phi}_{p}(\bm{k})\,\big|\rho^{(p)}(\bm{k})\big|^2
+\sum_{p\in\mathcal{S}}E_{\bm{\theta}_{\text{lr}}}^{\bm{0},(p)},
\end{equation}
where $V=L_xL_yL_z$ is the cell volume, $\bm{k}=(2\pi m_x/L_x,\,2\pi m_y/L_y,\,2\pi m_z/L_z)$ are reciprocal lattice vectors, $\bm{m}=(m_x,m_y,m_z)\in\mathbb{Z}^3$, and $\widehat{\Phi}_{p}(\bm{k})$ is the Fourier transform of $\mathcal{G}_{p}(r)/r^p$. The latent structure factor is
\begin{equation}
\rho^{(p)}(\bm{k}) := \sum_{i=1}^{N} q_{i}^{(p)} e^{\mathrm{i}\bm{k}\cdot\bm{r}_{i}},
\end{equation}
and $E_{\bm{\theta}_{\text{lr}}}^{\bm{0},(p)}$ is the zero‑frequency correction. Explicit forms of $\widehat{\Phi}_{p}(\bm{k})$ and $E_{\bm{\theta}_{\text{lr}}}^{\bm{0},(p)}$ for general $p$ are given in Methods. In practice, the reciprocal-space sum is truncated at $\|\bm{k}\|\le K_{\text{max}}$. Only the latent variables are learned. The remaining LR hyperparameters, including $\mathcal{S}$, $c$, $r_c$, and $K_{\text{max}}$, are selected automatically from the target tolerance described in Methods. We further combine these kernels with PSWF-based atom-grid spreading and a single batched FFT, yielding an $O(N\log N)$ framework (see Methods and Fig.~1).

PSWF-LR provides an exponent-aware mechanism for representing LR interactions in MLIPs. PSWF-LR can be used either as a complete SR-plus-LR potential or as a replacement for Ewald-style LR components in existing LR MLIPs and ML-parameterized molecular-mechanics (MM) workflows. In the benchmarks below, we keep the upstream SR representation, latent-variable predictor or force-field parameters fixed whenever possible. This design allows us to test whether the PSWF-based LR mechanism itself provides a more compact and accurate representation of nonlocal interactions.

\subsection*{Random particle systems}
\begin{figure}[!ht]
\centering
\includegraphics[width=0.82\linewidth]{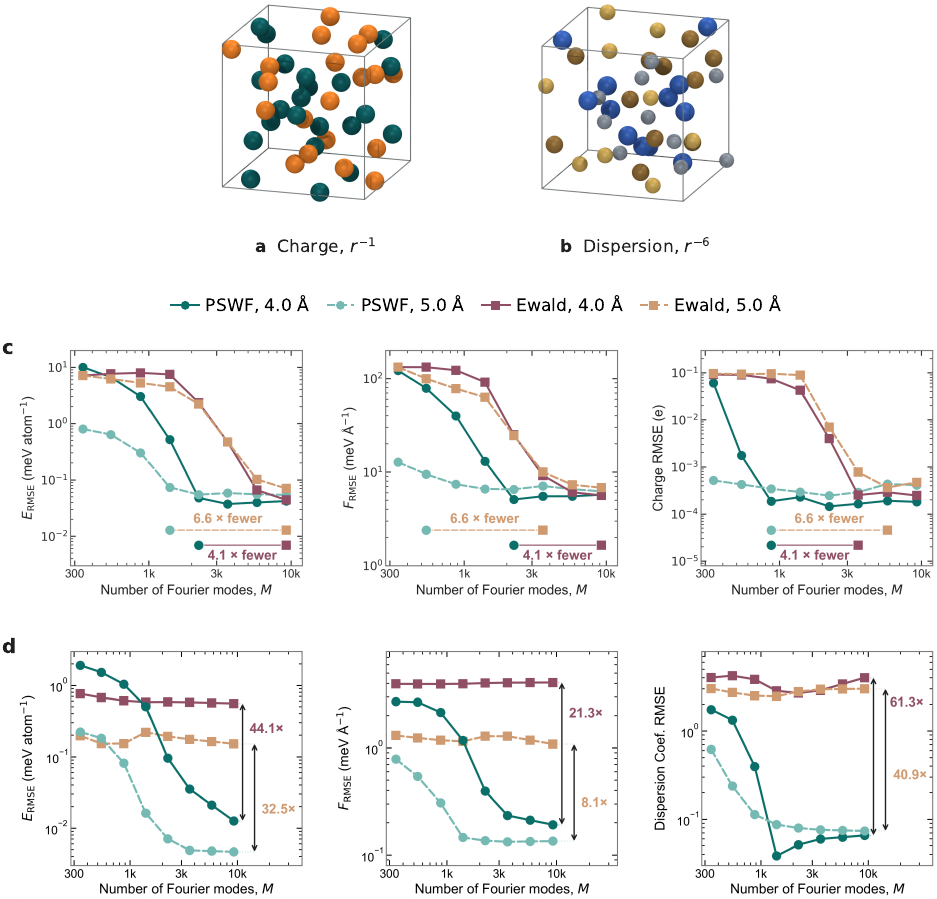}
\vspace{5mm}
\caption{\sf Benchmarking the PSWF-LR framework on model particle systems with long-range charge-charge ($r^{-1}$) and dispersion ($r^{-6}$) interactions. {\bf a,b}, Representative configurations of the charge-charge and dispersion systems, respectively; different colours denote different particle species. {\bf c}, Charge-charge benchmark: from left to right, the energy ($E_{\mathrm{RMSE}}$), force ($F_{\mathrm{RMSE}}$) and predicted-charge RMSEs as a function of the number of Fourier modes, $M$, for models built on the same short-range descriptor and augmented with either a PSWF or an Ewald long-range module. {\bf d}, Dispersion benchmark: from left to right, $E_{\mathrm{RMSE}}$, $F_{\mathrm{RMSE}}$ and the RMSE of the predicted dispersion coefficients, plotted against $M$ for the same model variants. Results are shown for short-range cutoffs $r_c = 4.0~\mathring{\mathrm{A}}$ and $5.0~\mathring{\mathrm{A}}$. Annotated arrows indicate the matched-error mode-count reduction ({\bf c}) or matched-mode error reduction ({\bf d}) of PSWF over Ewald.}
\label{fig:MLIPRandomCharge}
\end{figure}

We first isolated the spectral efficiency of the LR mechanism using two synthetic particle benchmarks with known LR physics: a Coulombic ($1/r$) system and a dispersion-like ($1/r^6$) system (Fig.~\ref{fig:MLIPRandomCharge}a,b). These systems remove most chemical complexity while retaining the central question: how many reciprocal-space modes are needed for an LR-augmented MLIP to recover both the nonlocal interaction and the latent variables that generate it?

Using the same SR descriptor and matched training settings, PSWF-LR reaches the plateau energy, force and charge-prediction errors with about $4$--$7\times$ fewer Fourier modes than the Ewald baseline in the Coulombic benchmark.
The advantage is even more pronounced in the dispersion benchmark, where, once the reciprocal-space sum is sufficiently resolved, PSWF-LR is consistently more accurate at matched reciprocal resolution and the matched-error reduction ranges from about $8\times$ to $64\times$, depending on the observable and cutoff. The same trend persists across training-set sizes (Extended Data Fig.~\ref{fig:MLIPRandomParticleSample}), indicating that the gain is not simply due to data efficiency but reflects a more compact reciprocal-space representation of the LR interaction itself. 

These results support the basic design principle of PSWF-LR. By replacing Gaussian screening with PSWF mollification, the smooth LR component can be represented with fewer reciprocal modes, especially for larger inverse-power exponents. This spectral compression reflects a lower-dimensional reciprocal-space representation of the LR interaction, rather than a loss of physical content. The optimal cutoff is not strictly monotonic, because the total MLIP error also contains SR representation error and statistical fitting error. In practice, cutoffs of about $4$--$5~\mathring{\mathrm{A}}$ provide an effective balance for these LR-augmented models.

\subsection*{Matching the asymptotic exponent improves extrapolation}
We next ask whether the inverse-power exponent assigned to PSWF-LR affects extrapolation accuracy in molecular systems. Molecular dimers from the BioFragment Database (BFDb)~\autocite{burns2017biofragment} provide such an interesting test because charged, polar and apolar monomers generate six classes of dimers with idealized asymptotic decays from $1/r$ to $1/r^6$: charge-charge (CC, $p=1$), charge-polar (CP, $p=2$), polar-polar (PP, $p=3$), charge-apolar (CA, $p=4$), polar-apolar (PA, $p=5$) and apolar-apolar (AA, $p=6$). The center-of-mass separations of the dimers span $5$--$15~\mathring{\mathrm{A}}$. We use dimers with separations between $5$ and $12~\mathring{\mathrm{A}}$ for training and those with separations between $12$ and $15~\mathring{\mathrm{A}}$ for testing, so that the test set probes true LR extrapolation. 

\begin{figure}[!htbp]
\centering
\includegraphics[width=0.75\linewidth]{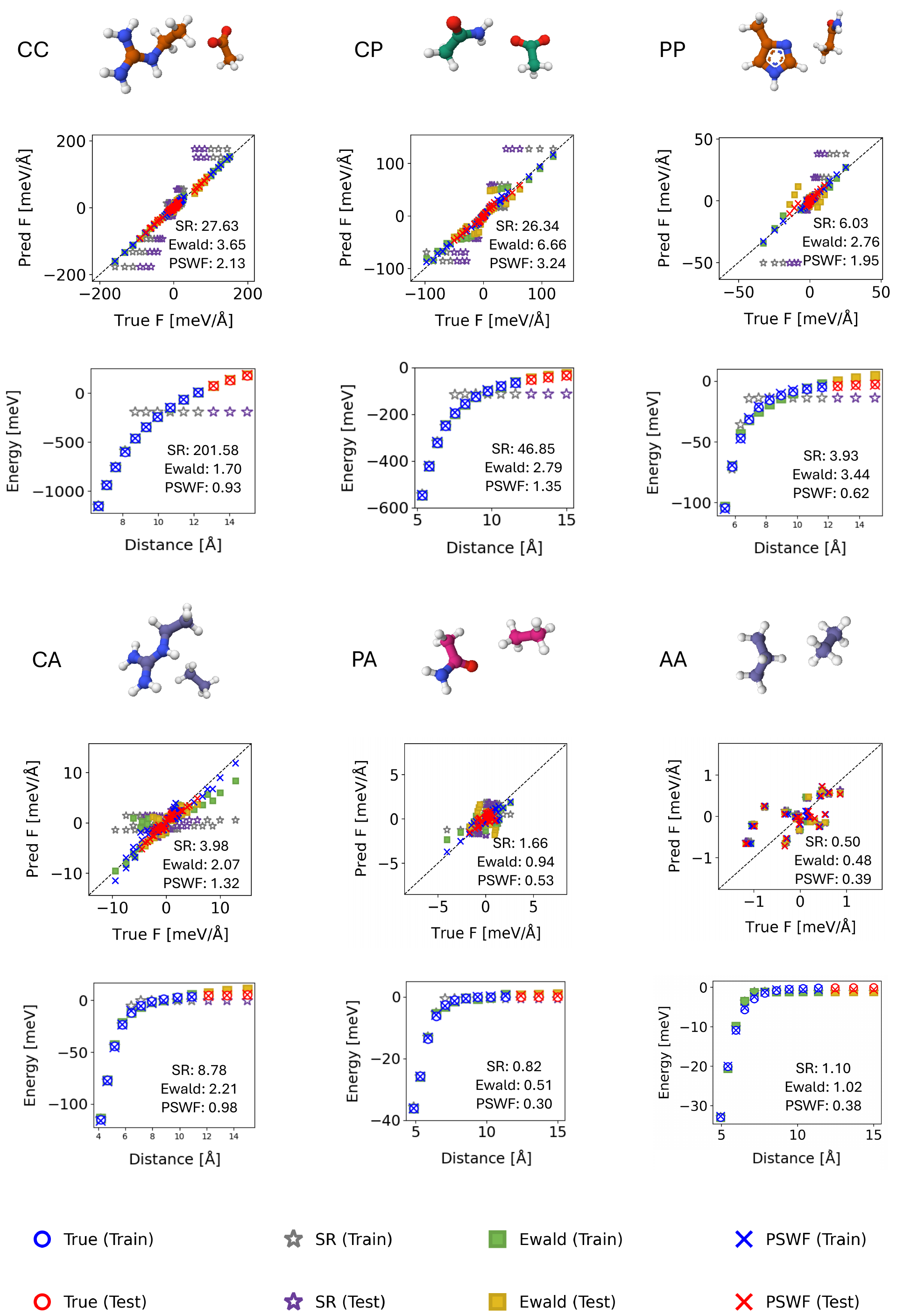}
\vspace{5mm}
\caption{\sf PSWF-LR improves binding-energy and force predictions across six molecular-dimer classes. The dimer classes are charge-charge (CC), charge-polar (CP), polar-polar (PP), charge-apolar (CA), polar-apolar (PA)
and apolar-apolar (AA), with ideal long-range decay rates ranging from $1/r$ to $1/r^6$. For each class, the top panel shows a representative configuration, the middle panel shows parity plots of Cartesian force components predicted by the short-range-only (SR), Ewald-augmented and PSWF-augmented models against the DFT reference, and the bottom panel shows the binding-energy curve as a function of intermolecular separation. The binding energy is defined as the energy difference between the dimer and the two isolated monomers. RMSEs for energies and force components are reported in the corresponding panels.}
\label{fig:MLIPDimer}
\end{figure}

Across all six classes, adding an LR term improves binding-energy curves and force predictions relative to the SR-only model (Fig.~\ref{fig:MLIPDimer}). PSWF-LR gives the lowest errors throughout the benchmark, with the largest gains in the extrapolative separation regime. These results show that explicitly matching the asymptotic decay improves LR generalization rather than only fitting the training separations. Extended Data Fig.~\ref{fig:MLIPDimerSI} further shows that validation scans over $p$ and mesh spacing provide a practical route for selecting the LR prior in new systems.

The polar-dipeptide benchmark further shows that the best exponent depends on the target observable. Among $p=1,~2,~3$, $p=2$ best matches common atomic-charge partitions, whereas $p=1$ better reproduces dipoles, quadrupoles and Born effective charges (Extended Data Figs.~\ref{fig:MLIPDipeptide} and \ref{fig:MLIPDipole}). This separation is important because atomic charges are model-dependent quantities, whereas multipoles and Born effective charges are tied more directly to molecular response. This result highlights the flexibility of PSWF-LR: it allows the LR exponent to be chosen according to the physical observable of interest, rather than being fixed by a default Coulombic ansatz or by a chosen charge partition.

\subsection*{Nonlocal charge transfer}
We next evaluate PSWF-LR on five benchmarks with pronounced nonlocal charge redistribution~\autocite{ko2021fourth,maruf2025learning}: NaCl cluster ions, protonated carbon chains, Au$_2$--MgO(001), dry Cu--BTA and solvated Cu--BTA interfaces. These systems span finite and periodic settings, charged and neutral environments, and cases where the key physics is a LR reshaping of charge rather than only local bonding. 

Table~\ref{tab:charge_transfer_rmse} compares PSWF-LR with representative LR MLIPs, including charge-constrained ACE~\cite{rinaldi2025charge}, 3G-HDNNP~\cite{morawietz2012neural}, 4G-HDNNP~\cite{ko2021fourth}, CACE-LR~\cite{king2025machine} and NequIP-LR~\cite{maruf2025learning}. PSWF-LR attains the lowest test errors for both energies and forces on all five benchmarks. Table~\ref{tab:charge_transfer_rmse} provides a broad benchmark comparison, whereas the same-backbone replacements provide the cleanest attribution: for solvated Cu--BTA interfaces, replacing the Ewald-based LR module in NequIP-LR or CACE-LR by PSWF-LR improves predictive accuracy without changing the SR model (Extended Data Fig.~\ref{fig:MLIPCuBTAHOS}), where the largest per-atom error reductions occur in the adsorbate and solvent regions where charge rearrangement is strongest. This indicates that the gain arises from the PSWF-based LR formulation itself rather than from a different local representation. Additional benchmark-specific tests show the same trend: along Na displacement pathways, PSWF-LR gives the smallest energy and force errors (Extended Data Fig.~\ref{fig:MLIPNaClSI}); for Au$_2$--MgO(001), it better captures both the wetting--non-wetting energy splitting and the doping-induced shift in equilibrium Au--O distance (Extended Data Figs.~\ref{fig:MLIP4GMgO} and~\ref{fig:MLIPAu2MgO}).

Additional scans reveal why the LR mechanism matters. In NaCl cluster ions, increasing the PSWF exponent from $p=1$ to $p=3$ progressively distorts both the charge-transfer landscape and the associated energy surface (Extended Data Fig.~\ref{fig:MLIPNaClPC}). Representative high-energy structures show the same trend: $p=1$ remains closest to the DFT charge distribution, whereas larger exponents over-localize the charge redistribution (Extended Data Fig.~\ref{fig:MLIPNaClCharge}). Broader component-selection tests show that adding higher-order LR channels can improve force accuracy, but the dominant channels are system-dependent (Extended Data Fig.~\ref{fig:MLIPDiffP}). These results show that LR accuracy depends not only on including a nonlocal term, but also on representing its asymptotic decay at the correct rate.

\begin{table*}[t]
\centering
\small
\setlength{\tabcolsep}{4.5pt}
\renewcommand{\arraystretch}{1.12}

\caption{
Test errors on five charge-transfer benchmarks.
Cutoff radii $r_c$ are reported in $\mathring{\mathrm{A}}$.
Energy and force errors are reported as $E_{\mathrm{RMSE}}$ in meV/atom and
$F_{\mathrm{RMSE}}$ in meV/$\mathring{\mathrm{A}}$, respectively.
}
\label{tab:charge_transfer_rmse}

\begin{tabular*}{\textwidth}{@{\extracolsep{\fill}}lcccccc@{}}
\toprule
& $\chi+\eta$ (ACE)
& 3G-HDNNP
& 4G-HDNNP
& CACE-LR
& NequIP-LR
& PSWF-LR \\
\midrule

\multicolumn{7}{@{}l}{\textbf{Na$_{8/9}$Cl$_8^{+}$}} \\
$r_c$
& 6.00 & 5.29 & 5.29 & 5.29 & -- & 5.29 \\
$E_{\mathrm{RMSE}}$
& 0.71 & 2.04 & 0.48 & 0.21 & -- & \textbf{0.15} \\
$F_{\mathrm{RMSE}}$
& 12.35 & 76.67 & 32.78 & 9.78 & -- & \textbf{4.43} \\
\midrule

\multicolumn{7}{@{}l}{\textbf{C$_{10}$H$_2$/C$_{10}$H$_3^{+}$}} \\
$r_c$
& 6.00 & 4.23 & 4.23 & 4.23 & 5.00 & 4.23 \\
$E_{\mathrm{RMSE}}$
& 0.75 & 2.05 & 1.19 & 0.73 & 1.01 & \textbf{0.67} \\
$F_{\mathrm{RMSE}}$
& 35.16 & 231.00 & 78.00 & 36.90 & 48.90 & \textbf{31.62} \\
\midrule

\multicolumn{7}{@{}l}{\textbf{Au$_2$--MgO}} \\
$r_c$
& 6.00 & -- & 4.23 & 5.50 & 5.50 & 5.50 \\
$E_{\mathrm{RMSE}}$
& 1.63 & -- & 0.22 & 0.073 & 0.17 & \textbf{0.068} \\
$F_{\mathrm{RMSE}}$
& 50.27 & -- & 66.00 & 7.91 & 21.99 & \textbf{4.93} \\
\midrule

\multicolumn{7}{@{}l}{\textbf{Cu--BTA}} \\
$r_c$
& -- & -- & -- & 5.00 & 5.00 & 5.00 \\
$E_{\mathrm{RMSE}}$
& -- & -- & -- & 0.074 & 0.11 & \textbf{0.028} \\
$F_{\mathrm{RMSE}}$
& -- & -- & -- & 3.23 & 3.48 & \textbf{2.25} \\
\midrule

\multicolumn{7}{@{}l}{\textbf{Cu--BTA(H$_2$O)}} \\
$r_c$
& -- & -- & -- & 5.00 & 5.00 & 5.00 \\
$E_{\mathrm{RMSE}}$
& -- & -- & -- & 0.038 & 0.18 & \textbf{0.030} \\
$F_{\mathrm{RMSE}}$
& -- & -- & -- & 4.05 & 7.58 & \textbf{2.11} \\

\bottomrule
\end{tabular*}

\vspace{3pt}
\begin{minipage}{\textwidth}
\footnotesize
Except for NequIP-LR, no message passing was used in the models in Table 1 where applicable.
PSWF-LR results are obtained using the training protocol described in Methods.
Results for $\chi+\eta$ (ACE) and CACE-LR are taken from Ref.~\cite{king2025machine};
3G-HDNNP and 4G-HDNNP from Ref.~\cite{ko2021fourth};
and NequIP-LR from Ref.~\cite{maruf2025learning}.
Bold numbers mark the lowest error for each benchmark and metric.
\end{minipage}

\end{table*}

\subsection*{PSWF-LR accelerates differentiable long-range evaluations}
\begin{figure}[!htbp]
\centering
\includegraphics[width=0.87\linewidth]{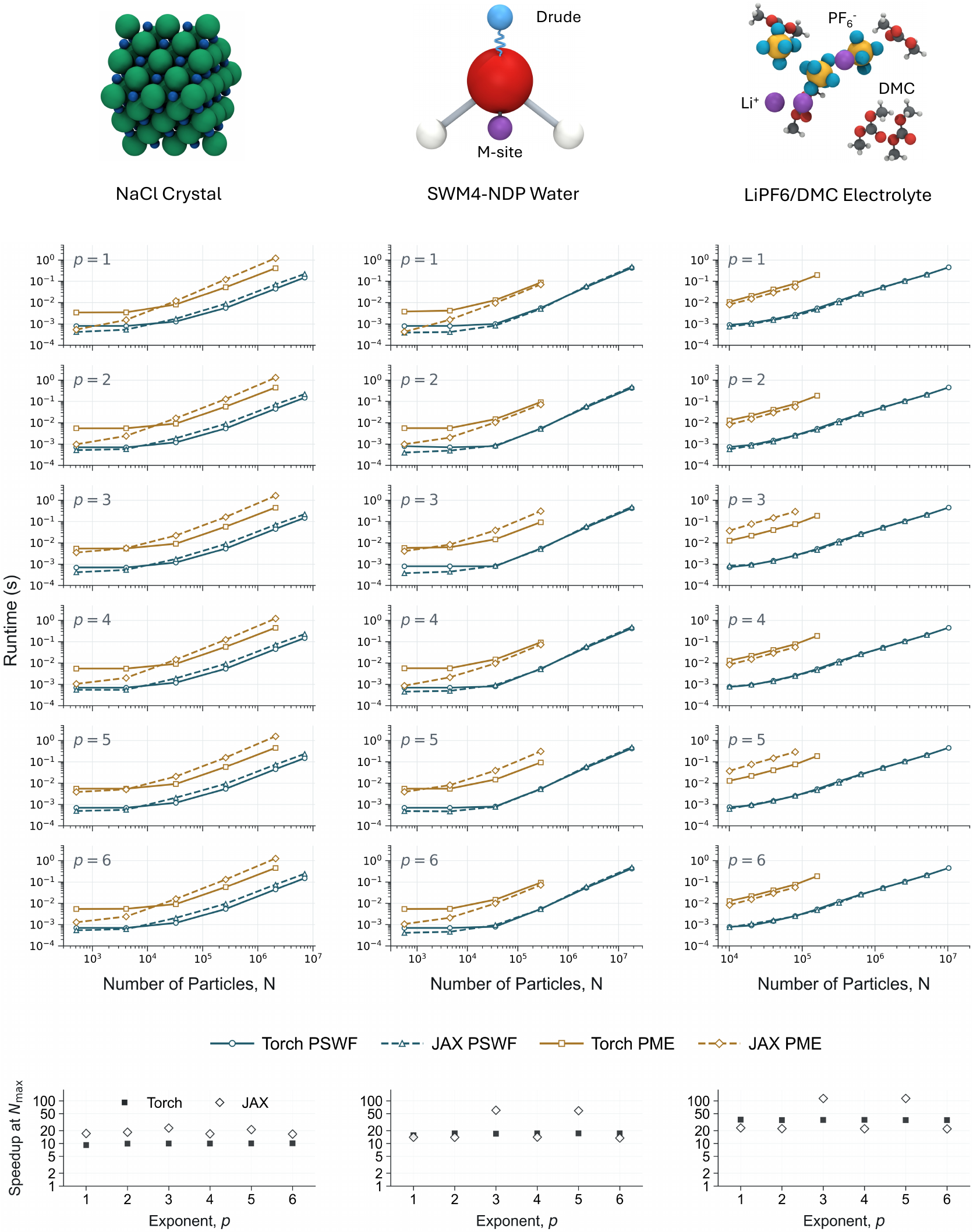}
\vspace{5mm}
\caption{\sf Performance comparison between PSWF- and PME-based long-range treatments across ionic, polarizable and electrolyte systems. The short-range cutoff is set to $4~\mathring{\mathrm{A}}$ for all methods. Top row, model systems used in the benchmark: an NaCl crystal, the SWM4-NDP polarizable water model and a LiPF$_6$/DMC electrolyte. Lower runtime panels show the full forward-and-backward runtime as a function of particle number, $N$, for exponents $p=1$ to $6$ (top to bottom). Blue curves denote PSWF and yellow curves denote PME; solid and dashed lines denote Torch and JAX implementations, respectively. Bottom panels summarize the PSWF speed-up over PME at the largest PME-reachable particle number, $N_{\max}$, for each exponent, with filled squares and open diamonds denoting Torch and JAX implementations, respectively. Missing PME data points indicate calculations that became GPU-memory limited.}
\label{fig:MLIPComparison}
\end{figure}

We next compare PSWF-LR with differentiable PME-based treatments in periodic systems. The benchmarks include a NaCl crystal, a SWM4-NDP polarizable water model, and a LiPF$_6$/DMC electrolyte (Fig.~\ref{fig:MLIPComparison}). Together, these systems span ionic, polarizable and dense electrolyte settings, and therefore test LR evaluation in increasingly crowded atomistic environments. The NaCl crystal is representative of systems used in phonon-dispersion calculations~\cite{zhang2022deep}; SWM4-NDP is a classical analogue of the DPLR water model used for phase-diagram prediction~\cite{zhang2021phase}; and LiPF$_6$/DMC is a representative liquid-electrolyte system for MLIP studies of solvation, transport, and electrolyte degradation chemistry~\cite{gong2025predictive,zheng_bridging_2026}. We evaluate PSWF-LR in both PyTorch and JAX against Torch-PME and JAX-PME~\cite{loche2025fast} at matched LR tolerance. For each system, we measure the end-to-end forward-and-backward runtime for exponents $p=1,\ldots,6$ as a function of the number of particles, $N$, while keeping the number density fixed. Details of the benchmark construction are provided in Methods, and the parameter settings and accuracy checks for each subfigure are also provided in the SM. 

PSWF-LR is faster than PME across all three systems, both software stacks and all tested exponents. For $r_\mathrm{cut}=4~\mathring{\mathrm{A}}$, the maximum speed-ups reach $10.67\times$, $18.50\times$ and $36.06\times$ in PyTorch for NaCl, SWM4-NDP water and LiPF$_6$/DMC, respectively, and $23.01\times$, $60.21\times$ and $114.20\times$ in JAX. The $r_\mathrm{cut}=5~\mathring{\mathrm{A}}$ results show the same trend (Extended Data Fig.~\ref{fig:MLIPComparisonSI}). At the largest tested sizes, several PME runs become GPU-memory limited, whereas the corresponding PSWF-LR calculations remain tractable and reach the $10^7$-particle regime in the largest differentiable benchmarks. This improved scalability arises because PSWF mollification, when paired with PSWF atom-grid spreading and FFT evaluation, translates the reduced reciprocal bandwidth into nearly $64\times$ fewer three-dimensional grid points at matched tolerance. These runtime benchmarks show that the spectral compression observed in the accuracy tests directly reduces memory use and accelerates differentiable LR evaluation and backpropagation.

\subsection*{MD simulations}
\begin{figure}[!htbp]
\centering
\includegraphics[width=0.95\linewidth]{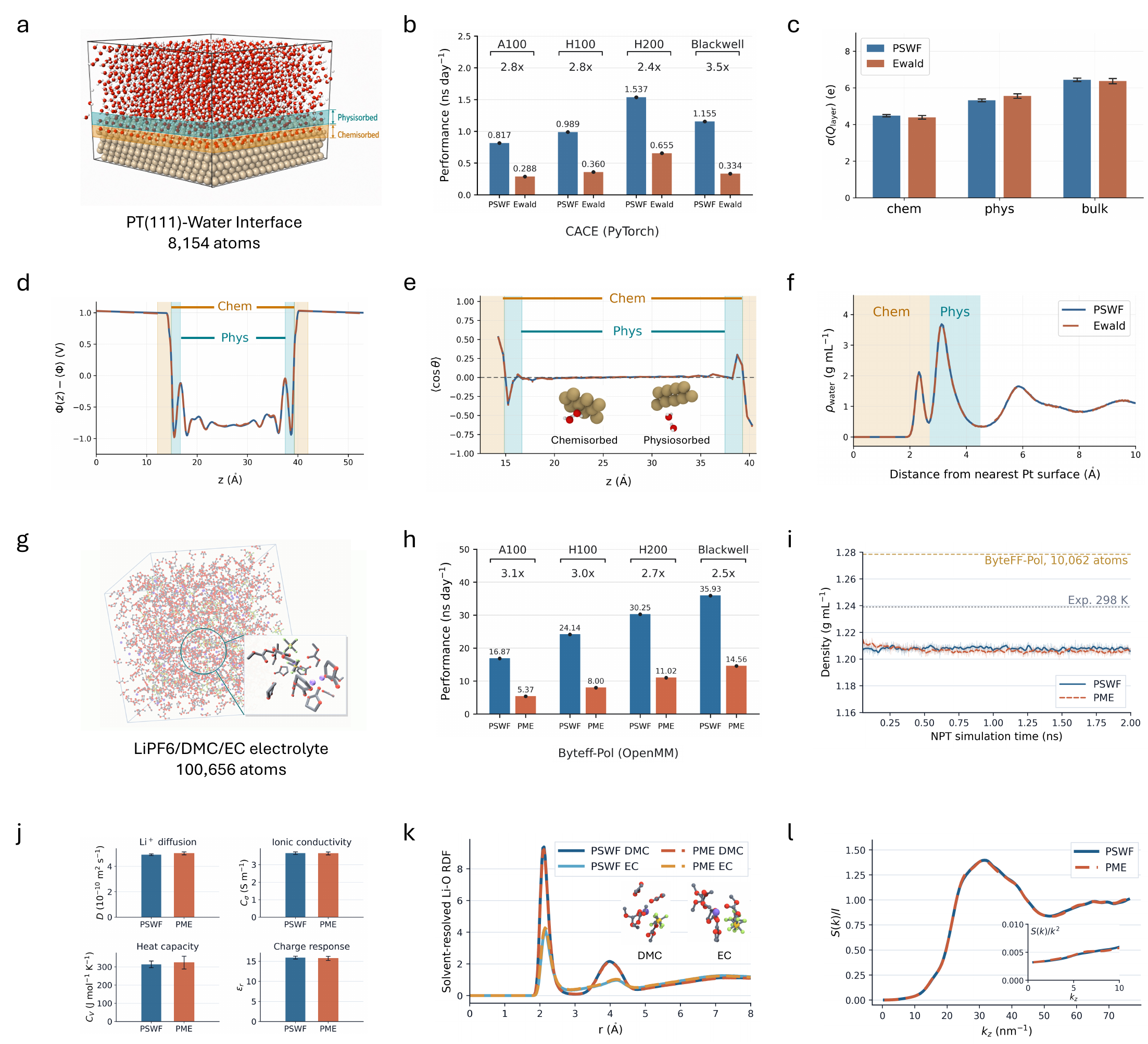}
\vspace{5mm}
\caption{\sf PSWF-LR accelerates machine learning molecular dynamics (MD) while preserving interfacial and electrolyte observables. {\bf a,} Representative Pt(111)-water interface used for the CACE benchmark, with chemisorbed and physisorbed water regions highlighted. {\bf b,} CACE (PyTorch) MD throughput for the Pt(111)-water system across A100, H100, H200 and Blackwell GPUs. Bars compare the PSWF-LR ($p\in\{1,6\}$) with the native Ewald-based framework; bracketed values denote the speed-up. {\bf c,} Surface-normal water mass-density profile relative to the nearest Pt surface, resolving the chemisorbed and physisorbed interfacial regions. {\bf d,} Planar-averaged electrostatic potential profile, plotted as the deviation from the cell average. 
{\bf e,} Water orientational order, quantified by the layer-resolved average molecular orientation, with representative chemisorbed and physisorbed motifs shown in the inset. {\bf f,} Layer-resolved charge response for chemisorbed, physisorbed and bulk-like water environments. {\bf g,} Representative LiPF6/DMC/EC electrolyte cell used for the ByteFF-Pol benchmark, with a local Li-centred solvation motif shown as an inset. {\bf h,} ByteFF-Pol (OpenMM) MD throughput for the 100,656-atom electrolyte system across the same GPU classes. Bars compare PSWF-LR ($p\in\{1,2,3,4,6\}$) with the PME reference; bracketed values denote the speed-up. {\bf i,} NPT density trajectories for PSWF and PME over the first 2 ns, compared with the experimental 298 K density and the original ByteFF-Pol 10,062-atom density anchor. {\bf j,} Extended electrolyte observables computed from PSWF and PME trajectories, including Li$^+$ diffusion, Nernst-Einstein ionic conductivity, heat capacity and charge response; error bars denote block standard errors. {\bf k,} Solvent-resolved Li-O radial distribution functions for DMC and EC, with representative local Li-O coordination motifs. {\bf l,} Charge structure factor, $S(\bm{k})$, along the simulation-cell $z$ direction; the inset shows the low-frequency scaling of $S(\bm{k})/\bm{k}^2$.}
\label{fig:MLIPMD}
\end{figure}

Finally, we test whether PSWF-LR can replace Ewald-style mechanism in production MD without degrading the simulated observables. We consider two chemically distinct settings (Fig.~\ref{fig:MLIPMD}a,g): an $8,154$-atom Pt(111)--water interface described by a CACE MLIP with an Ewald LR treatment, and a $100,656$-atom LiPF$_6$/DMC/EC electrolyte described by a ByteFF-Pol/OpenMM~\cite{zheng_bridging_2026,eastman2023openmm} workflow with PME. In both cases, the upstream model, learned parameters, bonded terms, initial configurations and simulation protocol are kept fixed; only the LR framework is changed.

In the Pt(111)--water benchmark, PSWF-LR accelerates CACE-based MD by $2.4$--$3.5\times$ across the tested GPU architectures (Fig.~\ref{fig:MLIPMD}b). The corresponding trajectories reproduce the Ewald surface-normal water density, planar electrostatic potential, orientational ordering and layer-resolved charge fluctuations (Fig.~\ref{fig:MLIPMD}c--f). Additional size-scaling tests show that the PSWF-LR speed-up increases with system size and extends the accessible single-GPU system size by about $4$--$8\times$ to the $5.8\times 10^4$-atom scale (Extended Data Fig.~\ref{fig:MLIPCombineScaling}a). The two-channel setting used in this benchmark, $\mathcal{S}=\{1,6\}$, adds only a small overhead relative to the single-channel $p=1$ PSWF-LR model, decreasing from about $10\%$ at smaller sizes to about $2\%$ at larger sizes. Even the more demanding six-channel $\mathcal{S}=\{1,\ldots,6\}$ model remains affordable, with the overhead decreasing from about $75\%$ to about $10\%$ as the system size grows (Extended Data Fig.~\ref{fig:MLIPCombineScaling}b).

In the electrolyte benchmark, PSWF-LR accelerates ByteFF-Pol MD by $2.5$--$3.1\times$ relative to PME (Fig.~\ref{fig:MLIPMD}h). The paired NPT trajectories remain stable and close to one another (Fig.~\ref{fig:MLIPMD}i), while the Li$^+$ diffusion, Nernst--Einstein ionic conductivity, heat capacity and charge-response diagnostics agree within block uncertainties (Fig.~\ref{fig:MLIPMD}j). Solvent-resolved Li--O radial distribution functions and charge structure factors further show that PSWF-LR preserves local solvation structure and collective charge-density fluctuations (Fig.~\ref{fig:MLIPMD}k,l). Additional viscosity, precision and scaling tests show that this advantage persists across simulation modes and larger system sizes, reaching the $10^7$-atom regime (Extended Data Fig.~\ref{fig:MLIPByteff2}). Backend-paired enthalpies of vaporization across 36 QUBEKit liquids further support the preservation of molecule-wise thermodynamic trends (Extended Data Fig.~\ref{fig:MLIPHVAPSI}). Together, these MD tests show that PSWF-LR provides production-level acceleration while preserving the tested structural, transport and charge-response observables.

\section*{Discussion}
A central conclusion of this work is that the LR mechanism is itself an important modeling choice in MLIPs. Much recent progress has focused on improving the upstream representation, such as learning better charges, dipoles, virtual sites or descriptors. Our results show that, even when the upstream model is held fixed, replacing Gaussian Ewald screening and B-spline particle–mesh spreading with PSWF mollification and spreading can substantially change the accuracy--efficiency trade-off. The reason is that PSWFs are compactly supported functions whose Fourier transforms are maximally concentrated within a finite band, allowing the mollified LR component to be represented with fewer Fourier grids. This spectral compression explains the lower Fourier-mode requirements in accuracy benchmarks, the speed-ups over differentiable Torch-PME and JAX-PME, and the production-MD acceleration observed in CACE and OpenMM workflows while preserving the tested observables.

PSWF-LR also reframes how LR physics can be built into MLIPs. Instead of fixing the LR term to a Coulombic Ewald form, the framework supports inverse-power channels $1/r^p$, so the decay exponent becomes a physical prior that can be selected according to the target system and observable. The molecular-dimer benchmark shows that matching the asymptotic exponent improves LR extrapolation, while the dipeptide benchmark shows that the best exponent for atomic charge partitions need not be the best exponent for dipoles, quadrupoles or Born effective charges. The charge-transfer benchmarks further show that using multiple exponent channels can improve accuracy, but the gain is not simply monotonic with the number of channels. Different channels can overlap or redistribute the learned LR contribution. Thus, the exponent set should be viewed as a structured modeling choice, selected using physical knowledge and validation performance, rather than as a default fixed by the Ewald ansatz or as a rule that more channels are always better.

A practical advantage of PSWF-LR is that it separates the representation of local chemistry from the evaluation of nonlocal interactions. As a complete framework, it couples an SR representation to suitable LR variables and evaluates the resulting inverse-power channels. As an LR evaluation layer, it can replace Ewald-style components in latent-variable MLIPs, charge-equilibration models, descriptor-based LR representations and ML-parameterized molecular-mechanics workflows, provided that the upstream model supplies suitable charges, features, latent coefficients or force-field parameters. This separation allows PSWF-LR to be inserted without redesigning the local model or changing learned parameters, while preserving the downstream MD workflow. The same modularity should also make PSWF-LR compatible with foundation-model pipelines~\cite{batatia2025foundation}, active learning~\cite{kulichenko2023uncertainty}, ensemble distillation~\cite{gong2025predictive} and learning-from-models~\cite{zheng2025learning} workflows, although these directions remain to be tested systematically.  The PyTorch, JAX and OpenMM/CUDA implementations further support its use in both differentiable training and production simulation settings.

Several limitations remain. The present framework focuses on pairwise inverse-power channels with suitable LR variables. It does not yet include explicit three-body LR terms or fully self-consistent charge-equilibration loops. Its accuracy is also limited by the upstream SR representation: an efficient LR layer cannot correct errors caused by local descriptors, local chemistry or insufficient training data. The benefit of PSWF-LR is therefore system-dependent. It is expected to be strongest when LR physics is a leading source of error or cost, as in ionic clusters, charge-transfer interfaces, polar liquids and dense electrolytes, and more modest when errors are dominated by local bonding, steric packing or hydrogen bonding. Future work should couple PSWF-LR with global charge-equilibration, density-based LR descriptors and higher-order polarization models, so that the LR variables themselves can be determined more globally while retaining the compact PSWF representation.

%TC:ignore
\section*{Methods}
\subsection*{Prolate spheroidal wave functions}
The PSWF is an eigenfunction of the compact integral operator $\mathscr{F}_c:L^{2}[-1,1]\rightarrow L^2[-1,1]$ defined by
\begin{equation}\label{eq::formula}
\mathscr{F}_c[\varphi](x)=\int_{-1}^{1}\varphi(t)\,e^{icxt}\,dt,
\end{equation}
where $c>0$ is a real parameter. We denote the eigenvalues by $\lambda_0,\lambda_1,\ldots$ and order them so that $|\lambda_n|\ge |\lambda_{n+1}|$ for all $n\ge 0$. Let $\psi_n^{c}$ be an eigenfunction associated with $\lambda_n$, i.e.,
\begin{equation}\label{eq::psidefinit}
\lambda_n\psi_{n}^{c}(x)=\int_{-1}^{1}\psi_{n}^{c}(t)\,e^{icxt}\,dt,
\qquad x\in[-1,1],\quad n\ge 0.
\end{equation}
The eigenfunctions may be chosen real-valued and form an orthogonal complete family in $L^2[-1,1]$~\cite{osipov2013prolate}. Throughout this work we use only the order-zero function and fix its scale by the convention $\psi_0^c(0)=1$; with this convention $\psi_0^c$ is not $L^2$-normalized unless explicitly rescaled.

A key feature of PSWFs is their joint time-frequency concentration. In particular, among all $L^2$ functions supported on $[-1,1]$ with unit $L^2$ norm, the unit-norm rescaling of the order-zero PSWF $\psi_0^{c}$ uniquely maximizes the fraction of Fourier energy contained in the band $[-c,c]$ (equivalently, it minimizes the $L^2$ energy outside $[-c,c]$); higher-order PSWFs provide subsequent maximizers subject to orthogonality constraints~\cite{osipov2013prolate,slepian1983sirev}. This optimal concentration property motivates using $\psi_0^{c}$ as a near-optimal compactly supported window.

PSWFs also satisfy a simple Fourier relation on $[-c,c]$. If $\psi_n^{c}$ is extended by zero outside $[-1,1]$, then its Fourier transform restricted to $|k|\le c$ obeys
\begin{equation}\label{eq::pswf_fourier}
\widehat{\psi}_{n}^{c}(k)=\lambda_n\,\psi_{n}^{c}(k/c),
\qquad |k|\le c.
\end{equation}
In particular, for $n=0$ this shows that the band-limited Fourier transform of the compactly supported window $\psi_0^{c}$ reproduces the same function (up to the scalar factor $\lambda_0$) under the scaling $k\mapsto k/c$, analogous to the self-reproducing property of Gaussians under the Fourier transform. For the conventional Gaussian screening of Eq.~\eqref{eq::screening}, as used in standard mesh-Ewald methods, achieving an error level $\varepsilon$ typically requires
\begin{equation}\label{eq:KmaxPPPM}
K_{\max} =  2\log(1/\varepsilon).
\end{equation}
Here $K_{\max}$ denotes the maximum Fourier frequency retained in the FFT-based evaluation in each dimension. For a PSWF mollifier, the analogous bandwidth parameter is $c$, and to reach the same small $\varepsilon$ one typically chooses $c$ such that
\begin{equation}\label{eq::c}
K_{\max}= c \approx \log(1/\varepsilon).
\end{equation}
The approximation becomes more accurate in the high-precision regime (small $\varepsilon$). Consequently, at high accuracy, PSWF splitting requires roughly $8\times$ fewer Fourier modes in three dimensions than Gaussian-based splitting. Equivalently, for fixed $K_{\max}$, PSWF achieves higher accuracy. Due to these nice properties, PSWFs have been used in compressed sensing~\cite{davenport2012compressive}, radar communications~\cite{chen2008mimo}, fast convolution~\cite{jiang2025dmk} and classical Coulomb solvers~\cite{liang2025accelerating}. Their use in LR MLIPs, however, has remained unexplored. Here we use the same concentration principle to construct an exponent-aware LR mechanism for MLIPs, extending the bandwidth advantage of PSWFs to learnable inverse-power channels $1/r^p$.

\subsection*{PSWF-based mollifier for general inverse power laws}
For a general inverse‑power kernel $1/r^p$ with $p>0$, the classical Ewald splitting employs incomplete‑gamma screening functions,
\begin{equation}\label{eq::screening}
\mathcal{G}_p^{\text{Ewald}}(r)=\frac{\Gamma\!\left(p/2,\alpha^2 r^2\right)}{\Gamma\!\left(p/2\right)},
\end{equation}
where $\Gamma(\cdot,\cdot)$ is the upper incomplete gamma function; the complementary part is given by the lower incomplete gamma function. The Coulomb ($p=1$) and dispersion ($p=6$) cases reduce to $\mathcal{G}_{1}^{\text{Ewald}}(r)=\mathrm{erfc}(\alpha r)$ and $\mathcal{G}_{6}^{\text{Ewald}}(r)=e^{-\alpha^2 r^2}\!\left(1+\alpha^2 r^2+\alpha^4 r^4/2\right)$, respectively. In our PSWF‑based framework, we construct mollifiers for $p\in\{2,\ldots,5\}$ (with $p=1$ and $p=6$ given in Eqs.~\eqref{eq::G1} and \eqref{eq::G6}) as
\begin{equation}\label{eq::G}
\begin{split}
\mathcal{G}_{2}(r)=1-\psi_{0}^{c}(r/r_c),&\qquad 
\mathcal{G}_{3}(r)=\frac{1}{C_{0}}\left(\int_{0}^{r/r_c}\psi_{0}^{c}(x)\,dx-\frac{r}{r_c}\psi_{0}^{c}(r/r_c)\right),\\[2em]
\mathcal{G}_{4}(r)=1-\psi_{0}^{c}(r/r_c)+\frac{r}{2r_c}\psi_{0}^{c}{}^{\prime}(r/r_c),&\qquad 
\mathcal{G}_{5}(r)=\frac{1}{C_0}\left(\int_{0}^{r/r_c}\psi_{0}^{c}(x)\,dx-\frac{r}{r_c}\psi_{0}^{c}(r/r_c)+\frac{r^2}{3r_c^2}\psi_{0}^{c}{}^{\prime}(r/r_c)\right),
\end{split}
\end{equation}
where $\lim\limits_{r\rightarrow \infty}\mathcal{G}_{p}(r)/r^p=1/r^p$ for all $p$.
Let $\widehat{C}_0:=\int_{0}^{\infty}\lambda_{0}\psi_{0}^{c}(x/c)dx$. The corresponding Fourier transforms of $\mathcal{G}_{p}(r)/r^p$, used in Eq.~\eqref{eq::Fourier}, are 
\begin{equation}\label{eq::PhiH}
\begin{gathered}
\widehat{\Phi}_{1}(\bm{k})=\frac{4\pi\psi_{0}^{c}(kr_c/c)}{k^2\psi_{0}^{c}(0)},\qquad
\widehat{\Phi}_{2}(\bm{k})=\frac{2\pi^2}{k}\left(1-\frac{1}{\widehat{C}_0}\int_{0}^{kr_c}\lambda_0\psi_{0}^{c}(x/c)\,dx\right),\\[2em]
\widehat{\Phi}_{3}(\bm{k})=\frac{2\pi}{C_0}\int_{kr_c}^{\infty}\frac{\lambda_0\psi_{0}^{c}(x/c)}{x}\,dx,\qquad
\widehat{\Phi}_{4}(\bm{k})=-\frac{\pi^2}{\widehat{C}_0}\left(\widehat{C}_{0}k-k\int_{0}^{kr_c}\lambda_{0}\psi_{0}^{c}(x/c)\,dx-\frac{1}{r_c}\int_{kr_{c}}^{\infty}x\lambda_{0}\psi_{0}^{c}(x/c)\,dx\right),\\[2em]
\widehat{\Phi}_{5}(\bm{k})=-\frac{\pi}{3C_0}\left(k^2\int_{kr_c}^{\infty}\frac{\lambda_{0}\psi_{0}^{c}(x/c)}{x}\,dx-\frac{1}{r_c^2}\int_{kr_c}^{\infty}x\lambda_{0}\psi_{0}^{c}(x/c)\,dx\right),\\[2em]
\widehat{\Phi}_{6}(\bm{k})=\frac{\pi^2}{12\widehat{C}_{0}}\left(\widehat{C}_{0}k^3-k^3\int_{0}^{kr_c}\lambda_{0}\psi_{0}^{c}(x/c)\,dx-\frac{3k^2}{2r_c}\int_{kr_c}^{\infty}x\lambda_{0}\psi_{0}^{c}(x/c)\,dx+\frac{1}{2r_c^3}\int_{kr_c}^{\infty}x^3\lambda_{0}\psi_{0}^{c}(x/c)\,dx\right),
\end{gathered}
\end{equation}
where $k=\|\bm{k}\|$ denotes the wavenumber. For $p>6$, we propose a recursive formula to derive $\mathcal{G}_{p}(r)$ and $\widehat{\Phi}_{p}(\bm{k})$; the construction is described in the Supplementary Materials (SM).

\subsection*{Implementation details}
In practice, we do not evaluate the explicit PSWF functions or the integral forms in Eqs.~\eqref{eq::G}–\eqref{eq::PhiH} during training or MD simulations. Instead, we accelerate the evaluations using a fast kernel auto-differentiation (FKAD) procedure based on online monomial approximation, Horner's rule and operator fusion. This procedure is an implementation acceleration and does not change the mathematical definition of the PSWF-LR kernel. We approximate the smooth parts of $\mathcal{G}_p(r)$ and $\widehat{\Phi}_p(\bm{k})$ by low-degree polynomials in real and reciprocal space, respectively. These polynomials are evaluated with Horner’s rule and fused into downstream operators to avoid unnecessary memory traffic. The resulting approximation substantially reduces runtime while maintaining controllable accuracy. Full algorithmic details are provided in the SM.

When integrating the PSWF framework into models that use neural networks for MM parametrization, the LR Fourier sum in Eq.~\eqref{eq::Fourier} should be paired with a real-space term and a self-correction term for a complete decomposition. The real-space term, $(1-\mathcal{G}_{p}(r))/r^p$, is the complementary part of $1/r^p$. The self-correction term, $E_{\text{self}}^{(p)}$, removes the unwanted self-energy contribution and is given by
\begin{equation}
\begin{gathered}
E_{\text{self}}^{(1)}=\frac{\psi_{0}^{c}(0)}{2C_0r_c}Q^{(1)}_{\mathrm{self}},\qquad E_{\text{self}}^{(2)}=-\frac{\psi_{0}^{c}{}^{\prime\prime}(0)}{4r_c^2}Q^{(2)}_{\mathrm{self}},\qquad
E_{\text{self}}^{(3)}=-\frac{\psi_{0}^{c}{}^{\prime\prime}(0)}{6C_0r_c^3}Q^{(3)}_{\mathrm{self}},\\[2em] E_{\text{self}}^{(4)}=\frac{\psi_{0}^{c}{}^{(4)}(0)}{48r_c^4}Q^{(4)}_{\mathrm{self}},\qquad
E_{\text{self}}^{(5)}=\frac{\psi_{0}^{c}{}^{(4)}(0)}{90C_0r_c^5}Q^{(5)}_{\mathrm{self}},\qquad E_{\text{self}}^{(6)}=-\frac{\psi_{0}^{c}{}^{(6)}(0)}{1440r_c^6}Q^{(6)}_{\mathrm{self}}.
\end{gathered}
\end{equation}
where $Q^{(p)}_{\mathrm{self}}:=\sum\limits_{i}|q_{i}^{(p)}|^2$ for $p=1,\cdots,6$ and $\psi_{0}^{c}{}^{(n)}(x)$ denotes the $n$th derivative of $\psi_{0}^{c}(x)$ for $n=4,6$.

The zero-frequency correction in Eq.~\eqref{eq::Fourier} depends on the exponent $p$. Define the net channel weight
\begin{equation}
Q_{\mathrm{net}}^{(p)}
:=
\sum_{i=1}^{N}q_i^{(p)} .
\end{equation}
For $p>3$, $\widehat{\Phi}_{p}(\bm{k})$ is nonsingular at $\bm{k}=\bm{0}$, and thus
\begin{equation}
E_{\bm{\theta}_{\mathrm{lr}}}^{\bm{0},(p)}=\frac{1}{2V}\widehat{\Phi}_{p}(\bm{0})\left|Q_{\mathrm{net}}^{(p)}\right|^2,\qquad p>3.
\end{equation}
For $p\le 3$, the zero-frequency term requires separate treatment because it diverges when $Q_{\mathrm{net}}^{(p)}\neq 0$. For $p\le 2$, a standard way is to introduce a uniform compensating background, $-Q_{\mathrm{net}}^{(p)}/V$. Under this ``latent-variable neutrality'' condition, the $p=1$ case remains conditionally convergent, and the chosen summation order introduces an additional infinite-boundary (IB) term. The resulting zero-frequency corrections for $p\le 2$ are
\begin{equation}\label{eq::Correction}
\begin{gathered}
E_{\bm{\theta}_{\mathrm{lr}}}^{\bm{0},(1)}=-\frac{2\pi |Q_{\mathrm{net}}^{(1)}|^2}{V}\left(\frac{r_c^2}{2}-\int_{0}^{r_c}r\,\psi_{0}^{c}(r/r_c)\,dr\right)+E_{\mathrm{IB}},\\[2em]
E_{\bm{\theta}_{\mathrm{lr}}}^{\bm{0},(2)}=-\frac{2\pi |Q_{\mathrm{net}}^{(2)}|^2}{V}\int_{0}^{r_c}\psi_{0}^{c}(r/r_{c})\,dr.
\end{gathered}
\end{equation}
For $p=1$, $E_{\mathrm{IB}}$ depends on the summation order and macroscopic boundary condition, consistent with the Coulomb case since $\lim_{r\to\infty}\mathcal{G}_{1}(r)/r=1/r$~\cite{hu2014jctc}. The case $p=3$ is more subtle: introducing the compensating background $-Q_{\mathrm{net}}^{(3)}/V$ produces a divergent correction that depends on both the splitting factor $c$ and the cutoff $r_c$. To avoid this ambiguity, we define
\begin{equation}\label{eq::Correction3r}
E_{\bm{\theta}_{\mathrm{lr}}}^{\bm{0},(3)}=\frac{2\pi \left|Q_{\mathrm{net}}^{(3)}\right|^2}{V}\left(\log(c/r_c)+\int_{0}^{1}\frac{\psi_{0}^{c}(r)-1}{r}dr\right),
\end{equation}
which is equivalent to setting the additive constant in the zero mode of the Fourier transform of the $1/r^3$ kernel to zero. With this normalization, the resulting energy is independent of the specific algorithm. The derivation of Eqs.~\eqref{eq::Correction}-\eqref{eq::Correction3r} is provided in the SM.

\subsection*{Accelerating atom-grid spreading with PSWFs}
Evaluating Eq.~\eqref{eq::Fourier} can require a large number of Fourier modes for two reasons: (i) MLIPs often use a short SR cutoff ($4$--$6~\mathring{\mathrm{A}}$), which increases high-frequency Fourier contributions; and (ii) MD simulation boxes are often much larger than the training configurations. Direct summation scales as $O(NM)$, and for fixed $r_c$ one typically has $M\sim N$. We therefore adopt FFT-based acceleration.

We define a uniform Cartesian grid on $\Omega$ with spacing $h$ and $M_d=L_d/h$ points per direction. The spreading kernel is separable:
\begin{equation}
W(\bm{r})=\prod_{\alpha\in\{x,y,z\}}W_{\mathrm{pswf}}(\alpha),
\end{equation}
with
\begin{equation}
W_{\mathrm{pswf}}(\alpha)=
\begin{cases}
\psi_0^c(\alpha/\omega), & |\alpha|\le \omega,\\
0, & |\alpha|>\omega,
\end{cases}
\end{equation}
where $\omega=Ph/2$ and $P\in\mathbb Z^+$ is the number of points in the grid coupled to each atom per dimension. Its Fourier transform is
\begin{equation}
\widehat W_{\mathrm{pswf}}(k_x)=\omega\lambda_0\psi_0^c(\omega k_x/c),\qquad
\widehat W(\bm{k})=\prod_{\alpha\in\{x,y,z\}}\widehat W_{\mathrm{pswf}}(k_\alpha).
\end{equation}
Because $\psi_0^c$ is evaluated by piecewise polynomials, the cost is comparable to B-spline windows~\cite{darden1993jcp,essmann1995jcp}. Inserting $1\equiv \widehat W(\bm{k})^{-2}\widehat W(\bm{k})^2$ into Eq.~\eqref{eq::Fourier} gives
\begin{equation}\label{eq::EthetaLR}
E_{\boldsymbol\theta_{\mathrm{lr}}}
=\frac{1}{2V}\sum_{p\in\mathcal S}\sum_{\bm{k}\neq \mathbf 0}
\widehat W(\bm{k})^{-2}\widehat\Phi_{p}(\bm{k})
\left|\sum_{i=1}^N q_i^{(p)}\widehat W(\bm{k})e^{i\bm{k}\cdot\bm{r}_i}\right|^2
+\sum_{p\in\mathcal S}E_{\boldsymbol\theta_{\mathrm{lr}}}^{\mathbf 0,(p)}.
\end{equation}
The first term is evaluated with four steps: (i) atom-grid spreading, we evaluate
\begin{equation}
\rho_{\mathrm{grid}}^{(p)}(\bm{r})=\sum_{j=1}^N q_j^{(p)}\,[W(\bm{r}-\bm{r}_j)]_*
\end{equation}
on a uniform grid with mesh size $h$, where $[\cdot]_{*}$ denotes periodization of $W$; (ii) apply a forward 3D FFT (batched over $p$) to obtain $\widehat{\rho}_{\text{grid}}^{(p)}(\bm{k})$; (iii) diagonal scaling, i.e. for each Fourier mode and $p\in\mathcal{S}$ we compute
\begin{equation}
\widehat\chi^{(p)}(\bm{k})=\frac{1}{2V}\widehat W(\bm{k})^{-2}\widehat\Phi_{p}(\bm{k})\left|\widehat\rho_{\mathrm{grid}}^{(p)}(\bm{k})\right|^2;
\end{equation}
(iv) Fourier-space reduction, i.e. we sum over Fourier modes and exponents
\begin{equation}
E_{\boldsymbol\theta_{\mathrm{lr}}}
=\sum_{p\in\mathcal S}\sum_{\bm{k}\neq\mathbf 0}\widehat\chi^{(p)}(\bm{k})
+\sum_{p\in\mathcal S}E_{\boldsymbol\theta_{\mathrm{lr}}}^{\mathbf 0,(p)}.
\end{equation}

No inverse FFT is required in the above procedure, because the energy is obtained directly by Fourier-space reduction. Since the PSWF kernel is smooth and compactly supported, the method exhibits spectral convergence with the increase of grid size. The total cost is near-optimal, $O(P^3N+M\log M)$, with $O(P^3N)$ from spreading and $O(M\log M)$ from FFT and Fourier-space reduction. Using PSWF (or its approximation) as the spreading kernel achieves the target accuracy with a near-minimal number of spreading points $P$, compared with the B-spline windows used in PME and PPPM. This further improves the efficiency of our PSWF-based framework for large-scale simulations. As shown in the upper-right panel of Fig.~\ref{fig:MLIPSketchMap}, the required number of Fourier modes can be reduced further.  

\subsection*{Parameter selection}
We select PSWF/FFT hyperparameters from a target LR tolerance $\varepsilon_{\mathrm{tot}}$ and decompose the total error into three components:
\begin{equation}
\varepsilon_{\mathrm{tot}}
\lesssim
\varepsilon_{\mathrm{band}}
+\varepsilon_{\mathrm{grid}}
+\varepsilon_{\mathrm{spread}},
\end{equation}
corresponding to the bandwidth truncation of PSWF, grid discretization, and atom--grid spreading. Unless otherwise stated, we use a simple uniform allocation,
$\varepsilon_{\mathrm{band}}
=\varepsilon_{\mathrm{grid}}
=\varepsilon_{\mathrm{spread}}
=\varepsilon_{\mathrm{tot}}/3$.

Given an SR cutoff $r_c$, we set the PSWF bandwidth as
\begin{equation}
c \approx \log(1/\varepsilon_{\mathrm{band}}),
\end{equation}
following Eq.~\eqref{eq::c}. For the same target accuracy, this is approximately half of the Gaussian/Ewald requirement in Eq.~\eqref{eq:KmaxPPPM}, implying about $8\times$ fewer 3D Fourier modes at the same precision. The effective Fourier cutoff is $k_c=c/r_c$. For this $k_c$, the mesh spacing $h$ is chosen from the Nyquist–Shannon sampling criterion
\begin{equation}
h \le \eta\frac{\pi}{k_c}=\eta\frac{\pi r_c}{c}, \qquad \eta\in(0,1],
\end{equation}
with grid size $M_d=\lceil L_d/h\rceil$ for $d\in\{x,y,z\}$, so that the target $\varepsilon_{\mathrm{grid}}$ is achieved. The spreading width $P$ controls the $O(P^3N)$ prefactor; we choose the smallest $P$ such that spreading error is less than $\varepsilon_{\mathrm{spread}}$ (typically $P=4$--$6$ for MLIP settings).

Finally, for the polynomial approximations of $\mathcal{G}_p$ and $\widehat{\Phi}_{p}$, we select the lowest degree such that the maximum approximation error on the real- and Fourier-space working intervals is below $\varepsilon_{\mathrm{poly}}$. This degree selection is performed automatically via Chebyshev-coefficient filtering: each target kernel is first expanded on a fixed high-order Chebyshev basis (up to degree $40$), then truncated by discarding terms $|a_n|\leq 0.01\,\varepsilon\max\limits_m|a_m|$, and finally converted to monomial coefficients. In practice, the $0.01\,\varepsilon$ threshold ensures that the polynomial approximation error is negligible.

\subsection*{Plug‑in compatibility with existing LR MLIPs}
PSWF-LR defines a flexible LR formulation for MLIPs: it reformulates screening, reciprocal-space evaluation and atom-grid spreading within a unified PSWF framework, while remaining modular enough to interface with several established classes of LR MLIPs. Its practical point of insertion depends on how a given model represents nonlocal physics. In latent-variable models such as LES~\cite{cheng2025latent} and DPLR~\cite{zhang2022deep}, the local network that predicts charges, dipoles or virtual sites can be left unchanged, while the reciprocal-space Ewald module is replaced by Eq.~\eqref{eq::Fourier} with PSWF-LR. In descriptor-based schemes such as LODE, the same idea applies one level earlier: the Gaussian smoothing used to generate nonlocal density descriptors~\cite{grisafi2019incorporating,huguenin2023physics} can be replaced by PSWF mollifiers while retaining the subsequent projection onto radial and angular bases. In charge-equilibration pipelines such as 4G-HDNNP~\cite{ko2021fourth} and BAMBOO~\cite{gong2025predictive}, PSWFs can replace Gaussian charge clouds in the electrostatic stage without altering the upstream electronegativity model or the global charge-conservation step.

The same logic extends to ML-parameterized MM frameworks, which are increasingly used in biomolecular and organic-liquid applications. sGNN~\cite{wang2021scalable} learns transferable intramolecular energy terms for large flexible molecules while retaining a physics-based nonbonded model. ByteFF~\cite{D4SC06640E} extends this parameter-assignment paradigm to an Amber-compatible fixed-charge force field for drug-like molecules, whereas ByteFF-Pol~\cite{zheng_bridging_2026} introduces a polarizable variant aimed at condensed-phase liquids and electrolytes.
In such models, PSWF-LR would not replace the learned bonded or valence terms. Instead, it replaces the LR terms, including the Gaussian Ewald splitting kernel and, in PME-based implementations, the B-spline spreading window. The corresponding real-space complement and self-correction terms are given in Methods, preserving both the downstream MD workflow and the interpretability of the learned force-field parameters. 

This modularity also provides a useful methodological control for the benchmarks included in this work. In the random particle, molecular-dimer and charge-transfer tests, the SR descriptor and latent-variable predictor are fixed, so any accuracy or efficiency gain reflects the change of LR formulation alone. In the runtime and production MD tests, only the splitting kernel, atom-grid spreading and particle--mesh backend are replaced, while the upstream model or force-field parameters are kept fixed. These comparisons show that PSWF-LR reduces reciprocal-space resolution requirements, improves energy, force and charge-derived observables, and accelerates simulations without changing the downstream MLIP workflow.

\subsection*{Experimental design} 
{\bf Random particle systems.} The charge-charge and dispersion benchmarks each comprise $1{,}000$ configurations. $900$ configurations were used for training and 100 for testing. The charge-charge dataset was introduced in Ref.~\cite{king2025machine}; each configuration contains $128$ particles, including $64$ with charge $+1e$ and $64$ with charge $-1e$, and the reference energies and forces were generated with LAMMPS~\cite{thompson2021lammps}. The dispersion dataset follows the structural setup of Ref.~\cite{huguenin2023physics}, but the reference energies and forces were recomputed here using direct summation in LAMMPS to machine precision. Each dispersion configuration contains $64$ particles with heterogeneous dispersion coefficients: $16$ particles have coefficient $2$, $16$ have coefficient $3$, $16$ have coefficient $5$, and $16$ have coefficient $7$. 

For both benchmarks, we use the same CACE SR frontend in the PSWF-LR and Ewald-based models. The CACE descriptor uses six Bessel radial functions, a radial embedding dimension of $12$, angular basis order $\ell_{\max}=3$, many-body order $\nu_{\max}=3$, atomic embedding dimension $N_{\rm embedding}=12$, and no message-passing layer. We test SR cutoffs $r_c=4~\mathring{\mathrm{A}}$ and $5~\mathring{\mathrm{A}}$. The charge benchmark uses one latent scalar per atom for the $p=1$ channel, while the dispersion benchmark uses one latent scalar per atom for the $p=6$ channel. For the Ewald baseline, we use the screening parameter $\alpha=1/\sqrt{2}$, following Ref.~\cite{cheng2025latent}. For PSWF-LR, we use $c=12.024$, corresponding to the tolerance used in the parameter-selection rule above. Because the public LES implementation evaluates the reciprocal-space contribution by direct summation and supports only $p=1$ and $p=6$, this benchmark is used to compare spectral efficiency rather than FFT implementation details.

\vspace{2mm}

\noindent{\bf Molecular dimers.} Reference DFT energies and forces were computed using the HSE06 hybrid density functional~\cite{heyd2003hybrid} together with a nonlocal many-body dispersion correction. For all dimer classes, we use the same CACE SR frontend with a $5~\mathring{\mathrm{A}}$ cutoff, six Bessel radial functions, radial embedding dimension $8$, $\ell_{\max}=2$, $\nu_{\max}=2$, $N_{\rm embedding}=3$, and one message-passing layer. We compare three variants: the SR-only CACE model, the same SR model augmented with the Ewald-based CACE-LR backend of Ref.~\cite{king2025machine}, and the same SR model augmented with PSWF-LR. In both CACE-LR and PSWF-LR, the LR component uses a one-dimensional latent variable. The Ewald baseline uses $\alpha=1/\sqrt{2}$, while PSWF-LR uses $c=12.024$. Unless otherwise stated, the mesh spacing for PSWF-LR is fixed at $h=1.5~\mathring{\mathrm{A}}$. Because the available CACE-LR implementation supports only Coulombic and dispersion channels, we train the two supported Ewald-based variants, corresponding to $p=1$ and $p=6$, and report the better-performing variant as a favorable Ewald-based baseline for each dimer class. In the matched-exponent experiments, PSWF-LR assigns $p=1,\ldots,6$ to the CC, CP, PP, CA, PA and AA dimer classes, respectively.
\vspace{2mm}

\noindent{\bf Polar dipeptides.} We use polar dipeptides from the SPICE dataset, selecting 12 oppositely charged dipeptides formed from one positively charged residue (Arg, Lys or HIP) and one negatively charged residue (Glu or Asp), including both sequence orders, with 50 conformers per dipeptide. We retain the conformers of one of the 12 dipeptides as a test set and $10\%$ of the remaining structures as a validation set. For these calculations, we use CACE as the SR descriptor, with a $4~\mathring{\mathrm{A}}$ cutoff, six trainable Bessel radial functions, $12$ radial embedding dimensions, $\ell_{\max}=3$, $\nu_{\max}=3$, one message-passing layer, and $N_{\rm embedding}=4$. Because the systems are aperiodic, PSWF-LR contributions are evaluated in real space. We use two channels $\mathcal{S}=\{p,6\}$, with $p\in\{1,2,3\}$. The $p$ channel captures the electrostatic-like LR contribution, whereas the $p=6$ channel accounts for dispersion. Model-predicted effective charges are compared with MBIS-, Hirshfeld- and Mulliken-like charge partitions. Dipoles, traceless quadrupoles and Born effective charges are computed from the predicted charge distributions and compared with the corresponding DFT-derived references. This setup is used to test whether the exponent that best reproduces an atomic charge partition is also the exponent that best reproduces molecular response observables.
\vspace{2mm}

\noindent{\bf Nonlocal charge transfer.} We evaluate nonlocal charge transfer on five benchmarks: Na$_9$Cl$_8^+$/Na$_8$Cl$_8^+$ ionic clusters, terminally protonated carbon chains (C$_{10}$H$_2$/C$_{10}$H$_3^+$), Au$_2$ adsorbed on MgO(001) with and without Al doping, dry Cu(111)--benzotriazole, and solvated Cu(111)--benzotriazole. Unless otherwise stated, a $90\%/10\%$ train/test split is used.

For Na$_{8/9}$Cl$_{8}^{+}$, we use PSWF-LR with a CACE SR descriptor, $r_c=5.29~\mathring{\mathrm{A}}$, six Bessel radial functions, radial embedding dimension $8$, $\ell_{\mathrm{max}}=3$, $\nu_{\mathrm{max}}=3$, no message passing, and $N_{\mathrm{embedding}}=2$. For C$_{10}$H$_2$/C$_{10}$H$_3^{+}$, we use a MACE SR descriptor with $r_c=4.23~\mathring{\mathrm{A}}$, eight radial basis functions, polynomial order $5$, highest angular momentum $\ell_{\mathrm{max}}=1$, $32$ channels, no message passing, and many-body order $4$. For Au$_2$-MgO(001), we use a CACE SR descriptor with $r_c=5.5~\mathring{\mathrm{A}}$, six Bessel radial functions, radial embedding dimension $12$, $\ell_{\mathrm{max}}=3$, $\nu_{\mathrm{max}}=3$, no message passing, and $N_{\mathrm{embedding}}=4$. For Cu-BTA and solvated Cu-BTA, we use two SR descriptors within PSWF-LR: CACE, with six Bessel radial functions, radial embedding dimension $12$, $\ell_{\mathrm{max}}=3$, $\nu_{\mathrm{max}}=3$, no message passing, and $N_{\mathrm{embedding}}=4$; and NequIP, with four interaction layers, $\ell_{\max}=1$, parity-enabled hidden features of multiplicity $32$, eight trainable Bessel radial basis functions, and polynomial cutoff order $6$. 

The PSWF-LR entries in Table~\ref{tab:charge_transfer_rmse} for the Cu--BTA benchmarks use the CACE SR backbone. Same-backbone NequIP+PSWF-LR and CACE+PSWF-LR comparisons are reported in Extended Data Fig.~\ref{fig:MLIPCuBTAHOS}. For the same-backbone comparisons, the SR architecture, train/test split, local descriptor settings and training protocol are kept fixed; only the LR evaluation layer is changed. This protocol is used to separate the effect of the PSWF-based LR mechanism from changes in the local representation.

In Table~\ref{tab:charge_transfer_rmse}, the results for $\chi+\eta$ (ACE) and CACE-LR are taken from Ref.~\cite{king2025machine}, results for 3G-HDNNP and 4G-HDNNP from Ref.~\cite{ko2021fourth}, and results for NequIP-LR from Ref.~\cite{maruf2025learning}. In Extended Data Figure~\ref{fig:MLIPNaClSI}, the CHGNet~\cite{deng2023chgnet} and MACE-MP-0~\cite{batatia2025foundation} results were obtained by fine-tuning the corresponding pretrained foundation-model checkpoints on the Na$_{8/9}$Cl$_8^{+}$ dataset using the same train/test split and evaluation protocol as for the models trained in this work. The 4G-HDNNP results are taken from Ref.~\cite{ko2021fourth}, and the CACE-LR results are computed using the model of Ref.~\cite{king2025machine}. In Extended Data Figs.~\ref{fig:MLIP4GMgO} and~\ref{fig:MLIPAu2MgO}, the DFT, 2G-HDNNP, and 4G-HDNNP results are taken from Ref.~\cite{ko2021fourth}, whereas the CACE results are generated using models trained in this work. In Extended Data Fig.~\ref{fig:MLIPCuBTAHOS}, the NequIP-LR model is trained from scratch using the input file provided in Ref.~\cite{maruf2025learning}, whereas the CACE-LR model is trained using a one-dimensional latent variable with Ewald parameter $\alpha=1/\sqrt{2}$ and $h=0.5~\mathring{\mathrm{A}}$. For PSWF-LR, we use $c=12.024$, $h=1.5~\mathring{\mathrm{A}}$, and $\mathcal{S}=\{1,2,3,4,5,6\}$, except in the NaCl exponent-scan experiments (Extended Data Figs.~\ref{fig:MLIPNaClPC} and~\ref{fig:MLIPNaClCharge}), where we use $\mathcal{S}=\{p,6\}$ with $p\in\{1,2,3\}$. In these exponent-scan tests, the $p$-channel latent variable is used to predict the effective charge. The Na$_{8/9}$Cl$_8^{+}$ and C$_{10}$H$_2$/C$_{10}$H$_3^{+}$ systems are aperiodic, so PSWF-LR contributions are evaluated in real space. The Au$_2$-MgO(001), dry Cu-BTA, and solvated Cu-BTA systems are periodic, so PSWF-LR contributions are evaluated in reciprocal space.
\vspace{2mm}

\noindent{\bf Performance comparison in differentiable long-range evaluations.} The differentiable runtime benchmarks use three periodic systems: a NaCl crystal, a SWM4-NDP polarizable water model and a LiPF$_6$/DMC electrolyte. For NaCl, we use an analytically generated cubic rocksalt reference cell implemented directly in Torch-PME and JAX-PME, containing four NaCl formula units in a normalized $2\times2\times2$ cubic cell. This system serves as a controlled charge-only reference rather than a density-matched atomistic model. For water, we use a SWM4-NDP benchmark derived from an OpenMM-generated periodic PDB, with an orthorhombic $15.0\times15.0\times15.0~\mathring{\mathrm{A}}^3$ box containing $112$ five-site water molecules. Electrostatics are assigned from the SWM4-NDP site-charge map. For LiPF$_6$/DMC, we use a local EC/DMC+LiPF$_6$ base cell generated through the ByteFF parameterization workflow~\cite{zheng_bridging_2026}. The resulting cubic $53.4\times53.4\times53.4~\mathring{\mathrm{A}}^3$ cell contains $10,062$ atoms and is evaluated with the generated fixed-charge force field and standard 1--4 scaling parameters.

Larger systems are generated by deterministic supercell replication. The tested ranges span $512$ to $7,077,888$ sites for NaCl, $560$ to $18,350,080$ sites for SWM4-NDP, and $10,062$ to $10,303,488$ sites for LiPF$_6$/DMC. We report full forward-plus-backward wall-clock times, including both the LR calculation and the shared automatic-differentiation overhead. Each timing value is collected after five untimed warm-up evaluations and averaged over $20$ timed evaluations. For PSWF-LR and PME, six predefined exponent settings, $p=1,\ldots,6$, are evaluated at each system size. Reciprocal meshes are generated by the native setup routines of each framework and logged for every successful run. For matched-setting comparisons, PSWF-LR and PME are run at the same target LR tolerance whenever applicable, while the same particle configurations, charges, system sizes and automatic-differentiation pipelines are used. In all timing plots, the standard deviation across repeated measurements is smaller than the plotting marker size.

\vspace{2mm}

{\bf{MD simulations.}} MD validation is performed on two practical systems. For the Pt(111)-water benchmark, we use an $8,154$-atom periodic Pt slab/water configuration and the same trained CACE SR model for both PSWF-LR and Ewald-based LR treatments. The PSWF-LR and Ewald trajectories use identical model weights, starting structures, neighbor-list settings, cutoffs and thermostat/barostat settings; only the LR component is changed. For the PSWF version, we use $S=\{1,6\}$, where the $p=1$ channel represents the electrostatic-like LR contribution and the $p=6$ channel accounts for the dispersion-like tail. Water density profiles are computed by binning water oxygen positions according to their distance from the nearest Pt surface and converting the resulting number density to mass density. Chemisorbed and physisorbed regions are assigned using the same surface-distance windows in both backends. The orientational order is computed from the cosine of the angle between the water molecular axis and the surface normal, and the layer-resolved charge response is computed as the standard deviation of the total learned charge in each interfacial region over sampled frames. MD throughput is reported in ns$\cdot$day$^{-1}$ from repeated timed propagation runs on A100, H100, H200 and RTX-Blackwell GPU nodes.

For the electrolyte benchmark, we use a periodic LiPF$_6$/DMC/EC system with DMC:EC:LiPF$_6$ = $5,053$\,:\,$3,450$\,:\,$690$, corresponding to $100,656$ atoms. This system is generated by deterministic replication of the $10,062$-atom ByteFF-Pol electrolyte cell. PSWF and PME simulations use the same ByteFF-Pol checkpoint, force-field parameters, bonded terms, SR nonbonded functional forms, initial coordinates and temperature/pressure protocol; only the LR component is changed. In the PSWF version, permanent-charge interactions are evaluated with the $p=1$ channel; induced-dipole LR contributions are represented through $p=2$ and $p=3$ channels; and the asymptotic charge-transfer and dispersion tails are represented by $p=4$ and $p=6$ channels, respectively. The OpenMM MD benchmark uses a multiple-time-step integration scheme based on the BAOAB-RESPA Langevin algorithm~\cite{lagardere2019pushing}, with an outer timestep of $2$~fs and an inner timestep of $1$~fs. Simulations are conducted at $300$~K with a friction coefficient of ($0.1~\mathrm{ps}^{-1}$).

NPT density traces are computed from the instantaneous box volume and total system mass. The plotted density comparison uses the first $2$~ns of NPT production, with Gaussian smoothing applied only for visualization; the raw density trace is retained as a translucent background. Transport and thermodynamic observables are computed from paired NVT trajectories. Molecular centers of mass are unwrapped under periodic boundary conditions, and diffusion coefficients are obtained from the long-time Einstein slopes of the center-of-mass mean-squared displacements, fitted over $1.0$--$2.5$~ns lag times. The ionic conductivity is estimated with the Nernst--Einstein relation using the Li$^+$ and PF$_6^-$ diffusion coefficients and therefore does not include ion-correlation corrections. The heat capacity is estimated from total-energy fluctuations after discarding the first $1$~ns of NVT sampling. Error bars denote standard errors from five contiguous trajectory blocks. Solvent-resolved Li--O radial distribution functions are computed from NPT frames using separate carbonyl-oxygen selections for DMC and EC. For the charge-structure analysis, we use NVT configurations and wavevectors $\mathbf{k}=(0,0,k_z)$, with
\begin{equation}
S(k_z)=\frac{\langle |\rho_q(k_z)|^2\rangle}{\sum_{i} q_i^2},\qquad
\rho_q(k_z)=\sum_{i=1}^N q_i e^{ik_z z_i},
\end{equation}
where $q_i$ is the partial charge of site $i$. The low-frequency inset reports the corresponding quantity $\langle|\rho_q(k_z)|^2\rangle/(Vk_z^2)$, where $V$ is the mean simulation-cell volume.

\subsection*{Hardware and software}
The computations in this work are performed on the Flatiron Institute Rusty cluster, supported by the Scientific Computing Core at the Flatiron Institute. The GPU nodes used in this study were equipped with one of the following configurations: NVIDIA A100 GPUs with 80 GB of memory per device, 64 Ice Lake CPU cores, and 1 TB of system memory; NVIDIA H100 GPUs with 80 GB of memory per device, 64 Ice Lake CPU cores, and 1 TB of system memory; NVIDIA H200 GPUs with 141 GB of memory per device, 96 Emerald Rapids CPU cores, and 2 TB of system memory; or NVIDIA RTX 6000 Pro Blackwell GPUs with 96 GB of memory per device, 144 Granite Rapids CPU cores, and 1 TB of system memory. The nodes are connected by an HDR-200 InfiniBand network with a bandwidth of 200 Gb/s. The PSWF-LR were implemented in PyTorch, JAX, and CUDA to enable direct comparisons with Torch-PME, JAX-PME, and native OpenMM.

\section*{Data Availability}
The benchmark datasets used in this study are available from the original sources cited in Methods and Results. Processed train/validation/test splits, derived labels, trained-model predictions, timing logs, MD observables and figure source data will be deposited in a public repository upon publication. During peer review, the processed data and figure source files are available to editors and reviewers through a private repository.

\section*{Code Availability}
The CACE, NequIP, PyTorch, JAX and OpenMM/CUDA implementations of PSWF-LR, together with scripts used to reproduce the main benchmarks and figures, will be made publicly available upon publication. During peer review, the code is available to editors and reviewers through a private repository.

\bibliographystyle{naturemag}
\bibliography{hpcewald,methods}

\section*{Acknowledgments}
We thank the Scientific Computing Core at the Flatiron Institute for access to the Rusty cluster and technical support. We also thank Yutong Zhao for helpful discussions. J.L. acknowledges support from the National Natural Science Foundation of China (Grant No. 12401570) and the China Postdoctoral Science Foundation
(Grant No. 2024M751948). Y. Ji acknowledges support from the National Natural Science Foundation of China (No. 124B2023). The Flatiron Institute is a division of the Simons Foundation. 

\section*{Author contributions}
S.J. and J.L. conceived the project. S.J. and J.L. designed the experiments and supervised the research. J.L., L.L. and Y.J. contributed to implementing the PSWF-LR framework in CACE and NequIP. J.L. performed model training, benchmark evaluations and data analysis. J.L. and L.L. implemented PSWF-LR in Torch-PME, JAX-PME and ByteFF-Pol/OpenMM, and carried out the differentiable-runtime and MD simulations. J.L. and Y.J. designed the figures and prepared the visualizations. S.J. and J.L. wrote the initial manuscript draft. All authors discussed the results, revised the manuscript and approved the final version.

\subsection*{Correspondence}
Correspondence and requests for materials should be addressed to S.J..

\section*{Competing interests}
The authors declare no competing interests.

\renewcommand{\figurename}{Extended Data Figure}
\setcounter{figure}{0} 
\renewcommand{\tablename}{Extended Data Table}
\setcounter{table}{0}
\begin{figure}[!htbp]
\centering
\includegraphics[width=0.86\linewidth]{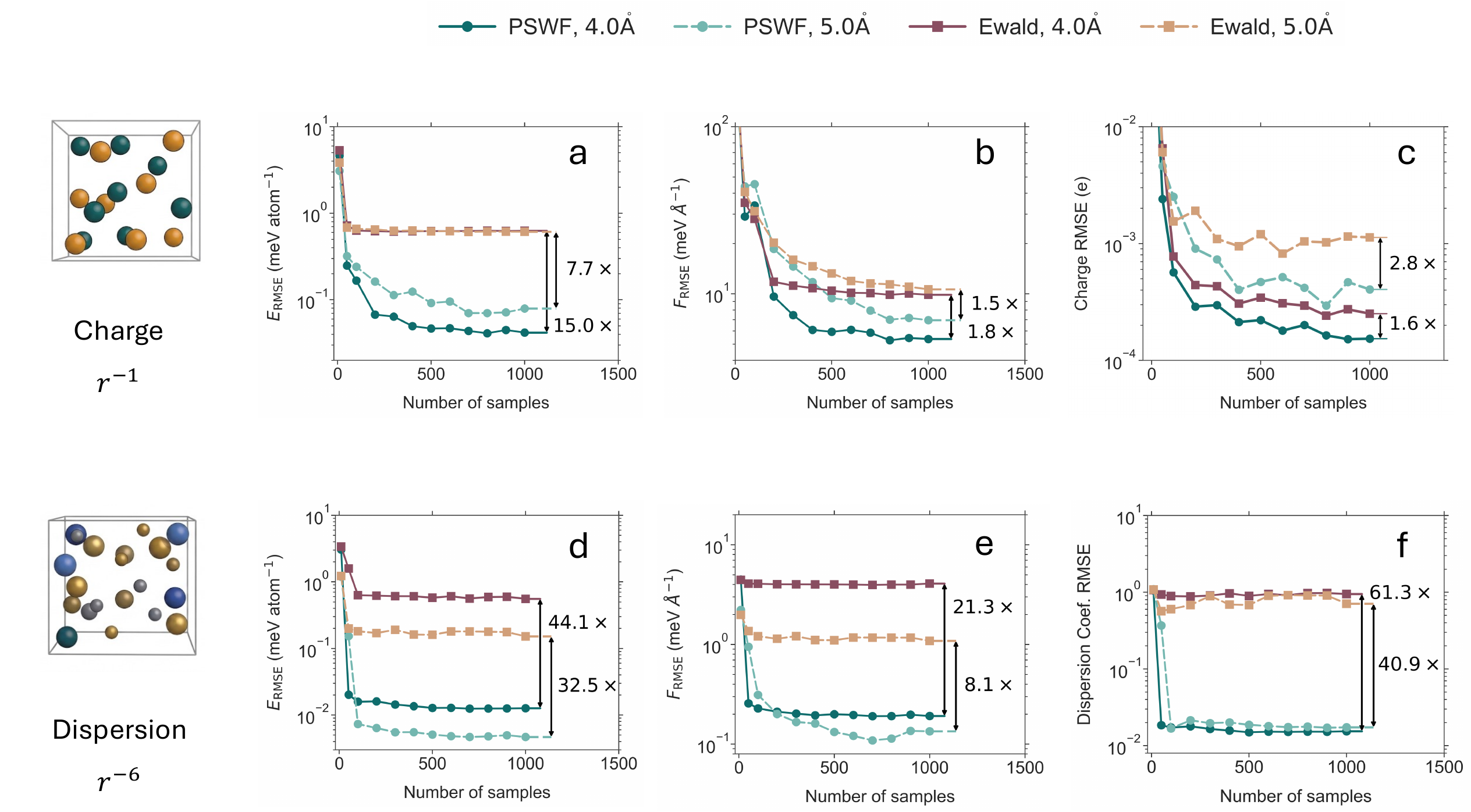}
\vspace{3mm}
\caption{\sf PSWF-LR retains its advantage over Ewald-based screening across training-set sizes in the synthetic particle benchmarks. {\bf a-c}, Charge-charge ($1/r$) benchmark. {\bf d-f}, Dispersion ($1/r^6$) benchmark. In each case, the model uses the same short-range baseline and differs only in whether the long-range module is PSWF-based or Ewald-based. Root-mean-square errors (RMSEs) of the energy ($E_{\mathrm{RMSE}}$; {\bf a,d}) and forces ($F_{\mathrm{RMSE}}$; {\bf b,e}) are shown as a function of the number of training samples. The rightmost panels report the RMSE of the predicted charges for $1/r$ benchmark ({\bf c}) and that of the predicted dispersion coefficients for the $1/r^6$ benchmark ({\bf f}). Results are reported for short-range cutoffs $r_c = 4.0~\mathring{\mathrm{A}}$ and $5.0~\mathring{\mathrm{A}}$.}
\label{fig:MLIPRandomParticleSample}
\end{figure}

\vspace{5mm}

\begin{figure}[!htbp]
\centering
\includegraphics[width=0.86\linewidth]{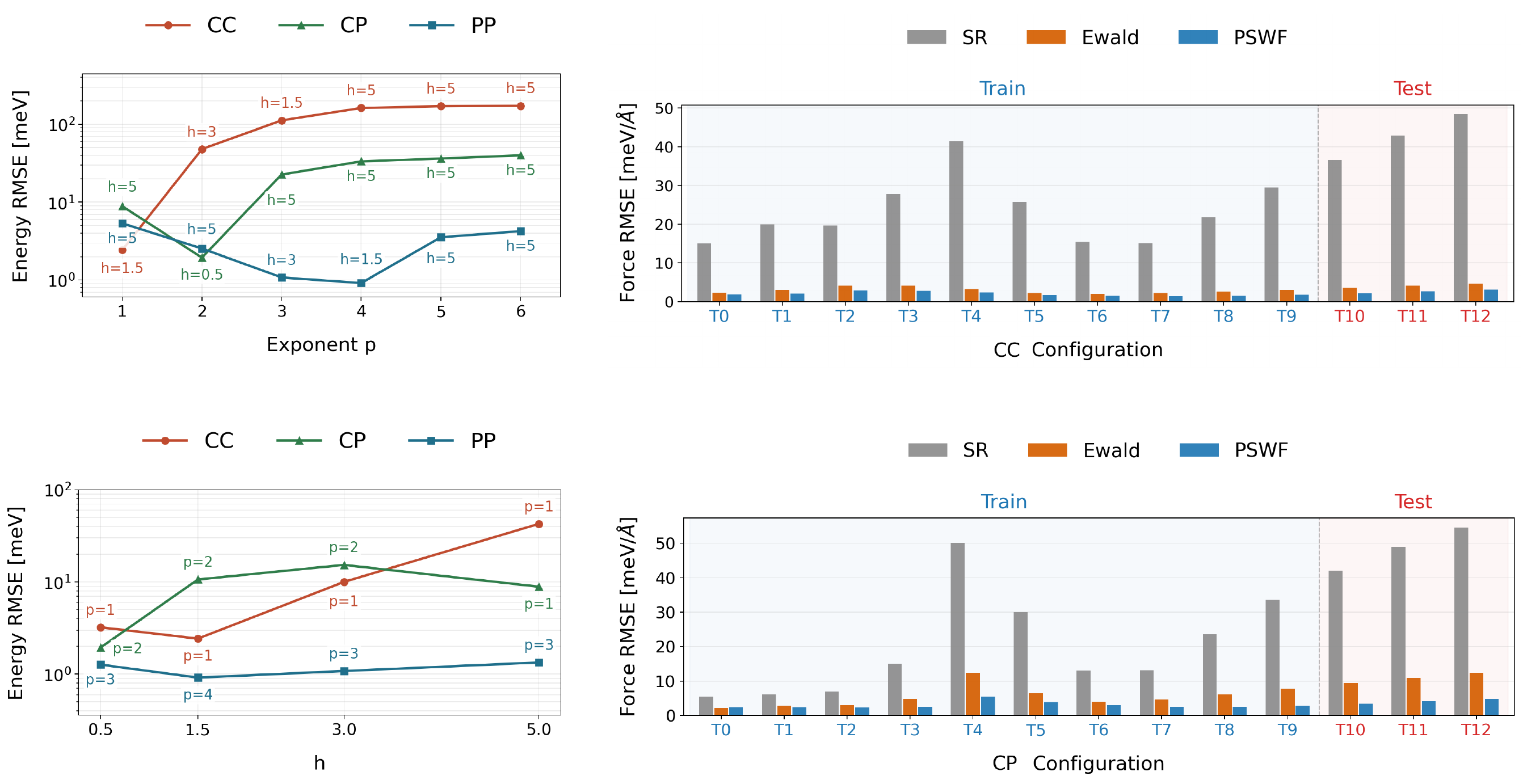}
\caption{\sf Supplementary analysis of the molecular-dimer benchmark. Left, dependence of the energy RMSE on the PSWF exponent $p$ and mesh size $h$ for the charge-charge (CC), charge-polar (CP) and polar-polar (PP) dimers. In the upper-left panel, the label next to each point gives the value of $h$ that minimizes the RMSE among the tested mesh sizes $h=0.5$, $1.5$, $3$ and $5~\mathring{\mathrm{A}}$; in the lower-left panel, the label gives the value of $p$ that minimizes the RMSE among the tested exponents $p=1,2,3,4,5,6$. Right, per-configuration force RMSEs for the CC (top) and CP (bottom) datasets, comparing the short-range-only model (SR) with the same baseline augmented by PSWF- or Ewald-based long-range modules. Blue and red shaded regions denote training and test configurations, respectively.}
\label{fig:MLIPDimerSI}
\end{figure}

\begin{figure}[!htbp]
\centering
\includegraphics[width=0.82\linewidth]{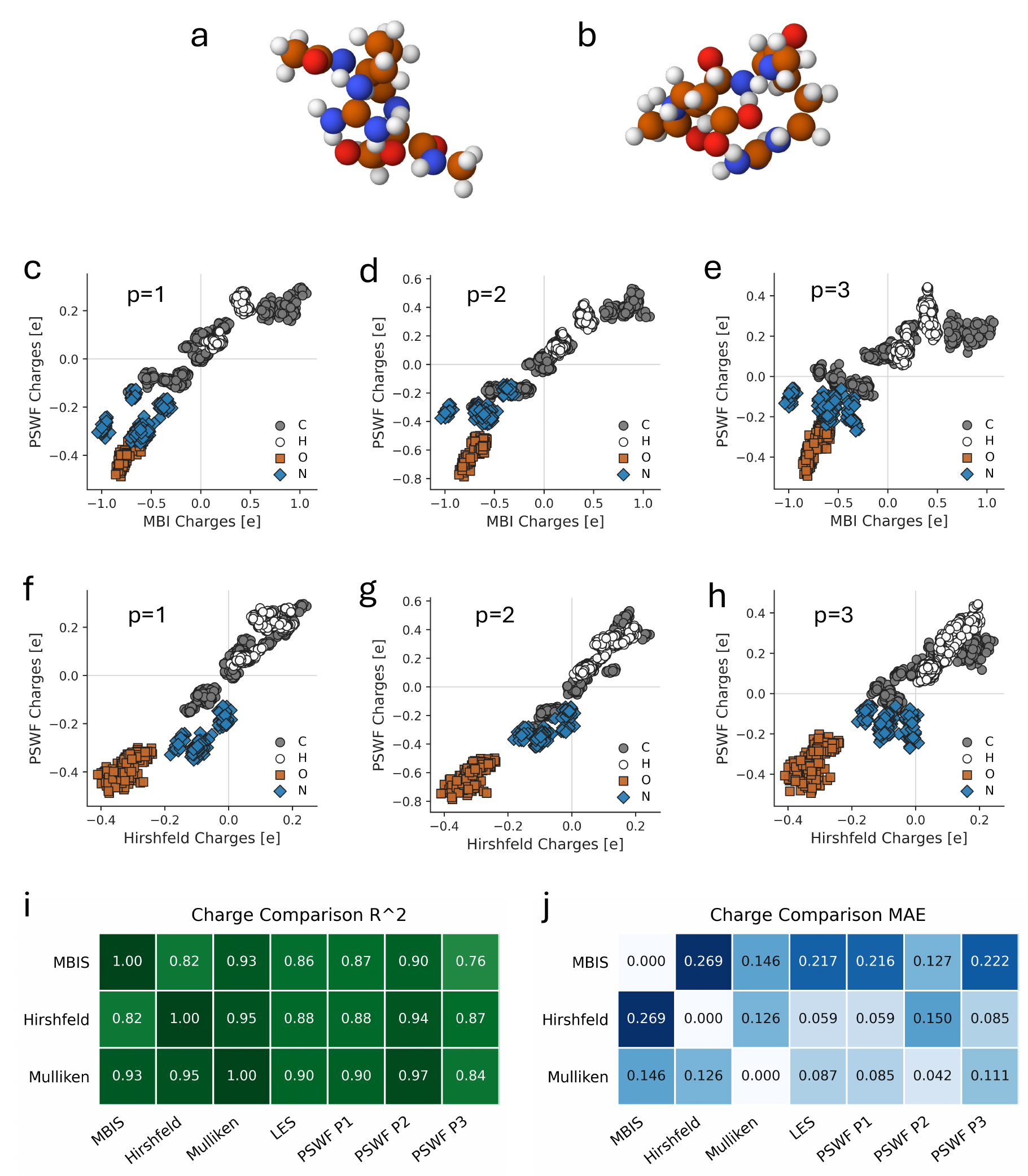}\vspace{5mm}
\caption{\sf The PSWF exponent that best reproduces reference atomic charges in the dipeptide benchmark differs across charge partitions. {\bf a,b}, Representative dipeptide conformers. {\bf c-e}, Scatter plots comparing PSWF-predicted effective charges with MBIS charges for $p=1$, $2$, and $3$. {\bf f-h}, Corresponding comparisons with Hirshfeld charges. Points in {\bf c-h} are colored by atomic species. {\bf i,j}, Pairwise comparison of charge-partitioning schemes and model-predicted charges, including MBIS, Hirshfeld, Mulliken, LES, and PSWF charges with $p=1$, $2$, and $3$, shown as the coefficient of determination ($R^2$; {\bf i}) and mean absolute error (MAE; {\bf j}). Among the PSWF exponents tested here, $p=2$ gives the best overall agreement with the reference charge schemes.}
\label{fig:MLIPDipeptide}
\end{figure}

\begin{figure}[!htbp]
\centering
\includegraphics[width=0.85\linewidth]{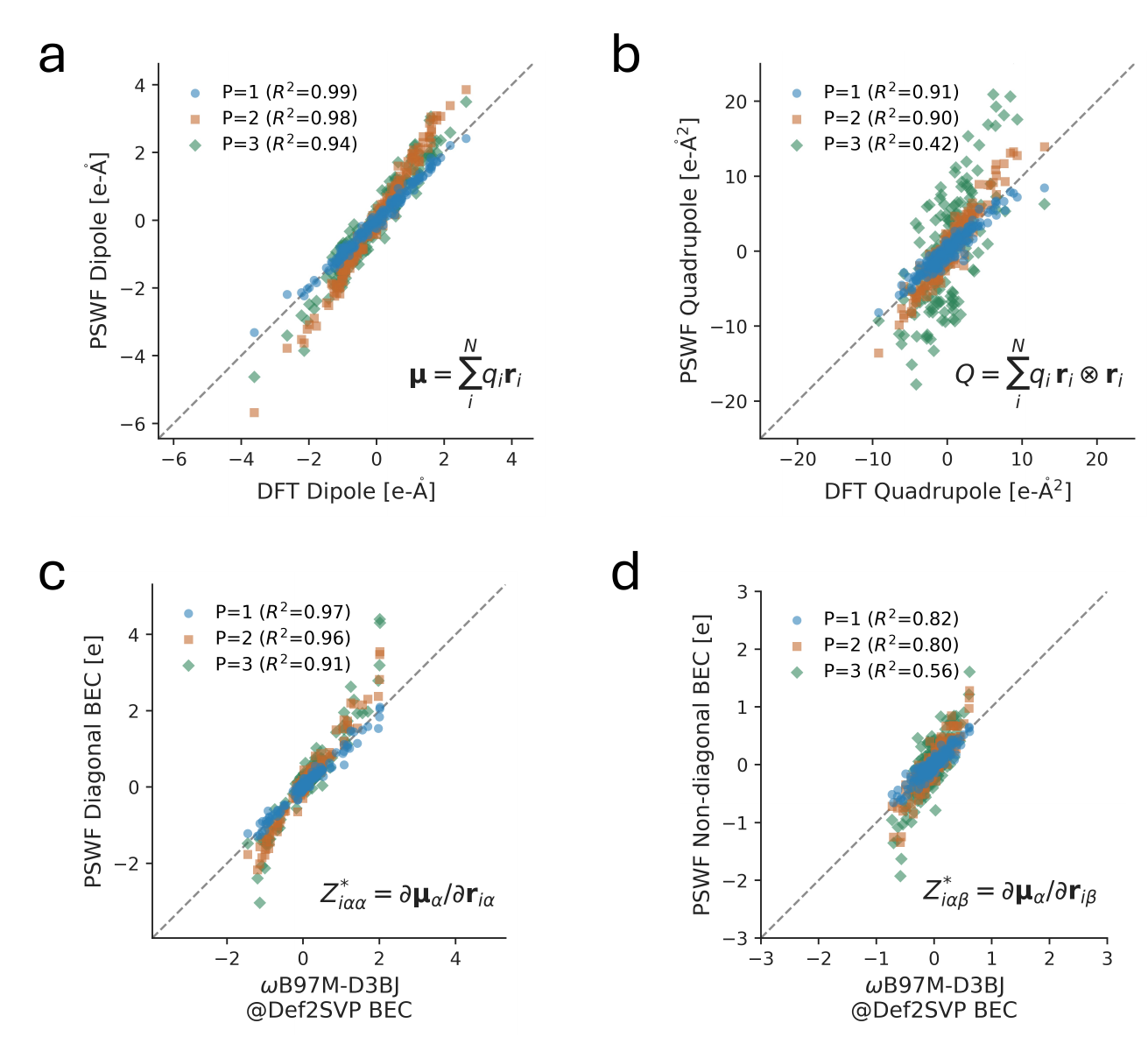}
\caption{\sf Lower PSWF exponents better reproduce charge-derived multipoles and Born effective charges in the dipeptide benchmark.  {\bf a}, Dipole moments computed from the PSWF-predicted charges and compared with the reference DFT dipoles from the SPICE dataset. {\bf b}, Traceless quadrupole tensor components computed from the PSWF-predicted charges and compared with the corresponding DFT quadrupole components. {\bf c}, Diagonal Born effective charge (BEC) components predicted by PSWF-LR and compared with reference BECs calculated using the $\omega$B97M-D3BJ functional with the def2-SVP basis set on the validation set reported in~\cite{king2025machine}. {\bf d}, Corresponding comparison for the off-diagonal BEC components. Data are shown for $p=1$, $2$, and $3$. Among the exponents tested here, $p=1$ gives the best overall agreement with the DFT dipole, quadrupole, and BEC data, although $p=2$ better matches the reference atomic charges (Extended Data Fig.~\ref{fig:MLIPDipeptide}).}
\label{fig:MLIPDipole}
\end{figure}

\begin{figure}[!htbp]
\centering
\includegraphics[width=0.95\linewidth]{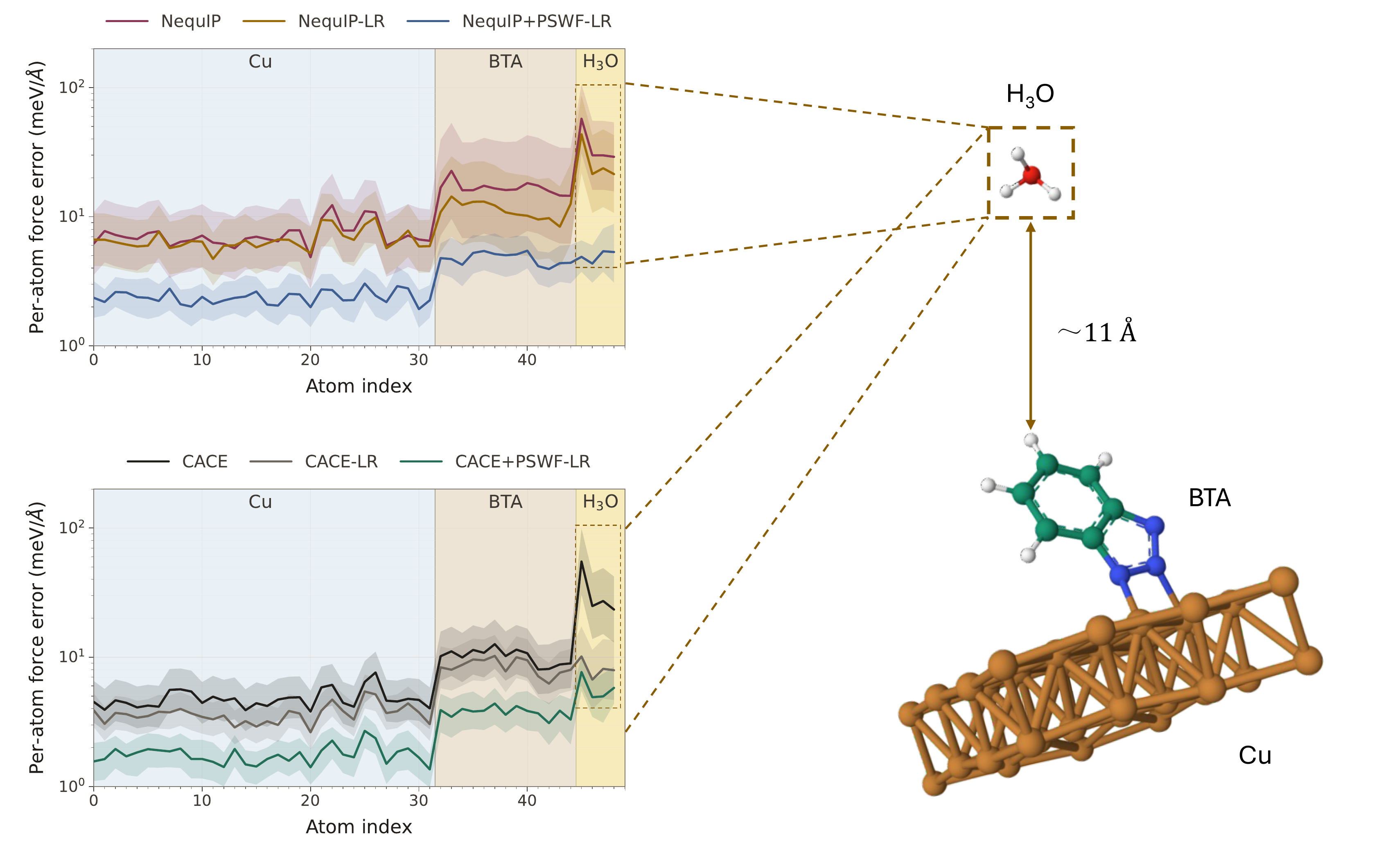}
\caption{\sf PSWF-LR lowers force errors in the adsorbate and solvent region of the solvated Cu-BTA interface. Upper left, comparison of NequIP, NequIP-LR and NequIP+PSWF-LR. Lower left, analogous comparison of CACE, CACE-LR and CACE+PSWF-LR. Force errors are plotted as a function of atom index, with Cu substrate atoms, BTA adsorbate atoms and solvent atoms distinguished by the shaded regions. Right, representative solvated interfacial configuration with the molecular region highlighted.}
\label{fig:MLIPCuBTAHOS}
\end{figure}

\begin{figure}[!htbp]
\centering
\includegraphics[width=0.9\linewidth]{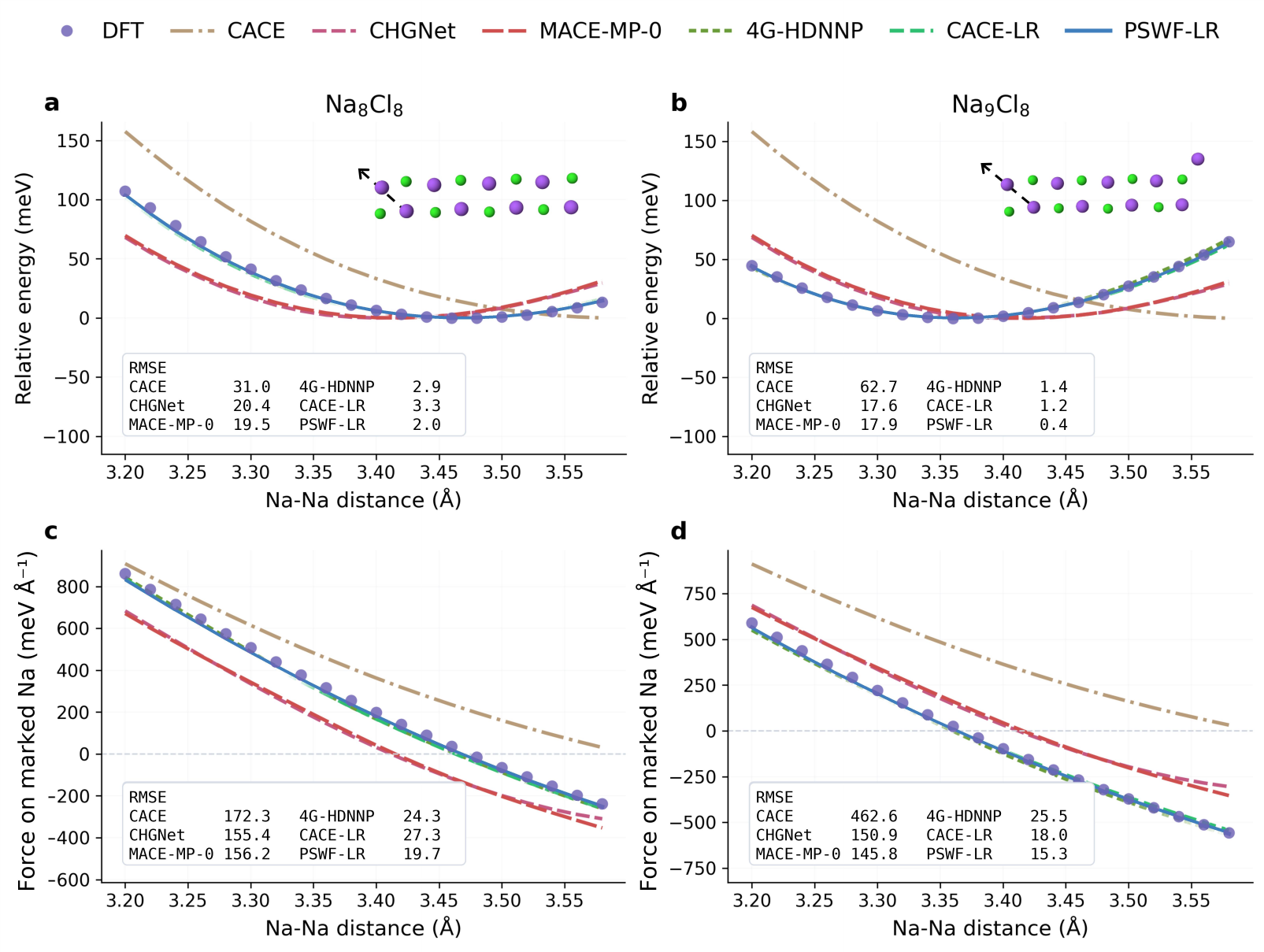}
\caption{\sf Comparison of DFT reference data with three short-range models (CACE, CHGNet and MACE-MP-0 finetuned on this dataset) and three long-range models (4G-HDNNP, CACE-LR and PSWF-LR), for the Na$_8$Cl$_8^+$ cluster. {\bf a,b}, Relative energies, referenced to the minimum DFT energy of the corresponding cluster, as a function of the Na-Na distance for two representative separation scans; the corresponding configurations are shown in the insets. The arrow indicates the direction along which the outermost sodium atom is displaced, and the dashed line indicates the measured Na-Na distance. {\bf c,d}, Force acting on the highlighted Na ion along the same scans. Root-mean-square errors (RMSEs) with respect to the DFT are reported in each panel.}
\label{fig:MLIPNaClSI}
\end{figure}

\begin{figure}[!htbp]
\centering
\includegraphics[width=0.8\linewidth]{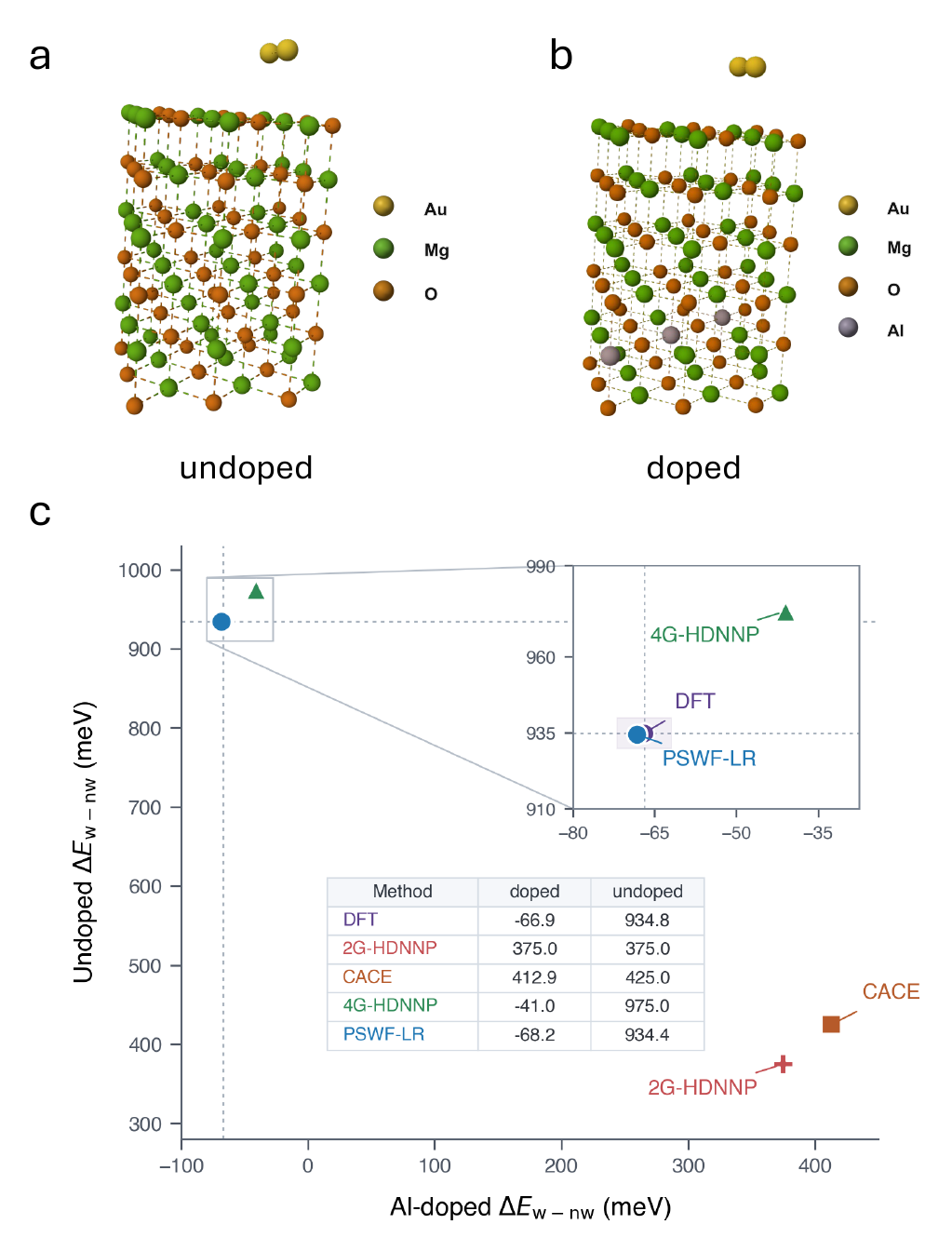}
\caption{\sf PSWF-LR reproduces the wetting-non-wetting energy splitting of Au$_2$ on MgO more accurately than the comparison models. {\bf a,b}, Representative undoped and Al-doped structures, respectively. Au, Mg, O, and Al atoms are shown in gold, green, orange, and grey, respectively. {\bf c}, Energy difference between the wetting and non-wetting configurations, $\Delta E_{\mathrm{w-nw}}$, for the Al-doped system ($x$ axis) and the undoped system ($y$ axis), as predicted by DFT and four MLIPs (2G-HDNNP, CACE, 4G-HDNNP, and PSWF-LR). The inset enlarges the region near the DFT reference, and the table reports the corresponding values in meV.}
\label{fig:MLIP4GMgO}
\end{figure}

\begin{figure}[!htbp]
\centering
\includegraphics[width=0.89\linewidth]{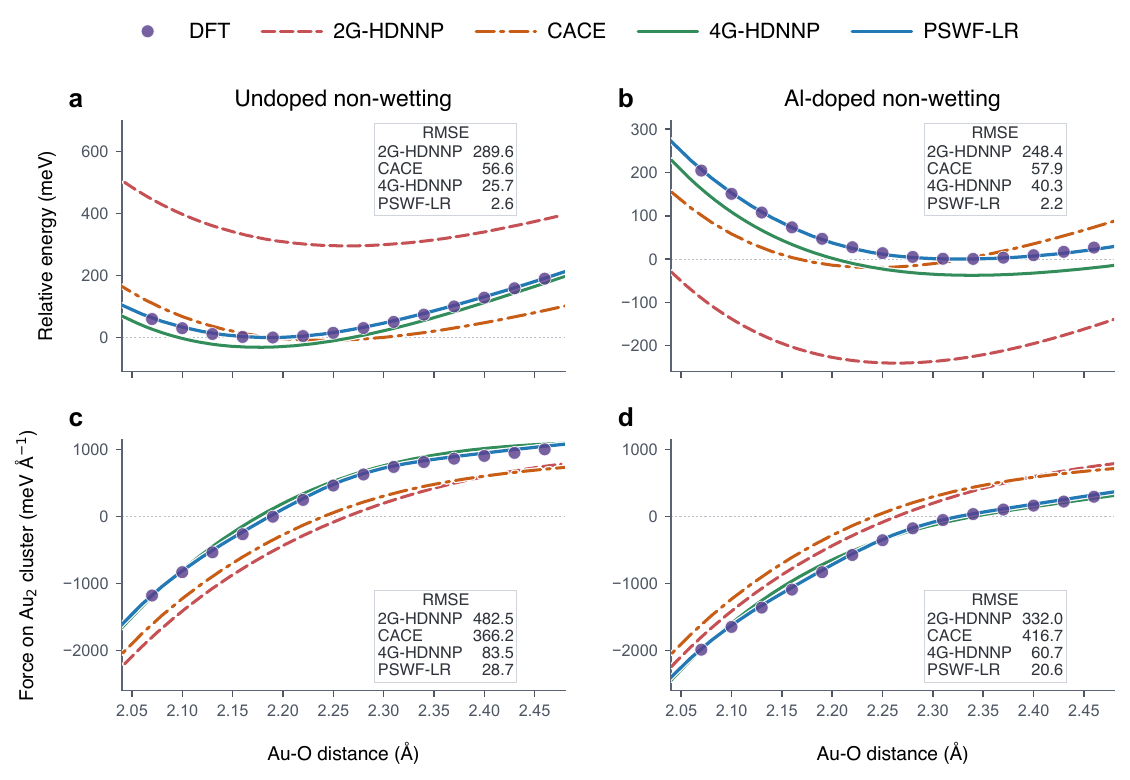}
\caption{\sf Comparison of MLIPs on capturing the Au-O equilibrium position and force profile for Au$_2$ on MgO(001). The horizontal axis is the Au-O bond length, defined as the distance between the Au atom closest to the surface and its neighboring oxygen atom. {\bf a,b}, Relative energy for the non-wetting geometry on undoped and Al-doped substrates. {\bf c,d}, Corresponding total force acting on the Au$_{2}$ cluster. Results are shown for two short-range models (2G-HDNNP and CACE) and two long-range models (4G-HDNNP and PSWF-LR).}
\label{fig:MLIPAu2MgO}
\end{figure}

\begin{figure}[!htbp]
\centering
\includegraphics[width=0.8\linewidth]{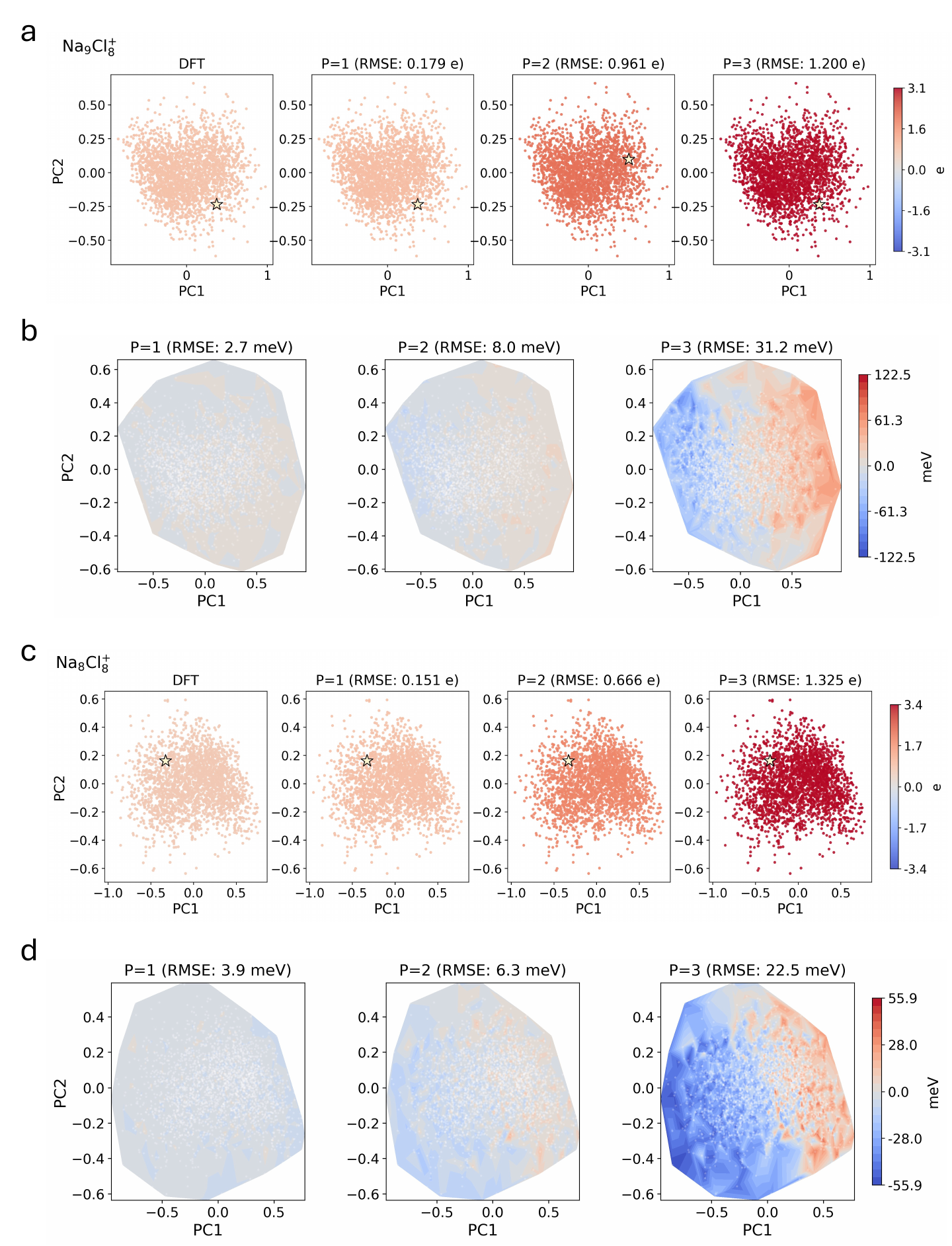}\vspace{5mm}
\caption{\sf Charge-transfer landscapes and relative-energy error surfaces as a function of the long-range decay exponent in Na$_9$Cl$_8^{+}$ and Na$_8$Cl$_8^{+}$. ({\bf a,c}) DFT and PSWF-LR charge-transfer patterns projected onto a shared PC1-PC2 structural embedding for Na$_9$Cl$_8^{+}$ and Na$_8$Cl$_8^{+}$, respectively, for $p=1$, $2$ and $3$ with all other settings fixed. PC1 and PC2 denote the first two principal components of a pairwise-distance descriptor constructed from all Na-Na, Na-Cl and Cl-Cl distances for each composition. Colors represent the charge-transfer descriptor, defined as the difference between the mean Na and Cl atomic charges, and the hollow star marks the representative highest-energy configuration identified from each model. ({\bf b,d}) Corresponding relative-energy error surfaces, $\Delta E = E_{\mathrm{MLIP}} - E_{\mathrm{DFT}}$, over the same configurational space. Increasing the PSWF exponent progressively distorts the charge-transfer and energy
landscapes of the NaCl cluster ions.}
\label{fig:MLIPNaClPC}
\end{figure}

\begin{figure}[!htbp]
\centering
\includegraphics[width=0.8\linewidth]{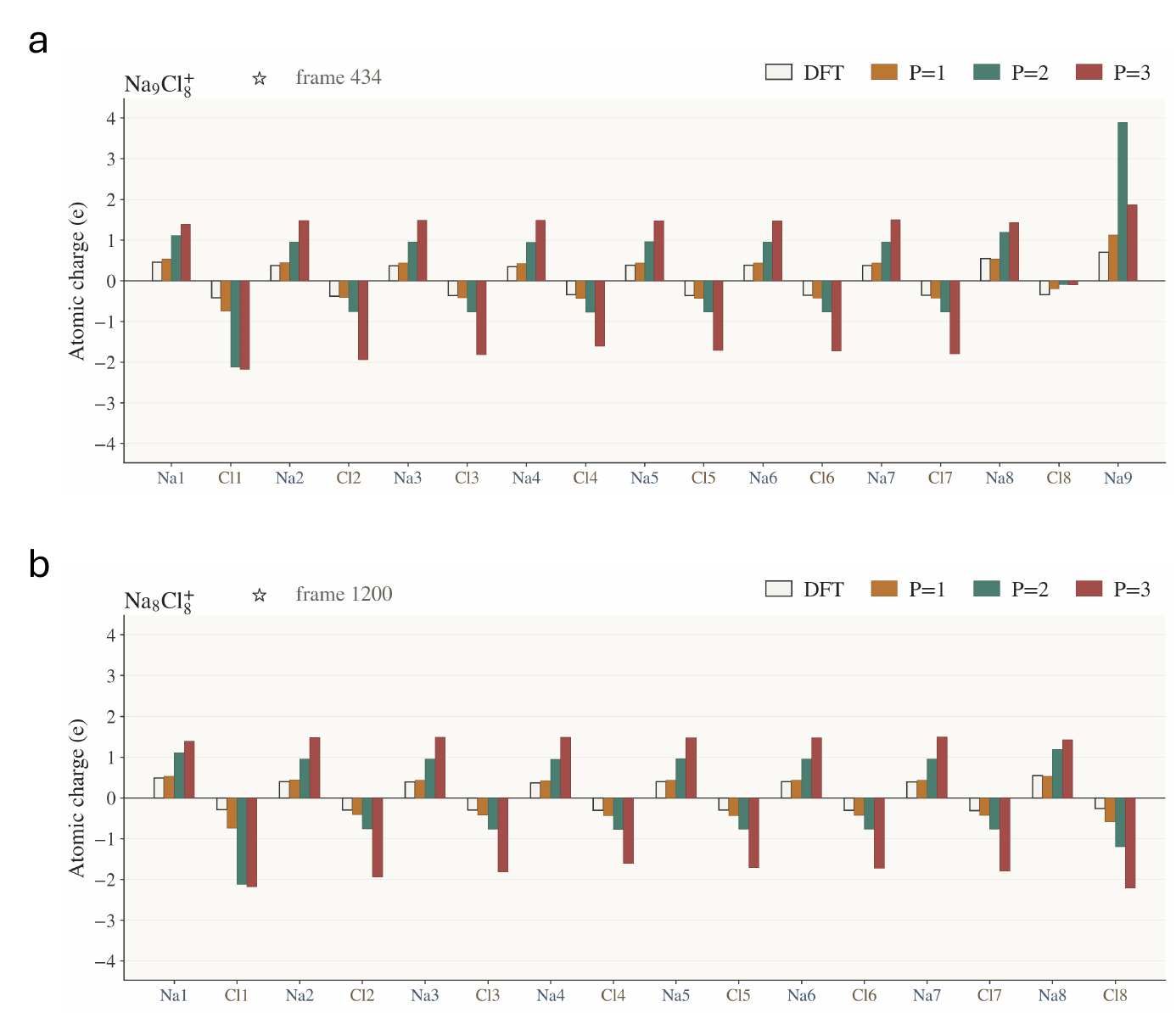}
\caption{\sf DFT and PSWF-LR-predicted effective charges are compared for the representative highest-energy configurations of Na$_9$Cl$_8^{+}$ and Na$_8$Cl$_8^{+}$ marked in panels {\bf a} and {\bf c} of Extended Data Figure~\ref{fig:MLIPNaClPC}, using $p=1, 2,$ and $3$ with all other settings fixed. In line with the charge-transfer landscapes and relative-energy error surfaces, $p=1$ shows the closest agreement with DFT, whereas larger exponents lead to progressively larger deviations in the atomic charge distribution. All PSWF-LR models nevertheless preserve the expected charge polarity, with electron density transferred from Na to Cl.}
\label{fig:MLIPNaClCharge}
\end{figure}

\begin{figure}[!htbp]
\centering
\includegraphics[width=0.68\linewidth]{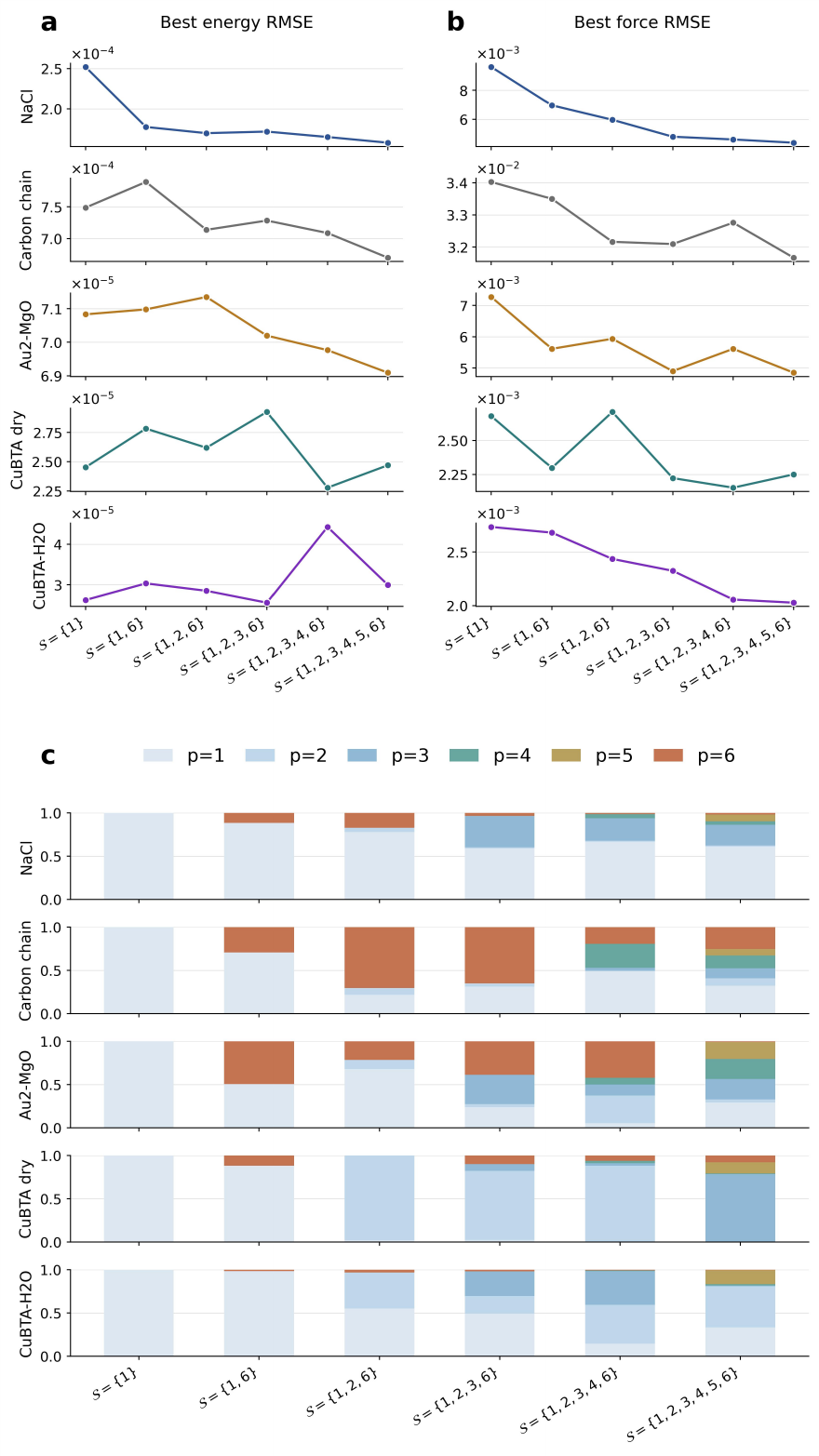}
\caption{\sf Effect of long-range PSWF component selection on model accuracy and learned long-range contributions.
$\mathbf{a,b,}$ Minimum validation energy RMSE (a) and best validation force RMSE (b) obtained for each system as the active long-range component set $\mathcal{S}$ is expanded from $\{1\}$ to $(\{1,2,3,4,5,6\})$. Stars mark the best-performing component set for each system. $\mathbf{c},$ Normalized contribution of each long-range component $p$, computed as $C_p=\langle |E_p| \rangle/
\sum_{q\in\mathcal{S}}\langle |E_q| \rangle$, where $E_p$ is the long-range energy contribution from component $p$ and $\langle \cdots\rangle$ denotes averaging over the validation set. Each stacked bar sums to one, with colors denoting the fractional contribution from $p=1$ to $p=6$.}
\label{fig:MLIPDiffP}
\end{figure}

\begin{figure}[!htbp]
\centering
\includegraphics[width=0.96\linewidth]{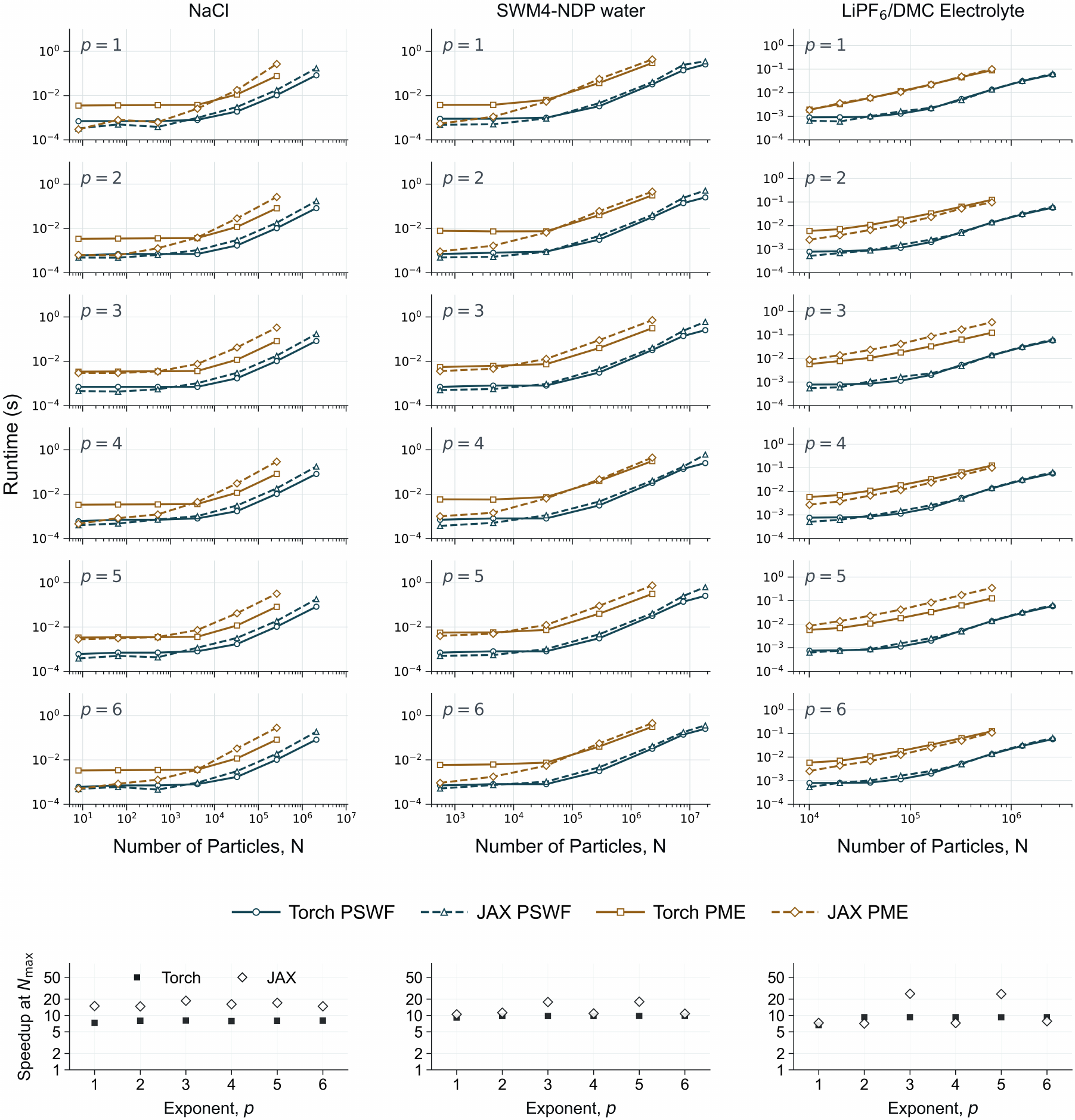}
\caption{\sf Performance comparison between PSWF- and PME-based long-range treatments across ionic, polarizable and electrolyte systems. The short-range cutoff is set to $5~\mathring{\mathrm{A}}$ for all methods. The benchmark includes an NaCl crystal, the SWM4-NDP polarizable water model and a LiPF$_6$/DMC electrolyte. Main panels show the full forward-and-backward runtime as a function of particle number, $N$, for exponents $p=1$ to $6$ (top to bottom). Blue curves denote PSWF and yellow curves denote PME; solid and dashed lines denote Torch and JAX implementations, respectively. Bottom panels summarize the corresponding PSWF speed-up over PME at the largest PME-reachable particle number, $N_{\max}$, for each exponent, with filled squares and open diamonds denoting Torch and JAX implementations, respectively. Missing PME data points indicate calculations that became GPU-memory limited.
}
\label{fig:MLIPComparisonSI}
\end{figure}

\begin{figure}[!htbp]
\centering
\includegraphics[width=0.59\linewidth]{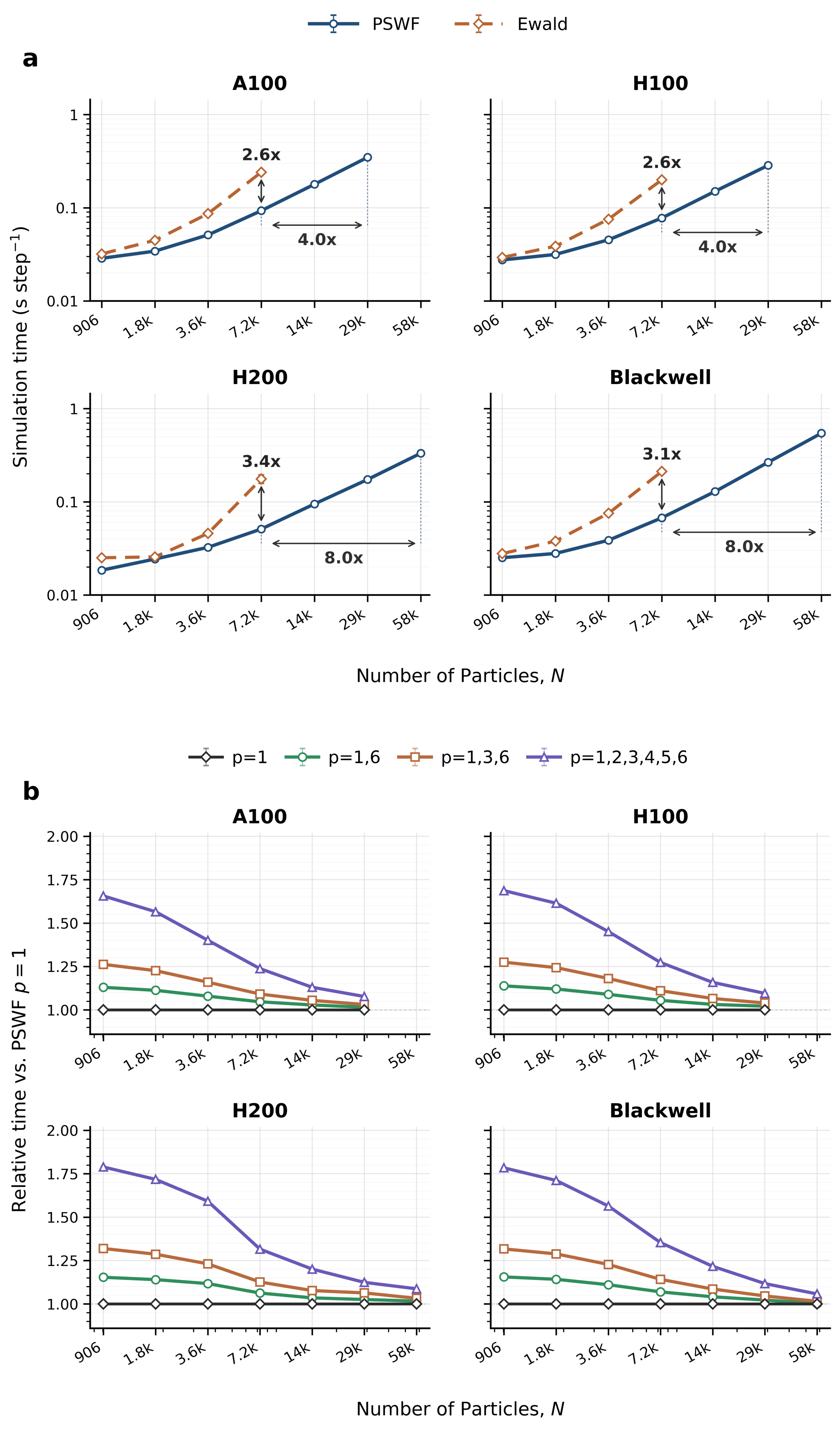}
\vspace{5mm}
\caption{\sf GPU scaling and multi-resolution overhead of PSWF-based long-range molecular dynamics.
{\bf a,} Strong-scaling benchmark of Pt(111)-water molecular dynamics on $\mathrm{A}100$, $\mathrm{H}100$, $\mathrm{H}200$ and RTX-Blackwell GPUs.
All simulations use the same short-range CACE model, neighbor list settings, cutoff and MD protocol; only the long-range electrostatic treatment is changed.
For PSWF, the plotted timings correspond to the short-range model coupled to a single $p=1$ PSWF long-range channel, whereas Ewald denotes the same short-range model coupled to the Fourier space Ewald electrostatic baseline.
The plotted quantity is the wall-clock simulation time per MD step as a function of system size.
Annotations indicate the PSWF speed-up at the largest system size accessible to Ewald and the increase in the maximum simulated system size before GPU-memory exhaustion.
{\bf b,} Relative overhead of adding multiple PSWF long-range channels on top of the same short-range model.
Timings are normalized by the runtime of the corresponding single-channel $p=1$ PSWF simulation on the same GPU and system size.
Curves compare $p=1$, $p=1,6$, $p=1,3,6$ and $p=1,2,3,4,5,6$, where each listed value denotes an additional PSWF long-range channel evaluated from the same short-range representation.
Error bars denote the standard error over repeated timing runs.}
\label{fig:MLIPCombineScaling}
\end{figure}

\begin{figure}[!htbp]
\centering
\includegraphics[width=0.8\linewidth]{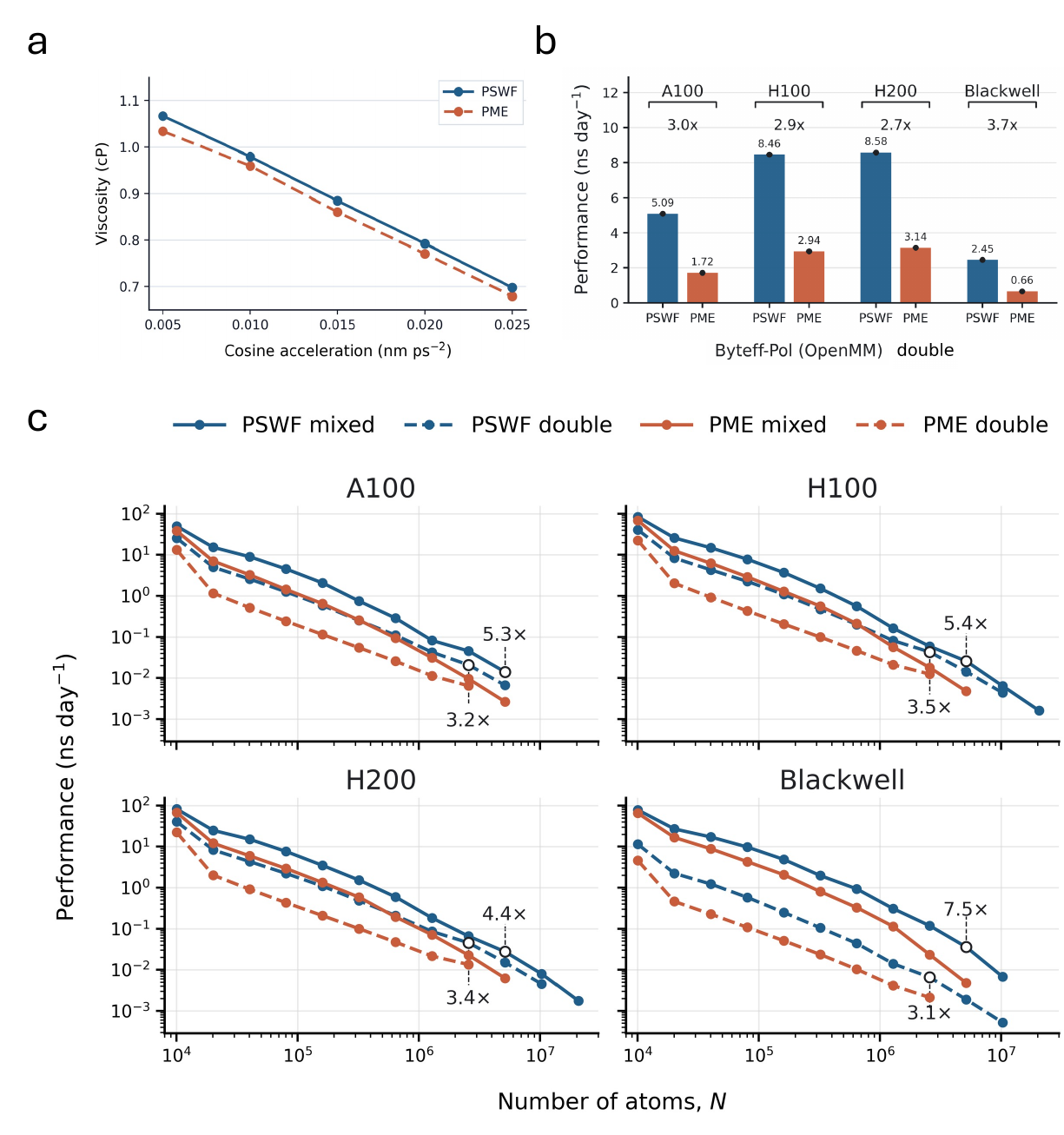}
\vspace{5mm}
\caption{\sf Additional accuracy and performance tests of PSWF-LR in ByteFF-Pol simulations.
{\bf a,} Nonequilibrium viscosity estimates for the 100,656-atom ByteFF-Pol benchmark as a function of cosine acceleration. PSWF and PME give closely matched viscosity responses across the tested field strengths, indicating that replacing PME electrostatics with PSWF does not measurably alter the transport trend.
{\bf b,} Wall-clock simulation performance for the 100,656-atom benchmark on $\mathrm{A} 100, \mathrm{H} 100, \mathrm{H} 200$ and Blackwell GPUs using double precision. PSWF consistently outperforms PME, with speed-ups marked in the figure. {\bf c,} Strong-size scaling of simulation throughput as the system size is increased from 10,062 atoms to 20,606,976 atoms. PSWF and PME are compared in both mixed and double precision. PSWF maintains higher throughput across GPU architectures and extends accessible system sizes relative to PME, with the largest-system speed-ups annotated for each GPU.}
\label{fig:MLIPByteff2}
\end{figure}

\begin{figure}[!htbp]
\centering
\includegraphics[width=0.96\linewidth]{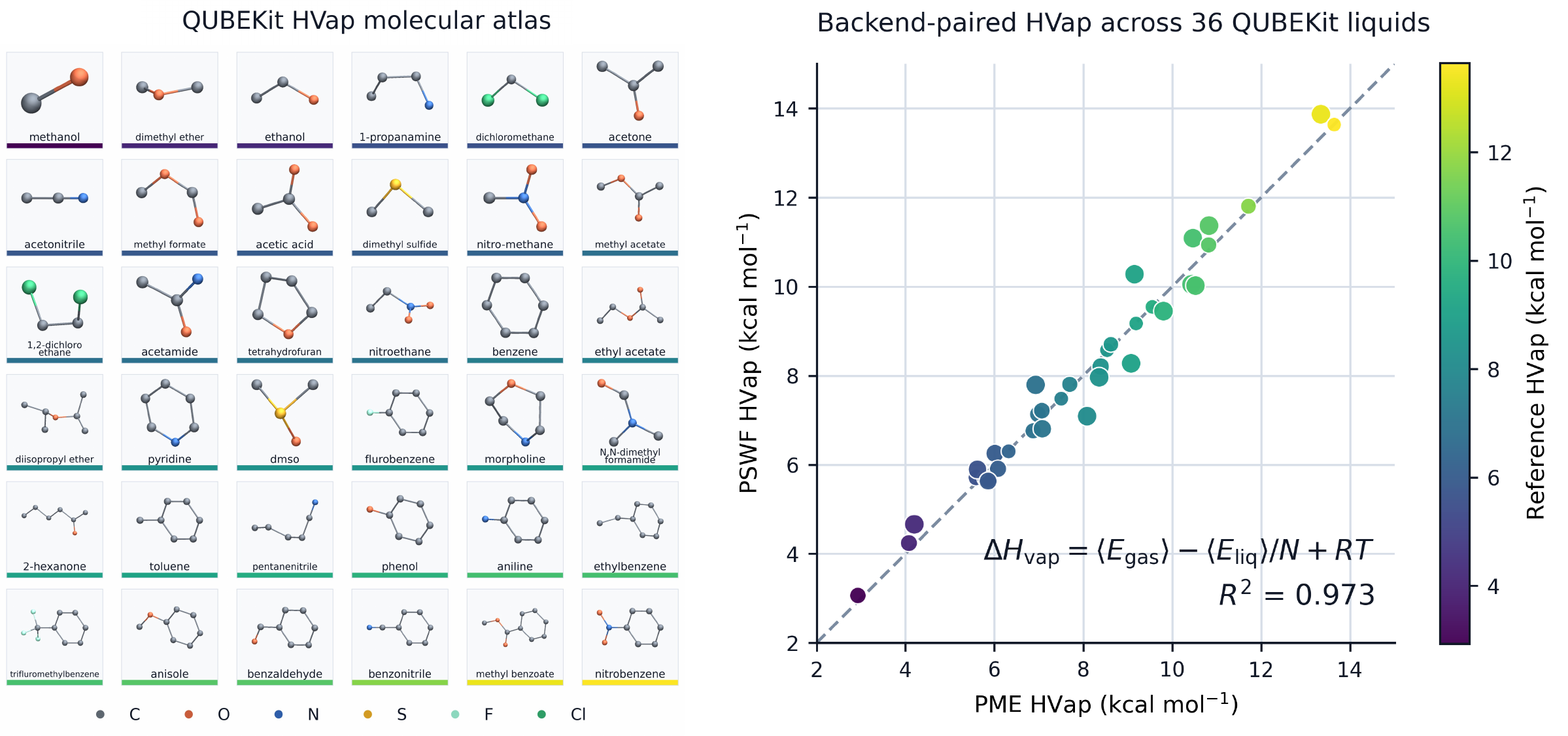}
\vspace{10mm}
\caption{\sf Backend-paired enthalpy of vaporization benchmark across 36 QUBEKit liquids. Left, molecular atlas of the 36 QUBEKit liquids included in the paired HVap benchmark. Molecules are ordered by increasing reference HVap, with the bottom colour strip in each tile using the same reference-HVap scale as in the right panel. Rendered structures show heavy atoms only for clarity; element colours are indicated in the legend. Right, comparison of enthalpies of vaporization computed with the PME and PSWF electrostatics backends for the same 36-molecule QUBEKit subset. Each point corresponds to one liquid; the x axis gives the PME reference value and the y axis gives the PSWF value. Marker colour reports the reference HVap computed with the higher-accuracy backend, and marker size encodes the absolute backend difference, $|\Delta H_{\mathrm{vap}}^{\mathrm{PSWF}}-\Delta H_{\mathrm{vap}}^{\mathrm{PME}}|$. The dashed line denotes parity. The calculation method for $H_\mathrm{vap}$ as well as the $R^2$ value is shown in the right panel. The high correlation indicates that PSWF preserves the molecule-wise HVap ranking of the PME reference while introducing only small backend-dependent deviations.}
\label{fig:MLIPHVAPSI}
\end{figure}

%TC:endignore

\clearpage


\section*{Supplementary material}

# 1 Introduction

Pairwise inverse-power interactions of the form $K_p(r) = 1/r^p$ with $p \geq 1$ arise throughout computational physics and chemistry: $p = 1$ governs Coulomb electrostatics, $p = 2, 3$ couple to charge–dipole and dipole–dipole channels, $p = 4, 5$ describe charge-induced-dipole and dipole-induced-dipole effects, and $p = 6$ controls dispersion. Their direct evaluation in a finite simulation cell of $N$ particles costs $O(N^2)$ operations, and a long line of acceleration techniques relies on splitting the kernel into a smooth, slowly decaying *long-range* component and a localized *short-range* component. Once such a splitting is fixed, the long-range part can be evaluated by spectral methods on a coarse Fourier grid, while the short-range part is summed by direct truncation in physical space. The accuracy and efficiency of the resulting algorithm depend critically on the choice of window function used to define the splitting.

In the main manuscript we use as the window the first prolate spheroidal wave function (PSWF) of order zero, $\psi_0^c$, i.e. the eigenfunction associated with the largest eigenvalue $\lambda_0^c$ of the truncated Fourier integral operator on $[-1, 1]$. We use the normalization $\psi_0^c(0) = 1$. After unit-norm rescaling, $\psi_0^c$ uniquely maximizes the fraction of $L^2$ energy contained in the band $[-c, c]$ among all unit-norm functions supported on $[-1, 1]$[1–3]. This extremal property makes $\psi_0^c$ a near-optimal window when the kernel splitting is to be evaluated by FFT-based methods.

The present supplement collects the mathematical material on which the main manuscript relies. The contributions are as follows.

1. In Section 2 we record the three-dimensional Fourier transforms of $1/r^p$ for $p = 1, \ldots, 6$. The borderline case $p = 3$ is handled rigorously by analytic continuation as a tempered distribution.

2. In Section 3 we introduce a general kernel-splitting framework based on an admissible window function $\phi$ and derive closed-form expressions for the long-range part $\mathcal{S}_p(r)$, the short-range part $\mathcal{L}_p(r)$, and the Fourier-space transform $\widehat{\mathcal{S}}_p(k)$ for $p = 1, \ldots, 6$.

3. In Section 4 we summarize the properties of $\psi_0^c$ that are needed in the sequel.

4. In Section 5 we specialize the generic splittings of Section 3 to $\phi = \psi_0^c(r)/\psi_0^c(0)$.

5. In Section 6 we establish a unified recurrence that extends the construction to arbitrary $p \geq 7$ and record explicit splittings for $p = 8$ and $p = 10$, which are relevant to polarizable force fields and machine-learning interatomic potentials with induction-like and anisotropic-dispersion channels.

6. In Section 7 we discuss the numerical evaluation of the splittings and the correspondence with the notation used in the main manuscript.

7. In Section 8 we derive the self-correction and zero-frequency terms required to define the periodic PSWF-LR energy.

8. In Section 9 we provide matched-tolerance accuracy checks for PSWF-LR and the PME comparisons, supporting the runtime and memory comparisons reported in the main text.

The cutoff radius $r_c$ of the main manuscript does not appear in Sections 2–6; it is reinstated by a dimensional substitution in Section 7 and is used in Sections 8-9.

Together, these sections provide the technical support for the main-text PSWF-LR framework. The Fourier-transform and kernel-splitting derivations justify the PSWF-based mollifiers $\mathcal{G}_p(r)$ and

reciprocal-space multipliers $\widehat{\Phi}^{(p)}(k)$ used in the inverse-power LR channels of main text and the Methods. The numerical-evaluation and notation section connects the dimensionless splittings $\mathcal{S}_p, \widehat{\mathcal{S}}_p$ to the physical-cutoff notation $\mathcal{G}_p, \widehat{\Phi}^{(p)}$ used in the main manuscript. The matched-tolerance accuracy checks support the differentiable PME comparisons in Fig. 4, showing that the reported speed-ups are not obtained by relaxing LR accuracy. The FKAD section explains how the same PSWF-LR kernels are evaluated efficiently and differentiated automatically in the PyTorch, JAX and OpenMM/CUDA implementations used in the runtime and MD benchmarks.

## 2 Three-dimensional Fourier transforms of inverse-power kernels

We adopt the convention

$$\widehat{f}(\mathbf{k}) = \int_{\mathbb{R}^3} e^{i\mathbf{k}\cdot\mathbf{r}} f(\mathbf{r})\,\mathrm{d}^3r, \qquad f(\mathbf{r}) = \frac{1}{(2\pi)^3}\int_{\mathbb{R}^3} e^{-i\mathbf{k}\cdot\mathbf{r}}\,\widehat{f}(\mathbf{k})\,\mathrm{d}^3k, \quad k = |\mathbf{k}|,\ r = |\mathbf{r}|, \tag{S1}$$

and recall that for a radial function $f(r)$ the transform reduces to

$$\widehat{f}(k) = \frac{4\pi}{k}\int_0^\infty r\,f(r)\,\sin(kr)\,\mathrm{d}r. \tag{S2}$$

The standard identity

$$\int_{\mathbb{R}^3} e^{i\mathbf{k}\cdot\mathbf{r}}\,r^{-\alpha}\,\mathrm{d}^3r = 2^{\,3-\alpha}\pi^{3/2}\frac{\Gamma\big(\frac{3-\alpha}{2}\big)}{\Gamma\big(\frac{\alpha}{2}\big)}k^{\alpha-3} \tag{S3}$$

is valid for $0 < \alpha < 3$. Beyond this range, $K_p$ fails to be locally integrable and the transform must be understood in the sense of tempered distributions. We will use repeatedly the elementary identity

$$\nabla^2\big(r^{-m}\big) = m(m-1)\,r^{-m-2}, \qquad r > 0, \tag{S4}$$

which translates into $-k^2$ multiplication on the Fourier side.

**Lemma 1** (3D Fourier transforms of $1/r^p$, $p = 1, \ldots, 6$)**.** *As tempered distributions on* $\mathbb{R}^3$,

$$\widehat{r^{-1}}(k) = \frac{4\pi}{k^2}, \tag{S5}$$

$$\widehat{r^{-2}}(k) = \frac{2\pi^2}{k}, \tag{S6}$$

$$\widehat{r^{-3}}(k) = -4\pi\ln k, \tag{S7}$$

$$\widehat{r^{-4}}(k) = -\pi^2 k, \tag{S8}$$

$$\widehat{r^{-5}}(k) = \frac{2\pi}{3}\,k^2\ln k, \tag{S9}$$

$$\widehat{r^{-6}}(k) = \frac{\pi^2}{12}\,k^3, \tag{S10}$$

*where* $k = |\mathbf{k}|$ *and the logarithmic terms are defined up to an additive constant arising from the choice of renormalization scale.*

*Proof.* For $p = 1$ and $p = 2$, set $\alpha = p$ in (S3). For $p \in \{4, 5, 6\}$, apply (S4) recursively to the cases already established. Concretely,

$$r^{-4} = \frac{1}{2}\nabla^2(r^{-2}), \quad r^{-5} = \frac{1}{6}\nabla^2(r^{-3}), \quad r^{-6} = \frac{1}{12}\nabla^2(r^{-4}), \tag{S11}$$

and the corresponding $k$-space transforms acquire factors of $-k^2/2$, $-k^2/6$, and $-k^2/12$, respectively. The case $p = 3$ is borderline and is treated in Proposition 1 below. □

**Proposition 1** (Distributional Fourier transform of $1/r^3$)**.** *The function $r^{-3}$ defines a tempered distribution on $\mathbb{R}^3$, and its Fourier transform satisfies*

$$\widehat{r^{-3}}(k) = -4\pi \ln k \tag{S12}$$

*modulo additive contact terms supported at the origin and a scale-dependent constant.*

*Proof.* Set $\alpha = 3 - \varepsilon$ for $0 < \varepsilon < 3$, apply (S3), and expand. With

$$\Gamma\left(\frac{\varepsilon}{2}\right) = \frac{2}{\varepsilon} - \gamma + O(\varepsilon), \qquad \Gamma\left(\frac{3}{2} - \frac{\varepsilon}{2}\right) = \Gamma\left(\frac{3}{2}\right)\left[1 - \frac{\varepsilon}{2}\psi\left(\frac{3}{2}\right) + O(\varepsilon^2)\right], \tag{S13}$$

$2^\varepsilon = 1 + \varepsilon \ln 2 + O(\varepsilon^2)$, $k^{-\varepsilon} = 1 - \varepsilon \ln k + O(\varepsilon^2)$, $\Gamma(3/2) = \sqrt{\pi}/2$, and $\psi(3/2) = 2 - \gamma - 2\ln 2$. The $\ln 2$ contributions from $2^\varepsilon$ and $\psi(3/2)$ cancel, and the convergent transform $\widehat{r^{-3+\varepsilon}}(k)$ expands as

$$\widehat{r^{-3+\varepsilon}}(k) = \frac{4\pi}{\varepsilon} - 4\pi \ln k + 4\pi(1 - \gamma) + O(\varepsilon).$$

The $\varepsilon^{-1}$ pole is momentum-independent and corresponds to a contact term proportional to $\delta(\mathbf{r})$, which can be absorbed by renormalization. The finite constant similarly depends on the renormalization scale through $-4\pi \ln k = -4\pi \ln(k/\mu) - 4\pi \ln \mu$. Discarding the contact counterterm and the arbitrary constant yields the stated formula. □

As a dimensional check, $[r^{-3}] = L^{-3}$ and the $\mathrm{d}^3 r$ measure contributes $L^3$, so $\widehat{r^{-3}}$ is dimensionless and $\ln k$ is appropriate.

# 3 A general framework for kernel splitting

## 3.1 Admissible windows and splitting conditions

We seek a decomposition

$$K_p(r) = \frac{1}{r^p} = \mathcal{S}_p(r) + \mathcal{L}_p(r), \qquad p = 1, \ldots, 6, \tag{S14}$$

where $\mathcal{S}_p$ is the smooth, long-range part whose Fourier transform decays rapidly, and $\mathcal{L}_p$ is the localized short-range part. Since the only singularity of $K_p$ is at the origin, $\mathcal{S}_p$ must satisfy

(i) $\mathcal{S}_p$ is smooth at $r = 0$,

(ii) $\lim_{r\to\infty} |\mathcal{S}_p(r) - K_p(r)| = 0$.

Such splittings underlie fast Ewald summation[4,5] and the dual-space multilevel kernel-splitting (DMK) framework[6], in which fast algorithms require explicit forms of both $\widehat{\mathcal{S}}_p$ and $\mathcal{L}_p$.

**Definition 1** (Admissible window function)**.** A function $\phi : [0,\infty) \to \mathbb{R}$ is an *admissible window* if it is positive, real, even, radially symmetric, monotonically decreasing on $[0,\infty)$, smooth on $[0,1]$, and either truncated to zero at $r = 1$ or rapidly decaying as $r \to \infty$. Without loss of generality we normalize $\phi$ by

$$\phi(0) = 1, \qquad \phi(1) = O(\varepsilon), \qquad \widehat{\phi}(c) = O(\varepsilon), \tag{S15}$$

where $\varepsilon > 0$ is a prescribed tolerance and $c > 0$ is the chosen Fourier bandwidth parameter.

For an admissible $\phi$ we write its Taylor expansion at the origin in the form

$$\phi(r) = \sum_{n\geq 0} c_{2n}\, r^{2n}, \qquad c_{2n} = \frac{\phi^{(2n)}(0)}{(2n)!}, \tag{S16}$$

so that $c_0 = 1$, $c_2 = \phi''(0)/2$, $c_4 = \phi^{(4)}(0)/24$, and in general $c_{2n}$ is the Taylor coefficient of $r^{2n}$. We further adopt the shorthand

$$\Phi(r) := \int_0^r \phi(x)\,\mathrm{d}x, \qquad C_0 := \int_0^\infty \phi(x)\,\mathrm{d}x,$$

and denote by $\widehat{\phi}$ the one-dimensional Fourier transform of $\phi$, with $\widehat{\Phi}(k) := \int_0^k \widehat{\phi}(s)\,\mathrm{d}s$, $\widehat{C}_0 := \int_0^\infty \widehat{\phi}(k)\,\mathrm{d}k$, and $\widehat{c}_0 := \widehat{\phi}(0)$. Note that $\Phi$ is odd, radially symmetric, and smooth on $[0,1]$.

### 3.2 Splittings for $p = 1$ and $p = 2$

**Proposition 2** (Coulomb splitting, $p = 1$)**.** *For any admissible window $\phi$, the splitting*

$$\mathcal{S}_1(r) = \frac{\Phi(r)}{C_0 r}, \qquad \mathcal{L}_1(r) = \frac{1 - \Phi(r)/C_0}{r}, \qquad \widehat{\mathcal{S}}_1(k) = \frac{4\pi}{\widehat{\phi}(0)\,k^2}\,\widehat{\phi}(k) \tag{S17}$$

*satisfies the splitting conditions and reduces the original DMK construction*[6] *to inverse-power form.*

The Fourier transform of $1/r$ is $4\pi/k^2$ by (S5). Substituting $r \leftrightarrow k$ formally in (S17) therefore yields a splitting of the Fourier-side kernel $1/k^2$, which in turn corresponds to a real-space splitting of $1/r^2$.

**Proposition 3** (Charge–dipole splitting, $p = 2$)**.** *For any admissible window $\phi$,*

$$\mathcal{S}_2(r) = \frac{1 - \phi(r)}{r^2}, \qquad \mathcal{L}_2(r) = \frac{\phi(r)}{r^2}, \qquad \widehat{\mathcal{S}}_2(k) = \frac{2\pi^2}{k}\left(1 - \frac{\widehat{\Phi}(k)}{\widehat{C}_0}\right) \tag{S18}$$

*satisfies the splitting conditions.*

### 3.3 Splitting for $p = 3$

**Proposition 4** (Dipole–dipole splitting, $p = 3$)**.** *For any admissible window $\phi$,*

$$\mathcal{S}_3(r) = \frac{\Phi(r) - r\phi(r)}{C_0\, r^3}, \qquad \mathcal{L}_3(r) = \frac{1 - (\Phi(r) - r\phi(r))/C_0}{r^3} \tag{S19}$$

*satisfies the splitting conditions, with*

$$\lim_{r\to 0} \mathcal{S}_3(r) = -\frac{\phi''(0)}{3\,C_0} = -\frac{2\,c_2}{3\,C_0}, \tag{S20}$$

*and if $\phi$ is truncated at $r = 1$ then $\mathcal{S}_3(r) = K_3(r)$ for $r > 1$, so $\mathcal{L}_3$ is compactly supported in the unit ball. The Fourier transform is*

$$\widehat{\mathcal{S}}_3(k) = \frac{2\pi}{C_0} \int_k^\infty \frac{\widehat{\phi}(\xi)}{\xi} \, \mathrm{d}\xi. \tag{S21}$$

*Proof.* Smoothness at $r = 0$ and the limit follow from the Taylor expansion of $\phi$ near the origin. The behavior $\mathcal{S}_3(r) = K_3(r)$ for $r > 1$ when $\phi$ is truncated is immediate from $\Phi(r) = C_0$ and $\phi(r) = 0$ in that range.

For the Fourier transform, observe that $(\Phi(r) - r\phi(r))/r^2 = -(\mathrm{d}/\mathrm{d}r)(\Phi(r)/r)$, so substituting the splitting into the radial transform formula (S2) gives

$$\widehat{\mathcal{S}}_3(k) = -\frac{4\pi}{C_0 k} \int_0^\infty \frac{\mathrm{d}}{\mathrm{d}r}\left(\frac{\Phi(r)}{r}\right) \sin(kr) \, \mathrm{d}r. \tag{S22}$$

Integrating by parts and noting that the boundary terms vanish (as $r \to 0$, $\sin(kr) \to 0$ and $\Phi(r)/r \to \phi(0) = 1$; as $r \to \infty$, $\Phi(r)$ approaches a constant while $\sin(kr)/r \to 0$), one obtains

$$\widehat{\mathcal{S}}_3(k) = \frac{4\pi}{C_0} \int_0^\infty \frac{\Phi(r)}{r} \cos(kr) \, \mathrm{d}r. \tag{S23}$$

Expressing $\Phi(r)$ through the 1D cosine transform $\phi(r) = \pi^{-1} \int_0^\infty \widehat{\phi}(\xi) \cos(\xi r) \, \mathrm{d}\xi$ gives $\Phi(r) = \pi^{-1} \int_0^\infty \widehat{\phi}(\xi) \sin(\xi r)/\xi \, \mathrm{d}\xi$. Substituting and applying the Dirichlet integral

$$\int_0^\infty \frac{\cos(kr)\sin(\xi r)}{r} \, \mathrm{d}r = \begin{cases} \pi/2, & \xi > k, \\ 0, & \xi < k, \end{cases}$$

yields (S21). □

### 3.4 Splitting for $p = 4$

For $p = 4$ we exploit the Fourier relation $\widehat{r^{-2}}(k) = 2\pi^2/k$ from (S6). Treating the splitting of $1/r^2$ in Proposition 3 as a Fourier-side splitting of $1/k^2$ and performing the formal substitution $k \leftrightarrow r$ gives a candidate splitting of $1/r^4$ in physical space.

**Proposition 5** (Charge–apolar splitting, $p = 4$)**.** *For any admissible window $\phi$,*

$$\mathcal{S}_4(r) = \frac{1 - \phi(r) + \frac{1}{2} r\phi'(r)}{r^4}, \qquad \mathcal{L}_4(r) = \frac{\phi(r) - \frac{1}{2} r\phi'(r)}{r^4}, \tag{S24}$$

*satisfies the splitting conditions, with*

$$\lim_{r \to 0} \mathcal{S}_4(r) = \frac{\phi^{(4)}(0)}{24}, \tag{S25}$$

*and if $\phi$ is truncated at $r = 1$ then $\mathcal{S}_4(r) = K_4(r)$ for $r > 1$. The Fourier transform is*

$$\widehat{\mathcal{S}}_4(k) = -\pi^2 \left( k - \frac{k}{\widehat{C}_0} \int_0^k \widehat{\phi}(\xi) \, \mathrm{d}\xi + \frac{1}{\widehat{C}_0} \int_0^k \xi \widehat{\phi}(\xi) \, \mathrm{d}\xi - \frac{1}{\widehat{C}_0} \int_0^\infty \xi \widehat{\phi}(\xi) \, \mathrm{d}\xi \right). \tag{S26}$$

*Proof.* Smoothness at $r = 0$, the limit, and the truncation property follow from Taylor expansion. For the Fourier transform, the identity $\big(1-\phi(r)+\frac{1}{2}r\phi'(r)\big)/r^3 = -\frac{1}{2}(\mathrm{d}/\mathrm{d}r)\big((1-\phi(r))/r^2\big)$ combined with (S2) gives

$$\widehat{\mathcal{S}}_4(k) = 2\pi \int_0^\infty \frac{1-\phi(r)}{r^2} \cos(kr)\, \mathrm{d}r. \tag{S27}$$

Differentiating in $k$ yields $\widehat{\mathcal{S}}_4'(k) = -\frac{1}{2}k\,\widehat{\mathcal{S}}_2(k)$, and integrating with $\widehat{\mathcal{S}}_4(\infty) = 0$ recovers (S26) using (S18). □

*Remark* 1. An alternative splitting for $p = 4$ is

$$\widetilde{\mathcal{S}}_4(r) = \frac{1-\phi(r)(1-c_2r^2)}{r^4}, \qquad \widetilde{\mathcal{L}}_4(r) = \frac{\phi(r)(1-c_2r^2)}{r^4}, \tag{S28}$$

which has the same boundary behavior but trades the explicit derivative term for a polynomial factor in $r$. The two are interchangeable for our purposes.

### 3.5 Splitting for $p = 5$

**Proposition 6** (Polar–apolar splitting, $p = 5$)**.** *For any admissible window* $\phi$*,*

$$\mathcal{S}_5(r) = \frac{\Phi(r) - r\phi(r) + \frac{1}{3}r^2\phi'(r)}{C_0\, r^5}, \qquad \mathcal{L}_5(r) = \frac{1-\big(\Phi(r) - r\phi(r) + \frac{1}{3}r^2\phi'(r)\big)/C_0}{r^5}, \tag{S29}$$

*satisfies the splitting conditions, with*

$$\lim_{r\to 0} \mathcal{S}_5(r) = \frac{\phi^{(4)}(0)}{45\, C_0}, \tag{S30}$$

*and if* $\phi$ *is truncated at* $r = 1$ *then* $\mathcal{S}_5(r) = K_5(r)$ *for* $r > 1$*. The Fourier transform is*

$$\widehat{\mathcal{S}}_5(k) = -\frac{\pi}{3C_0}\left(k^2\int_k^\infty \frac{\widehat{\phi}(\xi)}{\xi}\,\mathrm{d}\xi - \int_k^\infty \xi\widehat{\phi}(\xi)\,\mathrm{d}\xi\right). \tag{S31}$$

*Proof.* Smoothness and the limit follow from the Taylor expansion $\Phi(r) = r+\frac{1}{3}c_2r^3+\frac{1}{5}c_4r^5+O(r^7)$. For the Fourier transform, the identity $\big(\Phi(r)-r\phi(r)+\frac{1}{3}r^2\phi'(r)\big)/r^4 = -\frac{1}{3}(\mathrm{d}/\mathrm{d}r)\big((\Phi(r)-r\phi(r))/r^3\big)$ together with (S2) yields $\widehat{\mathcal{S}}_5'(k) = -\frac{1}{3}k\,\widehat{\mathcal{S}}_3(k)$, and integration with $\widehat{\mathcal{S}}_5(\infty) = 0$ recovers (S31). □

*Remark* 2. An alternative splitting for $p = 5$ is

$$\widetilde{\mathcal{S}}_5(r) = \frac{\Phi(r) - \phi(r)\big(r - \frac{2}{3}c_2r^3\big)}{C_0\, r^5}, \qquad \widetilde{\mathcal{L}}_5(r) = \frac{1-\big(\Phi(r) - \phi(r)(r - \frac{2}{3}c_2r^3)\big)}{C_0\, r^5}. \tag{S32}$$

### 3.6 Splitting for $p = 6$

**Proposition 7** (Dispersion splitting, $p = 6$)**.** *For any admissible window* $\phi$*,*

$$\mathcal{S}_6(r) = \frac{1-\phi(r)+\frac{5}{8}r\phi'(r) - \frac{1}{8}r^2\phi''(r)}{r^6}, \quad \mathcal{L}_6(r) = \frac{\phi(r) - \frac{5}{8}r\phi'(r) + \frac{1}{8}r^2\phi''(r)}{r^6} \tag{S33}$$

*satisfies the splitting conditions, with*

$$\lim_{r\to 0} \mathcal{S}_6(r) = -\frac{\phi^{(6)}(0)}{720}, \tag{S34}$$

*and if $\phi$ is truncated at $r = 1$ then $\mathcal{S}_6(r) = K_6(r)$ for $r > 1$. The Fourier transform is*

$$\widehat{\mathcal{S}}_6(k) = \frac{\pi^2}{12}\left(k^3 - \frac{k^3}{\widehat{C}_0}\int_0^k \widehat{\phi}(\xi)\,\mathrm{d}\xi - \frac{3k^2}{2\widehat{C}_0}\int_k^\infty \xi\widehat{\phi}(\xi)\,\mathrm{d}\xi + \frac{1}{2\widehat{C}_0}\int_k^\infty \xi^3\widehat{\phi}(\xi)\,\mathrm{d}\xi\right). \tag{S35}$$

*Proof.* Smoothness at $r = 0$ and the limit follow from Taylor expansion. For the Fourier transform, the identity $\big(1-\phi(r)+\frac{5}{8}r\phi'(r)-\frac{1}{8}r^2\phi''(r)\big)/r^5 = -\frac{1}{4}(\mathrm{d}/\mathrm{d}r)\big((1-\phi(r)+\frac{1}{2}r\phi'(r))/r^4\big)$ combined with (S2) gives $\widehat{\mathcal{S}}_6'(k) = -\frac{1}{4}k\,\widehat{\mathcal{S}}_4(k)$, and integration with $\widehat{\mathcal{S}}_6(\infty) = 0$ recovers (S35) using (S26). □

*Remark* 3. An alternative splitting for $p = 6$ is

$$\widetilde{\mathcal{S}}_6(r) = \frac{1-\phi(r)(1-c_2r^2-(c_4-c_2^2)r^4)}{r^6}, \qquad \widetilde{\mathcal{L}}_6(r) = \frac{\phi(r)(1-c_2r^2-(c_4-c_2^2)r^4)}{r^6}. \tag{S36}$$

# 4 Prolate spheroidal wave functions

The prolate spheroidal wave functions of order zero form a countable family $\{\psi_n^c\}_{n\geq 0}$ of eigenfunctions of the truncated Fourier integral operator $\mathcal{F}_c : L^2[-1,1] \to L^2[-1,1]$ defined by

$$\mathcal{F}_c[\sigma](x) = \int_{-1}^{1} \sigma(t)\, e^{icxt}\,\mathrm{d}t. \tag{S37}$$

The corresponding eigenvalues $\{\lambda_n^c\}_{n\geq 0}$ are conventionally ordered by decreasing magnitude, $|\lambda_0^c| > |\lambda_1^c| > \cdots$. In what follows we use only $\psi_0^c$, the first member of this family — equivalently, the eigenfunction associated with the largest eigenvalue $\lambda_0^c$ — normalized by $\psi_0^c(0) = 1$:

$$\lambda_0^c\,\psi_0^c(x) = \int_{-1}^{1} \psi_0^c(t)\, e^{icxt}\,\mathrm{d}t. \tag{S38}$$

Writing $\widehat{\psi_0^c}(k) := \int_{-1}^1 \psi_0^c(t)\,e^{ikt}\,\mathrm{d}t$ for the 1D Fourier transform of $\psi_0^c$ extended by zero outside $[-1,1]$, the substitution $k = cx$ in (S38) gives

$$\widehat{\psi_0^c}(k) = \lambda_0^c\,\psi_0^c(k/c), \qquad |k| \leq c. \tag{S39}$$

While no nontrivial function is exactly band-limited in both physical and Fourier spaces, $\psi_0^c$ is essentially band-limited in both: it is compactly supported on $[-1,1]$ and its Fourier transform decays rapidly outside $[-c,c]$. The fundamental Slepian extremal property states that the unit-norm rescaling of $\psi_0^c$ uniquely maximizes the $L^2$ norm of its Fourier transform on $[-c,c]$ among all unit-norm functions in $L^2[-1,1]$[1]. This optimality underpins the use of $\psi_0^c$ as a near-optimal window in signal processing, numerical quadrature, and kernel splitting.

The eigenvalue equation (S38) provides stable expressions for several quantities that we will need. Differentiating both sides $2n$ times and evaluating at $x = 0$ gives

$$\lambda_0^c\,\psi_0^{c\,(2n)}(0) = (-1)^n c^{2n}\int_{-1}^{1} \psi_0^c(t)\,t^{2n}\,\mathrm{d}t, \tag{S40}$$

so high-order derivatives of $\psi_0^c$ at the origin can be evaluated by integration of $\psi_0^c$ rather than by numerical differentiation. Setting $n = 0$ in (S40) yields

$$\lambda_0^c\,\psi_0^c(0) = \int_{-1}^{1} \psi_0^c(t)\,\mathrm{d}t. \tag{S41}$$

## 5 Specialization to prolate spheroidal wave functions

We now specialize the generic splittings of Section 3 to the window obtained from the first PSWF of order zero,

$$\phi(r) = \psi_0^c(r)/\psi_0^c(0), \tag{S42}$$

which is admissible in the sense of Definition 1 and is consistent with the normalization $\phi(0) = 1$. We adopt the truncation convention $\widehat{\phi}(k) \equiv 0$ for $|k| > c$, so that $\phi$ is exactly band-limited to $[-c, c]$. Under this convention the bandwidth admissibility condition $\widehat{\phi}(c) = O(\varepsilon)$ in Definition 1 is enforced exactly, and all integrals $\int_0^\infty \widehat{\phi}(\cdot)\,\mathrm{d}k$ that appear below reduce to integrals on $[0, c]$. The cutoff radius $r_c$ of the main manuscript does not appear in this section; it is reinstated through the variable substitution $r \to r/r_c$ in Section 7.

Under (S42), direct computation gives

$$C_0 = \int_0^1 \phi(x)\,\mathrm{d}x, \qquad \widehat{\phi}(k) = \frac{2C_0}{\psi_0^c(0)}\,\psi_0^c(k/c), \tag{S43}$$

together with the auxiliary identities

$$\Phi(r) = \frac{1}{\psi_0^c(0)} \int_0^r \psi_0^c(x)\,\mathrm{d}x, \qquad \widehat{\Phi}(k) = \frac{2C_0}{\psi_0^c(0)} \int_0^k \psi_0^c(\xi/c)\,\mathrm{d}\xi, \tag{S44}$$

$$\widehat{C}_0 = \int_0^c \widehat{\phi}(k)\,\mathrm{d}k = 2c\,C_0^2, \qquad \frac{\widehat{\Phi}(k)}{\widehat{C}_0} = \frac{1}{C_0} \int_0^{k/c} \phi(x)\,\mathrm{d}x, \tag{S45}$$

$$\frac{\widehat{\phi}(\xi)}{C_0} = 2\,\phi(\xi/c), \qquad \frac{\widehat{\phi}(\xi)}{\widehat{C}_0} = \frac{1}{c\,C_0}\,\phi(\xi/c). \tag{S46}$$

Substituting these identities into Propositions 2–7 produces compact closed-form expressions for the Fourier-side splittings.

**Theorem 1** (PSWF-based Fourier-side splittings)**.** *With $\phi = \psi_0^c/\psi_0^c(0)$, the long-range Fourier transforms of Section 3 take the form*

$$\widehat{\mathcal{S}}_1(k) = \frac{4\pi\,\phi(k/c)}{k^2}, \tag{S47}$$

$$\widehat{\mathcal{S}}_2(k) = \frac{2\pi^2}{k} - \frac{2\pi^2}{C_0\,k} \int_0^{k/c} \phi(x)\,\mathrm{d}x, \tag{S48}$$

$$\widehat{\mathcal{S}}_3(k) = -4\pi \ln(k/c) + 4\pi \int_{k/c}^1 \frac{\phi(x) - 1}{x}\,\mathrm{d}x, \tag{S49}$$

$$\widehat{\mathcal{S}}_4(k) = -\pi^2 \left( k - \frac{k}{C_0} \int_0^{k/c} \phi(x)\,\mathrm{d}x - \frac{c}{C_0} \int_{k/c}^1 x\,\phi(x)\,\mathrm{d}x \right), \tag{S50}$$

$$\widehat{\mathcal{S}}_5(k) = \frac{2\pi}{3} \left( k^2 \ln(k/c) + k^2 \int_{k/c}^1 \frac{1 - \phi(x)}{x}\,\mathrm{d}x + c^2 \int_{k/c}^1 x\,\phi(x)\,\mathrm{d}x \right), \tag{S51}$$

$$\widehat{\mathcal{S}}_6(k) = \frac{\pi^2}{12} \left( k^3 - \frac{k^3}{C_0} \int_0^{k/c} \phi(x)\,\mathrm{d}x - \frac{3c\,k^2}{2C_0} \int_{k/c}^1 x\,\phi(x)\,\mathrm{d}x + \frac{c^3}{2C_0} \int_{k/c}^1 x^3\,\phi(x)\,\mathrm{d}x \right). \tag{S52}$$

*Each $\widehat{\mathcal{S}}_p$ equals the unwindowed Fourier transform from Lemma 1 plus a smooth correction expressed as a finite combination of integrals of $\phi$, and satisfies $\widehat{\mathcal{S}}_p(c) = 0$ for $p = 1, \ldots, 6$.*

*Remark* 4 (Recurrences among the real-space splittings)*.* The splittings $\mathcal{S}_p$ obey the elementary recurrences

$$\mathcal{S}_3(r) = -\frac{1}{r}\frac{\mathrm{d}}{\mathrm{d}r}\mathcal{S}_1(r), \qquad \mathcal{S}_5(r) = -\frac{1}{3r}\frac{\mathrm{d}}{\mathrm{d}r}\mathcal{S}_3(r), \tag{S53}$$

$$\mathcal{S}_4(r) = -\frac{1}{2r}\frac{\mathrm{d}}{\mathrm{d}r}\mathcal{S}_2(r), \qquad \mathcal{S}_6(r) = -\frac{1}{4r}\frac{\mathrm{d}}{\mathrm{d}r}\mathcal{S}_4(r), \tag{S54}$$

which reflect the elementary identity $-\frac{1}{p\,r}\frac{\mathrm{d}}{\mathrm{d}r}(1/r^p) = 1/r^{p+2}$. These relations extend to a unified recurrence valid for arbitrary $p \geq 1$, as we now show.

## 6 Extension to higher inverse powers

The main text focuses on $p = 1, \cdots, 6,$, which cover the effective Coulombic, charge–polar, polar–polar, higher-order induction-like and dispersion-like channels used in the reported benchmarks. The recurrence below shows that the same splitting strategy formally extends to higher inverse powers. These higher-$p$ formulas are included for completeness and future extensions.

**Theorem 2** (Unified recurrence for inverse-power splittings)**.** *For any admissible window $\phi$ and any $p \geq 1$,*

$$\mathcal{S}_{p+2}(r) = -\frac{1}{p\,r}\frac{\mathrm{d}}{\mathrm{d}r}\mathcal{S}_p(r), \tag{S55}$$

$$\frac{\mathrm{d}}{\mathrm{d}k}\widehat{\mathcal{S}}_{p+2}(k) = -\frac{k}{p}\,\widehat{\mathcal{S}}_p(k), \qquad \widehat{\mathcal{S}}_{p+2}(k) = \frac{1}{p}\int_k^\infty s\,\widehat{\mathcal{S}}_p(s)\,\mathrm{d}s. \tag{S56}$$

*Proof.* Because $-\frac{1}{p\,r}\frac{\mathrm{d}}{\mathrm{d}r}(1/r^p) = 1/r^{p+2}$, the operator on the right of (S55) maps any valid splitting $\mathcal{S}_p$ to a function that (i) agrees with $1/r^{p+2}$ outside the support of $\phi$ and (ii) remains smooth at $r = 0$, hence is a valid splitting $\mathcal{S}_{p+2}$. Taking the three-dimensional radial Fourier transform of (S55) and using $\widehat{\mathcal{S}}_{p+2}(\infty) = 0$ yields (S56). □

The four splitting relations stated in Section 5 are the special cases $p = 1, 2, 3, 4$ of Theorem 2. Iterating once and twice more produces explicit formulas for $\mathcal{S}_8$ and $\mathcal{S}_{10}$.

**Corollary 1** (Explicit splittings for $p = 8$ and $p = 10$)**.** *For any admissible window $\phi$,*

$$\mathcal{S}_8(r) = \frac{1 - \phi(r) + \frac{11}{16}\,r\,\phi'(r) - \frac{3}{16}\,r^2\phi''(r) + \frac{1}{48}\,r^3\phi'''(r)}{r^8}, \tag{S57}$$

$$\mathcal{S}_{10}(r) = \frac{1 - \phi(r) + \frac{93}{128}\,r\,\phi'(r) - \frac{29}{128}\,r^2\phi''(r) + \frac{7}{192}\,r^3\phi'''(r) - \frac{1}{384}\,r^4\phi^{(4)}(r)}{r^{10}}, \tag{S58}$$

*with limits at the origin*

$$\lim_{r\to 0}\mathcal{S}_8(r) = \frac{\phi^{(8)}(0)}{8!}, \qquad \lim_{r\to 0}\mathcal{S}_{10}(r) = -\frac{\phi^{(10)}(0)}{10!}. \tag{S59}$$

*If $\phi$ is truncated at $r = 1$ then $\mathcal{S}_8(r) = K_8(r)$ and $\mathcal{S}_{10}(r) = K_{10}(r)$ for $r \geq 1$, and the corresponding short-range parts $\mathcal{L}_8 = K_8 - \mathcal{S}_8$ and $\mathcal{L}_{10} = K_{10} - \mathcal{S}_{10}$ are compactly supported in the unit ball.*

The Fourier-space transforms $\widehat{\mathcal{S}}_8(k)$ and $\widehat{\mathcal{S}}_{10}(k)$ follow from (S56) applied to $\widehat{\mathcal{S}}_6$ in (S52) and $\widehat{\mathcal{S}}_8$, respectively. The same procedure produces splittings for arbitrary $p \geq 7$. Channels with $p = 8$ (induction-like) and $p = 10$ (anisotropic dispersion) appear in polarizable force fields and in some machine-learning interatomic potentials; the construction continues unchanged to $p = 12$, relevant to Lennard-Jones-like short-range repulsion.

# 7 Numerical evaluation and notational correspondence

## 7.1 Numerical evaluation

The expressions for $\mathcal{S}_p(r)$ and $\widehat{\mathcal{S}}_p(k)$ derived in Sections 5–6 involve $\psi_0^c$ together with its derivatives, antiderivatives, and weighted integrals appearing in Theorem 1. All of these reduce to spectrally accurate operations once $\psi_0^c$ is represented by a Chebyshev or Legendre expansion of modest degree on $[0, 1]$; the relevant computational aspects of $\psi_0^c$, including stable evaluation, differentiation, antidifferentiation, and weighted quadrature, are developed in detail in Ref. [7].

For inner-loop evaluation in production molecular dynamics packages such as GROMACS[8] and LAMMPS[9], a piecewise polynomial approximation of $\mathcal{S}_p(r)$ and $\widehat{\mathcal{S}}_p(k)$ on a uniform partition of the working interval is typically preferred over repeated spectral evaluation. Such an approximation is itself a simple one-dimensional precomputation: once the spectral representation of $\psi_0^c$ is in hand, the coefficients of the piecewise polynomial approximation of each $\mathcal{S}_p$ and $\widehat{\mathcal{S}}_p$ are obtained at negligible cost and can be tabulated once per choice of bandwidth $c$, after which kernel and force evaluations reduce to Horner evaluation on the appropriate panel. Their runtime evaluation and automatic differentiation are described in Section 9.2.

## 7.2 Correspondence with the main-manuscript notation

The splittings $\mathcal{S}_p$ derived above are defined on the unit interval with the dimensionless window $\phi = \psi_0^c(r)/\psi_0^c(0)$. The main manuscript instead works on the physical cutoff $[0, r_c]$, with the long-range part written as $\mathcal{G}_p(r)/r^p$, where $\mathcal{G}_p(r)$ is the dimensionless mollifier defined in the main-manuscript Methods. The two notations are related by the substitution $r \to r/r_c$:

$$\frac{\mathcal{G}_p(r)}{r^p} = \frac{1}{r_c^p}\,\mathcal{S}_p(r/r_c), \qquad \widehat{\Phi}^{(p)}(\boldsymbol{k}) = r_c^{3-p}\,\widehat{\mathcal{S}}_p(\|\boldsymbol{k}\|\,r_c). \tag{S60}$$

Equivalently, $\mathcal{G}_p(r) = (r/r_c)^p\,\mathcal{S}_p(r/r_c)$. As a check for $p = 1$,

$$\mathcal{G}_1(r) = \frac{r}{r_c}\,\mathcal{S}_1(r/r_c) = \frac{1}{C_0}\int_0^{r/r_c} \phi(x)\,\mathrm{d}x, \tag{S61}$$

which reproduces the main-manuscript expression for $\mathcal{G}_1$. The corresponding Fourier-space form is

$$\widehat{\Phi}^{(1)}(\boldsymbol{k}) = r_c^2\,\widehat{\mathcal{S}}_1(kr_c) = \frac{4\pi\,\phi(kr_c/c)}{k^2}, \tag{S62}$$

which, on identifying $\phi$ with the main-manuscript $\psi_0^c$ (both normalized to unity at the origin), reproduces the $\widehat{\Phi}_1$ entry of the main-manuscript expression for $\widehat{\Phi}^{(p)}$. The remaining $\mathcal{G}_p$ and $\widehat{\Phi}^{(p)}$ formulas in the main manuscript follow from Propositions 2–7, Theorem 1, and (S60) in the same way.

# 8 Self-corrections and zero-frequency terms

The splittings derived above define the nonzero real- and reciprocal-space parts of each inverse-power channel. In a periodic particle system, two additional energy-level terms are needed to obtain a complete decomposition: a self-correction, which removes the interaction of each particle with its own smooth long-range cloud, and a zero-frequency term, which specifies the treatment of the $\boldsymbol{k} = \boldsymbol{0}$ mode. We derive both terms here using the same notation as in the main manuscript.

For a channel $p$, define the net latent charge and the self-weight

$$Q_{\text{net}}^{(p)} = \sum_{i=1}^{N} q_i^{(p)}, \qquad Q_{\text{self}}^{(p)} = \sum_{i=1}^{N} |q_i^{(p)}|^2. \tag{S63}$$

The physical long-range and short-range kernels are

$$\mathcal{S}_{p,r_c}(r) = \frac{1}{r_c^p}\mathcal{S}_p(r/r_c), \qquad \mathcal{L}_{p,r_c}(r) = \frac{1}{r_c^p}\mathcal{L}_p(r/r_c). \tag{S64}$$

Here $\mathcal{S}_{p,r_c}$ is the smooth LR kernel and $\mathcal{L}_{p,r_c}$ is the real-space complement. With these definitions, the periodic energy for channel $p$ can be written as

$$E^{(p)} = E_{\text{real}}^{(p)} + \frac{1}{2V}\sum_{\boldsymbol{k}\neq\boldsymbol{0}} \widehat{\Phi}^{(p)}(\boldsymbol{k}) \left|\rho^{(p)}(\boldsymbol{k})\right|^2 + E_0^{(p)} - E_{\text{self}}^{(p)}, \tag{S65}$$

where $E_{\text{real}}^{(p)}$ is the direct sum over the short-range complement $L_{p,r_c}$ excluding the singular self term, and the term $E_{\text{self}}^{(p)}$ is subtracted to remove the unwanted self-interaction of the smooth LR kernel.

**Self-correction.** The reciprocal-space representation of the smooth long-range kernel includes the term in which a particle interacts with the smooth cloud generated by itself. This unwanted contribution is $\frac{1}{2}\sum_i |q_i^{(p)}|^2 \mathcal{S}_{p,r_c}(0)$, and therefore

$$E_{\text{self}}^{(p)} = \frac{1}{2}\mathcal{S}_{p,r_c}(0) Q_{\text{self}}^{(p)} = \frac{1}{2r_c^p}\mathcal{S}_p(0) Q_{\text{self}}^{(p)}. \tag{S66}$$

The values of $\mathcal{S}_p(0)$ follow from the Taylor expansions used in Section 3:

| $p$ | $\mathcal{S}_p(0)$ |
|---|---|
| 1 | $C_0^{-1}$ |
| 2 | $-\phi''(0)/2$ |
| 3 | $-\phi''(0)/(3C_0)$ |
| 4 | $\phi^{(4)}(0)/24$ |
| 5 | $\phi^{(4)}(0)/(45C_0)$ |
| 6 | $-\phi^{(6)}(0)/720$ |

(S67)

Thus, for the PSWF $\phi(r) = \psi_0^c(r)/\psi_0^c(0)$, and with the normalization $\psi_0^c(0) = 1$ used in the main text, Eq. (S66) gives

$$\begin{aligned} E_{\text{self}}^{(1)} &= \frac{1}{2C_0 r_c} Q_{\text{self}}^{(1)}, \qquad E_{\text{self}}^{(2)} = -\frac{\psi_0^{c\,\prime\prime}(0)}{4r_c^2} Q_{\text{self}}^{(2)}, \qquad E_{\text{self}}^{(3)} = -\frac{\psi_0^{c\,\prime\prime}(0)}{6C_0 r_c^3} Q_{\text{self}}^{(3)}, \\ E_{\text{self}}^{(4)} &= \frac{\psi_0^{c\,(4)}(0)}{48 r_c^4} Q_{\text{self}}^{(4)}, \qquad E_{\text{self}}^{(5)} = \frac{\psi_0^{c\,(4)}(0)}{90 C_0 r_c^5} Q_{\text{self}}^{(5)}, \qquad E_{\text{self}}^{(6)} = -\frac{\psi_0^{c\,(6)}(0)}{1440 r_c^6} Q_{\text{self}}^{(6)}. \end{aligned} \tag{S68}$$

These terms remove only the artificial self-interaction of the smooth LR kernel and do not modify the pairwise interaction between distinct particles.

**Uniform-background correction for $p < 3$.** For $p \le 3$, a nonzero $Q_{\rm net}^{(p)}$ makes the periodic inverse-power sum divergent or convention-dependent. For $p < 3$, the standard neutralization procedure is to add a uniform background of density $-Q_{\rm net}^{(p)}/V$. Let

$$I_p = \int_{\mathbb{R}^3} \mathcal{L}_{p,r_c}(\boldsymbol{r})\, d^3\boldsymbol{r} = 4\pi \int_0^\infty r^2 \mathcal{L}_{p,r_c}(r)\, dr. \tag{S69}$$

Applying the Poisson summation formula gives

$$\sum_{\boldsymbol{n}} K_p(\boldsymbol{r}+\boldsymbol{n}) - \frac{1}{V}\int_{\mathbb{R}^3} K_p(\boldsymbol{r}')\, d^3\boldsymbol{r}' = \sum_{\boldsymbol{n}} \mathcal{L}_{p,r_c}(\boldsymbol{r}+\boldsymbol{n}) + \frac{1}{V}\sum_{\boldsymbol{k}\neq\boldsymbol{0}} \widehat{\Phi}^{(p)}(\boldsymbol{k}) e^{i\boldsymbol{k}\cdot\boldsymbol{r}} - \frac{I_p}{V}. \tag{S70}$$

Multiplying by the particle weights and including the usual factor 1/2 for double counting, the background correction is therefore

$$E_{\rm bg}^{(p)} = -\frac{\left(Q_{\rm net}^{(p)}\right)^2}{2V} I_p, \qquad p < 3. \tag{S71}$$

The integral $I_p$ is finite only for $p < 3$, because near the origin $\mathcal{L}_{p,r_c}(r) \sim r^{-p}$ and hence $r^2\mathcal{L}_{p,r_c}(r) \sim r^{2-p}$.

For $p = 1$, using $\mathcal{G}_1(r) = C_0^{-1}\int_0^{r/r_c} \phi(x)\, dx$, we obtain

$$\begin{aligned} I_1 &= 4\pi \int_0^{r_c} r\,[1-\mathcal{G}_1(r)]\, dr \\ &= 4\pi r_c^2 \int_0^1 u \left[1 - \frac{1}{C_0}\int_0^u \phi(x)\, dx\right] du \\ &= \frac{2\pi r_c^2}{C_0}\int_0^1 u^2 \phi(u)\, du. \end{aligned} \tag{S72}$$

Thus

$$E_0^{(1)} = -\frac{\left(Q_{\rm net}^{(1)}\right)^2}{2V} I_1 + E_{\rm IB} = -\frac{\pi r_c^2 \left(Q_{\rm net}^{(1)}\right)^2}{V C_0}\int_0^1 u^2\phi(u)\, du + E_{\rm IB}. \tag{S73}$$

The additional term $E_{\rm IB}$ denotes the infinite-boundary contribution associated with the conditionally convergent Coulomb sum; its value depends on the macroscopic boundary condition and summation convention[10].

For $p = 2$, since $\mathcal{G}_2(r) = 1 - \phi(r/r_c)$,

$$I_2 = 4\pi \int_0^{r_c} [1-\mathcal{G}_2(r)]\, dr = 4\pi r_c \int_0^1 \phi(u)\, du = 4\pi r_c C_0, \tag{S74}$$

and therefore

$$E_0^{(2)} = -\frac{2\pi r_c C_0}{V}\left(Q_{\rm net}^{(2)}\right)^2. \tag{S75}$$

**Zero mode for $p = 3$.** The case $p = 3$ is critical. A uniform compensating background is required to remove the non-neutrality of the periodic sum, but the background interaction $\int \mathcal{L}_{3,r_c}(\boldsymbol{r})\, d^3\boldsymbol{r}$ diverges logarithmically. We therefore define the zero mode by a finite-part convention in reciprocal space. From the PSWF-specialized Fourier multiplier,

$$\widehat{\Phi}^{(3)}(\boldsymbol{k}) = -4\pi \log\left(\frac{k r_c}{c}\right) + 4\pi \int_{k r_c/c}^1 \frac{\phi(u)-1}{u}\, du, \qquad k = \|\boldsymbol{k}\|, \tag{S76}$$

we separate the divergent $-4\pi \log k$ part and retain the finite constant:

$$\mathrm{FP}_{k\to 0}\, \widehat{\Phi}^{(3)}(\boldsymbol{k}) = 4\pi \left[\log\left(\frac{c}{r_c}\right) + \int_0^1 \frac{\phi(u)-1}{u}\, du\right]. \tag{S77}$$

The corresponding zero-frequency contribution is

$$E_0^{(3)} = \frac{2\pi \left(Q_{\mathrm{net}}^{(3)}\right)^2}{V} \left[\log\left(\frac{c}{r_c}\right) + \int_0^1 \frac{\phi(u)-1}{u}\, du\right]. \tag{S78}$$

This convention is equivalent to fixing the additive constant in the Fourier transform of the $1/r^3$ kernel. With this normalization, the PSWF-LR energy is independent of the particular reciprocal-space algorithm used to evaluate the nonzero modes.

**Finite zero mode for $p > 3$.** For $p > 3$, the inverse-power kernel is integrable at large distances in three dimensions, and the PSWF reciprocal-space multiplier is nonsingular at $\boldsymbol{k} = \boldsymbol{0}$. No uniform compensating background is required. The zero-frequency contribution is simply the finite $k = 0$ mode:

$$E_0^{(p)} = \frac{1}{2V} \widehat{\Phi}^{(p)}(\boldsymbol{0}) \left(Q_{\mathrm{net}}^{(p)}\right)^2, \qquad p > 3. \tag{S79}$$

Eqs. (S68), (S73)–(S78) and (S79) complete the periodic PSWF-LR energy decomposition used in the main manuscript.

# 9 Accuracy validation and implementation details

## 9.1 Accuracy and parameter setting for performance benchmarks

The performance benchmarks in the main text compare PSWF-LR with PME in evaluating differentiable long-range interactions. To verify that the runtime gains are not obtained by relaxing the long-range accuracy, we evaluated the force and energy errors against a high-accuracy PME reference. In this test, the short-range contribution is kept identical to the production calculation, so that the reported errors isolate the reciprocal-space long-range discretization error.

Let $\boldsymbol{F} \in \mathbb{R}^{3N}$ denote the concatenated force vector over all particles, and let $E$ denotes the energy. For a computed $\boldsymbol{F}$ and $E$, the relative errors are defined as

$$\varepsilon_F = \frac{\|\boldsymbol{F} - \boldsymbol{F}_{\mathrm{ref}}\|_2}{\|\boldsymbol{F}_{\mathrm{ref}}\|_2}, \qquad \varepsilon_E = \frac{|E - E_{\mathrm{ref}}|}{|E_{\mathrm{ref}}|}, \tag{S80}$$

where $\|\cdot\|_2$ denotes the Euclidean norm. For systems small enough for direct high-accuracy reference calculations, $\boldsymbol{F}_{\mathrm{ref}}$ and $E_{\mathrm{ref}}$ are obtained from double-precision PME calculations using the same real-space cutoff as the production runs. For larger systems, where such PME references become memory-limited, we instead use a stricter PSWF-LR calculation with target tolerance $\varepsilon_{\mathrm{target}} = 10^{-7}$ as the reference. On smaller systems where both references are affordable, we verify that the stricter PSWF-LR reference gives errors comparable to the double-precision PME reference. The production PME and PSWF-LR calculations use single-precision arithmetic and the reciprocal grids reported in Tables S1–S3. Following the Torch-PME and JAX-PME benchmark protocol[11], we use reciprocal-grid sizes that are powers of two to improve GPU efficiency. In Tables S1–S3, the real-space cutoff is set to 4 Å, the PSWF bandwidth parameter is set to $c = 13.75$, and the PME Gaussian screening parameter is set to $\alpha = 0.8$, corresponding to an estimated force error of $10^{-4}$

Table S1: Accuracy checks for differentiable PSWF-LR and PME on NaCl crystals with a cutoff of $r_c = 4$ Å. Errors are measured relative to high-accuracy double-precision reference calculations. The table reports the inverse-power exponent $p$ in $1/r^p$, the number of particles $N$, the reciprocal-grid sizes used by PSWF-LR and PME, and the relative energy $\varepsilon_E$ and force $\varepsilon_F$ errors.

| $p$ | $N$ | PSWF grid | PME grid | PSWF $\varepsilon_F$ | PME $\varepsilon_F$ | PSWF $\varepsilon_E$ | PME $\varepsilon_E$ |
|---|---|---|---|---|---|---|---|
| 1 | 512 | $16^3$ | $64^3$ | $5.466\times10^{-5}$ | $1.254\times10^{-4}$ | $3.793\times10^{-6}$ | $3.666\times10^{-6}$ |
| 1 | 4096 | $32^3$ | $128^3$ | $5.674\times10^{-5}$ | $1.270\times10^{-4}$ | $3.859\times10^{-6}$ | $3.644\times10^{-6}$ |
| 1 | 32768 | $64^3$ | $256^3$ | $6.011\times10^{-5}$ | $1.262\times10^{-4}$ | $3.868\times10^{-6}$ | $3.604\times10^{-6}$ |
| 1 | 262144 | $128^3$ | $512^3$ | $6.366\times10^{-5}$ | $1.265\times10^{-4}$ | $3.853\times10^{-6}$ | $3.581\times10^{-6}$ |
| 1 | 2097152 | $256^3$ | $1024^3$ | $6.386\times10^{-5}$ | $1.269\times10^{-4}$ | $3.865\times10^{-6}$ | $3.574\times10^{-6}$ |
| 2 | 512 | $16^3$ | $64^3$ | $1.329\times10^{-5}$ | $5.970\times10^{-5}$ | $1.189\times10^{-6}$ | $1.825\times10^{-6}$ |
| 2 | 4096 | $32^3$ | $128^3$ | $1.372\times10^{-5}$ | $6.007\times10^{-5}$ | $1.229\times10^{-6}$ | $1.882\times10^{-6}$ |
| 2 | 32768 | $64^3$ | $256^3$ | $1.463\times10^{-5}$ | $6.010\times10^{-5}$ | $1.178\times10^{-6}$ | $1.752\times10^{-6}$ |
| 2 | 262144 | $128^3$ | $512^3$ | $1.703\times10^{-5}$ | $6.055\times10^{-5}$ | $1.224\times10^{-6}$ | $1.798\times10^{-6}$ |
| 2 | 2097152 | $256^3$ | $1024^3$ | $2.246\times10^{-5}$ | $6.269\times10^{-5}$ | $1.199\times10^{-6}$ | $1.824\times10^{-6}$ |
| 3 | 512 | $16^3$ | $64^3$ | $2.401\times10^{-5}$ | $2.061\times10^{-5}$ | $1.065\times10^{-5}$ | $1.047\times10^{-6}$ |
| 3 | 4096 | $32^3$ | $128^3$ | $2.343\times10^{-5}$ | $2.097\times10^{-5}$ | $1.072\times10^{-5}$ | $1.044\times10^{-6}$ |
| 3 | 32768 | $64^3$ | $256^3$ | $2.415\times10^{-5}$ | $2.156\times10^{-5}$ | $1.072\times10^{-5}$ | $1.022\times10^{-6}$ |
| 3 | 262144 | $128^3$ | $512^3$ | $2.568\times10^{-5}$ | $2.345\times10^{-5}$ | $1.068\times10^{-5}$ | $9.819\times10^{-7}$ |
| 3 | 2097152 | $256^3$ | $1024^3$ | $3.026\times10^{-5}$ | $2.827\times10^{-5}$ | $1.074\times10^{-5}$ | $1.036\times10^{-6}$ |
| 4 | 512 | $16^3$ | $64^3$ | $3.109\times10^{-6}$ | $9.208\times10^{-6}$ | $2.041\times10^{-6}$ | $8.724\times10^{-7}$ |
| 4 | 4096 | $32^3$ | $128^3$ | $4.174\times10^{-6}$ | $9.381\times10^{-6}$ | $1.954\times10^{-6}$ | $9.618\times10^{-7}$ |
| 4 | 32768 | $64^3$ | $256^3$ | $6.442\times10^{-6}$ | $1.070\times10^{-5}$ | $2.013\times10^{-6}$ | $9.107\times10^{-7}$ |
| 4 | 262144 | $128^3$ | $512^3$ | $1.459\times10^{-5}$ | $1.640\times10^{-5}$ | $1.882\times10^{-6}$ | $1.044\times10^{-6}$ |
| 4 | 2097152 | $256^3$ | $1024^3$ | $2.137\times10^{-5}$ | $2.294\times10^{-5}$ | $1.792\times10^{-6}$ | $9.272\times10^{-7}$ |
| 5 | 512 | $16^3$ | $64^3$ | $4.201\times10^{-6}$ | $4.116\times10^{-6}$ | $5.870\times10^{-7}$ | $2.421\times10^{-7}$ |
| 5 | 4096 | $32^3$ | $128^3$ | $4.082\times10^{-6}$ | $4.515\times10^{-6}$ | $6.447\times10^{-7}$ | $3.499\times10^{-7}$ |
| 5 | 32768 | $64^3$ | $256^3$ | $6.644\times10^{-6}$ | $6.933\times10^{-6}$ | $6.450\times10^{-7}$ | $3.540\times10^{-7}$ |
| 5 | 262144 | $128^3$ | $512^3$ | $1.985\times10^{-5}$ | $1.994\times10^{-5}$ | $5.663\times10^{-7}$ | $4.337\times10^{-7}$ |
| 5 | 2097152 | $256^3$ | $1024^3$ | $2.449\times10^{-5}$ | $2.460\times10^{-5}$ | $5.510\times10^{-7}$ | $3.671\times10^{-7}$ |
| 6 | 512 | $16^3$ | $64^3$ | $4.436\times10^{-6}$ | $4.642\times10^{-6}$ | $2.943\times10^{-7}$ | $7.727\times10^{-7}$ |
| 6 | 4096 | $32^3$ | $128^3$ | $4.718\times10^{-6}$ | $4.766\times10^{-6}$ | $1.273\times10^{-7}$ | $9.379\times10^{-7}$ |
| 6 | 32768 | $64^3$ | $256^3$ | $7.282\times10^{-6}$ | $7.357\times10^{-6}$ | $2.269\times10^{-7}$ | $9.791\times10^{-7}$ |
| 6 | 262144 | $128^3$ | $512^3$ | $2.778\times10^{-5}$ | $2.766\times10^{-5}$ | $1.310\times10^{-7}$ | $1.076\times10^{-6}$ |
| 6 | 2097152 | $256^3$ | $1024^3$ | $2.907\times10^{-5}$ | $2.908\times10^{-5}$ | $9.720\times10^{-8}$ | $1.113\times10^{-6}$ |

for the Coulombic $p = 1$ case. The same parameter choices are used for the other exponents to provide a controlled comparison of the reciprocal-space treatments across $p$, rather than separately retuning each exponent.

For the NaCl crystal benchmark, high-accuracy references fit in memory for all tested accuracy-check sizes. Across $p = 1, \ldots, 6$, PSWF-LR gives force errors comparable to or lower than PME while using 64× fewer reciprocal-grid points. For SWM4-NDP water and the $LiPF_6$/DMC electrolyte, PSWF-LR preserves PME-level force and energy accuracy on the reference-accessible systems. In these replicated systems, partial cancellation of local force errors is observed across repeated cells, while accumulated single-precision rounding error becomes more visible as the replicated system size increases. We therefore use larger systems only for timing and memory-scaling comparisons, because high-accuracy PME and stricter PSWF-LR reference calculations become GPU-memory limited and the error estimates are no longer a clean measure of LR discretization error alone.

## 9.2 Fast kernel automatic-differentiation implementation

Fast kernel automatic differentiation (FKAD) is an implementation-level acceleration used to evaluate PSWF-LR kernels without changing their mathematical definition. At the code level, each special-function or quadrature-based kernel evaluation is replaced by a static piecewise-polynomial evaluation graph. The polynomial coefficients are precomputed for each PSWF bandwidth, inverse-power exponent and polynomial panel, stored as non-trainable buffers, and moved to the same device and data type as the atomic coordinates during execution. For an input scalar tensor $x$, representing a normalized pair distance, normalized Fourier magnitude or normalized spreading coordinate on a given polynomial panel, the kernel value is evaluated by Horner's rule,

$$P(x) = a_0 + x\big(a_1 + x(a_2 + \cdots + xa_m)\big), \tag{S81}$$

where the coefficients $\{a_j\}_{j=0}^{m}$ are fixed once the PSWF bandwidth, inverse-power exponent and panel are specified.

The same polynomial-evaluation routine is used for the real-space complementary pair kernel, the reciprocal-space diagonal multiplier and the atom-grid spreading window. Because these operations are expressed using differentiable tensor operations, the backward pass is generated by automatic differentiation. No separate analytical PSWF derivative, quadrature derivative or hand-written backward rule is required. The resulting forces and parameter gradients are therefore the exact derivatives of the polynomial kernels used in the forward pass. Their deviation from the derivatives of the analytical PSWF kernels is controlled by the polynomial approximation error.

In performance-critical paths, the Horner recurrence is fused with the surrounding scatter, diagonal-scaling or reduction operation. In the PyTorch and JAX implementations, this fusion is enabled through compiled computation graphs, while in the OpenMM/CUDA implementation it is implemented through fused CUDA kernels. As a result, polynomial evaluation is lowered together with the neighboring PSWF-LR operation rather than launched as a separate kernel. The implementation avoids explicitly materializing large intermediate arrays such as $\mathcal{G}_p(r), \widehat{\Phi}^{(p)}(\boldsymbol{k})$ or full spreading-weight tensors when these quantities are only needed locally for accumulation. Instead, the evaluated kernel values are accumulated directly into the real-space energy, reciprocal-space energy or grid density. This preserves end-to-end differentiability while reducing memory traffic, kernel-launch overhead and special-function evaluation cost compared with an implementation that evaluates special functions, quadratures or full kernel arrays as separate intermediate tensors.

Table S2: Accuracy checks for differentiable PSWF-LR and PME on SWM4–NDP water systems with a cutoff of $r_c = 4$ Å. Errors are measured relative to high-accuracy double-precision reference calculations. The table reports the inverse-power exponent $p$ in $1/r^p$, the number of particles $N$, the reciprocal-grid sizes used by PSWF-LR and PME, and the relative energy $\epsilon_E$ and force $\epsilon_F$ errors. Entries marked "—" were not available because the corresponding PME calculation or high-accuracy reference exceeded H200 GPU memory.

| $p$ | $N$ | PSWF grid | PME grid | PSWF $\varepsilon_F$ | PME $\varepsilon_F$ | PSWF $\varepsilon_E$ | PME $\varepsilon_E$ |
|---|---|---|---|---|---|---|---|
| 1 | 560 | $32^3$ | $128^3$ | $9.295\times10^{-6}$ | $9.322\times10^{-6}$ | $1.284\times10^{-7}$ | $5.042\times10^{-7}$ |
| 1 | 4480 | $64^3$ | $256^3$ | $2.027\times10^{-5}$ | $2.027\times10^{-5}$ | $2.475\times10^{-7}$ | $6.574\times10^{-8}$ |
| 1 | 35840 | $128^3$ | $512^3$ | $4.413\times10^{-5}$ | $4.417\times10^{-5}$ | $3.727\times10^{-7}$ | $6.574\times10^{-8}$ |
| 1 | 286720 | $256^3$ | $1024^3$ | $8.003\times10^{-5}$ | $8.007\times10^{-5}$ | $2.001\times10^{-6}$ | $1.563\times10^{-6}$ |
| 1 | 2293760 | $512^3$ | $2048^3$ | $1.598\times10^{-4}$ | — | $7.399\times10^{-8}$ | — |
| 2 | 560 | $32^3$ | $128^3$ | $1.231\times10^{-5}$ | $1.230\times10^{-5}$ | $6.041\times10^{-7}$ | $6.041\times10^{-7}$ |
| 2 | 4480 | $64^3$ | $256^3$ | $3.013\times10^{-5}$ | $3.012\times10^{-5}$ | $2.557\times10^{-7}$ | $2.557\times10^{-7}$ |
| 2 | 35840 | $128^3$ | $512^3$ | $7.016\times10^{-5}$ | $7.015\times10^{-5}$ | $1.307\times10^{-6}$ | $1.211\times10^{-6}$ |
| 2 | 286720 | $256^3$ | $1024^3$ | $1.133\times10^{-4}$ | $1.134\times10^{-4}$ | $4.173\times10^{-6}$ | $4.077\times10^{-6}$ |
| 2 | 2293760 | $512^3$ | $2048^3$ | $2.346\times10^{-4}$ | — | $8.493\times10^{-8}$ | — |
| 3 | 560 | $32^3$ | $128^3$ | $1.515\times10^{-5}$ | $1.512\times10^{-5}$ | $1.357\times10^{-6}$ | $1.357\times10^{-6}$ |
| 3 | 4480 | $64^3$ | $256^3$ | $4.312\times10^{-5}$ | $4.307\times10^{-5}$ | $2.753\times10^{-7}$ | $2.753\times10^{-7}$ |
| 3 | 35840 | $128^3$ | $512^3$ | $1.064\times10^{-4}$ | $1.065\times10^{-4}$ | $3.578\times10^{-6}$ | $3.645\times10^{-6}$ |
| 3 | 286720 | $256^3$ | $1024^3$ | $1.516\times10^{-4}$ | $1.516\times10^{-4}$ | $5.470\times10^{-6}$ | $5.403\times10^{-6}$ |
| 3 | 2293760 | $512^3$ | $2048^3$ | $3.337\times10^{-4}$ | — | $1.055\times10^{-8}$ | — |
| 4 | 560 | $32^3$ | $128^3$ | $1.733\times10^{-5}$ | $1.736\times10^{-5}$ | $2.599\times10^{-6}$ | $2.599\times10^{-6}$ |
| 4 | 4480 | $64^3$ | $256^3$ | $5.766\times10^{-5}$ | $5.776\times10^{-5}$ | $1.794\times10^{-6}$ | $1.794\times10^{-6}$ |
| 4 | 35840 | $128^3$ | $512^3$ | $1.487\times10^{-4}$ | $1.486\times10^{-4}$ | $7.958\times10^{-6}$ | $7.958\times10^{-6}$ |
| 4 | 286720 | $256^3$ | $1024^3$ | $1.900\times10^{-4}$ | $1.901\times10^{-4}$ | $4.648\times10^{-6}$ | $4.737\times10^{-6}$ |
| 4 | 2293760 | $512^3$ | $2048^3$ | $4.605\times10^{-4}$ | — | $3.808\times10^{-7}$ | — |
| 5 | 560 | $32^3$ | $128^3$ | $1.884\times10^{-5}$ | $1.890\times10^{-5}$ | $4.238\times10^{-6}$ | $4.238\times10^{-6}$ |
| 5 | 4480 | $64^3$ | $256^3$ | $7.276\times10^{-5}$ | $7.286\times10^{-5}$ | $4.904\times10^{-6}$ | $4.904\times10^{-6}$ |
| 5 | 35840 | $128^3$ | $512^3$ | $1.925\times10^{-4}$ | $1.925\times10^{-4}$ | $1.529\times10^{-5}$ | $1.540\times10^{-5}$ |
| 5 | 286720 | $256^3$ | $1024^3$ | $2.270\times10^{-4}$ | $2.271\times10^{-4}$ | $8.652\times10^{-7}$ | $7.543\times10^{-7}$ |
| 5 | 2293760 | $512^3$ | $2048^3$ | $6.213\times10^{-4}$ | — | $1.514\times10^{-6}$ | — |
| 6 | 560 | $32^3$ | $128^3$ | $1.991\times10^{-5}$ | $1.981\times10^{-5}$ | $6.648\times10^{-6}$ | $6.584\times10^{-6}$ |
| 6 | 4480 | $64^3$ | $256^3$ | $8.777\times10^{-5}$ | $8.770\times10^{-5}$ | $1.026\times10^{-5}$ | $1.033\times10^{-5}$ |
| 6 | 35840 | $128^3$ | $512^3$ | $2.360\times10^{-4}$ | $2.361\times10^{-4}$ | $2.639\times10^{-5}$ | $2.639\times10^{-5}$ |
| 6 | 286720 | $256^3$ | $1024^3$ | $2.630\times10^{-4}$ | $2.630\times10^{-4}$ | $7.294\times10^{-6}$ | $7.294\times10^{-6}$ |
| 6 | 2293760 | $512^3$ | $2048^3$ | $7.650\times10^{-4}$ | — | $4.247\times10^{-6}$ | — |

Table S3: Accuracy checks for differentiable PSWF-LR and PME on $LiPF_6$/DMC electrolyte systems with a cutoff of $r_c = 4.0$ Å. Errors are measured relative to high-accuracy double-precision reference calculations. The table reports the inverse-power exponent $p$ in $1/r^p$, the number of particles $N$, the reciprocal-grid sizes used by PSWF-LR and PME, and the relative energy $\varepsilon_E$ and force $\varepsilon_F$ errors. Entries marked "—" were not available because the corresponding PME calculation or high-accuracy reference exceeded H200 GPU memory.

| $p$ | $N$ | PSWF grid | PME grid | PSWF $\varepsilon_F$ | PME $\varepsilon_F$ | PSWF $\varepsilon_E$ | PME $\varepsilon_E$ |
|---|---|---|---|---|---|---|---|
| 1 | 10062 | $128^3$ | $512^3$ | $1.674\times10^{-4}$ | $5.782\times10^{-4}$ | $5.435\times10^{-6}$ | $1.723\times10^{-5}$ |
| 1 | 80496 | $256^3$ | $1024^3$ | $1.676\times10^{-4}$ | $5.784\times10^{-4}$ | $5.435\times10^{-6}$ | $1.723\times10^{-5}$ |
| 1 | 643968 | $512^3$ | $2048^3$ | $1.680\times10^{-4}$ | — | $5.435\times10^{-6}$ | — |
| 1 | 5151744 | $1024^3$ | $4096^3$ | $1.690\times10^{-4}$ | — | $5.435\times10^{-6}$ | — |
| 2 | 10062 | $128^3$ | $512^3$ | $3.083\times10^{-5}$ | $4.559\times10^{-4}$ | $1.345\times10^{-6}$ | $2.752\times10^{-5}$ |
| 2 | 80496 | $256^3$ | $1024^3$ | $3.173\times10^{-5}$ | $4.562\times10^{-4}$ | $1.255\times10^{-6}$ | $2.752\times10^{-5}$ |
| 2 | 643968 | $512^3$ | $2048^3$ | $3.472\times10^{-5}$ | — | $1.345\times10^{-6}$ | — |
| 2 | 5151744 | $1024^3$ | $4096^3$ | $4.183\times10^{-5}$ | — | $1.345\times10^{-6}$ | — |
| 3 | 10062 | $128^3$ | $512^3$ | $1.825\times10^{-5}$ | $3.157\times10^{-4}$ | $6.347\times10^{-6}$ | $3.020\times10^{-5}$ |
| 3 | 80496 | $256^3$ | $1024^3$ | $2.015\times10^{-5}$ | $3.159\times10^{-4}$ | $6.235\times10^{-6}$ | $3.009\times10^{-5}$ |
| 3 | 643968 | $512^3$ | $2048^3$ | $2.647\times10^{-5}$ | — | $6.347\times10^{-6}$ | — |
| 3 | 5151744 | $1024^3$ | $4096^3$ | $3.863\times10^{-5}$ | — | $6.459\times10^{-6}$ | — |
| 4 | 10062 | $128^3$ | $512^3$ | $1.515\times10^{-5}$ | $1.992\times10^{-4}$ | $9.535\times10^{-6}$ | $2.686\times10^{-5}$ |
| 4 | 80496 | $256^3$ | $1024^3$ | $1.315\times10^{-5}$ | $1.995\times10^{-4}$ | $9.535\times10^{-6}$ | $2.686\times10^{-5}$ |
| 4 | 643968 | $512^3$ | $2048^3$ | $2.394\times10^{-5}$ | — | $9.607\times10^{-6}$ | — |
| 4 | 5151744 | $1024^3$ | $4096^3$ | $4.115\times10^{-5}$ | — | $9.678\times10^{-6}$ | — |
| 5 | 10062 | $128^3$ | $512^3$ | $9.801\times10^{-6}$ | $1.155\times10^{-4}$ | $1.011\times10^{-5}$ | $2.092\times10^{-5}$ |
| 5 | 80496 | $256^3$ | $1024^3$ | $1.442\times10^{-5}$ | $1.161\times10^{-4}$ | $1.030\times10^{-5}$ | $2.083\times10^{-5}$ |
| 5 | 643968 | $512^3$ | $2048^3$ | $2.724\times10^{-5}$ | — | $1.011\times10^{-5}$ | — |
| 5 | 5151744 | $1024^3$ | $4096^3$ | $4.732\times10^{-5}$ | — | $1.011\times10^{-5}$ | — |
| 6 | 10062 | $128^3$ | $512^3$ | $1.037\times10^{-5}$ | $6.313\times10^{-5}$ | $1.937\times10^{-6}$ | $1.451\times10^{-5}$ |
| 6 | 80496 | $256^3$ | $1024^3$ | $1.617\times10^{-5}$ | $6.461\times10^{-5}$ | $1.693\times10^{-6}$ | $1.433\times10^{-5}$ |
| 6 | 643968 | $512^3$ | $2048^3$ | $3.077\times10^{-5}$ | — | $1.937\times10^{-6}$ | — |
| 6 | 5151744 | $1024^3$ | $4096^3$ | $5.362\times10^{-5}$ | — | $1.937\times10^{-6}$ | — |


\end{document}